\documentclass[useAMS,usenatbib,usegraphicx]{mn2e}
\usepackage{xspace}
\usepackage{lscape}
\usepackage{appendix}
\usepackage{amsfonts}
\usepackage{amsmath,amssymb,epsfig,natbib,bm}
\usepackage{subfloat}

\newcommand{\halpha}{H$_\alpha$\xspace}

\begin{document}

\title[Foreground Analysis of 7-year WMAP data]{Foreground Analysis Using Cross-Correlations of External
  Templates on the 7-year \emph{WMAP} data}
\author[T. Ghosh et al.]{Tuhin Ghosh$^{1}$\thanks{tuhin@iucaa.ernet.in}, 
  A.J. Banday$^{2,3}$,
  Tess Jaffe$^{2,3}$,
  Clive Dickinson$^{4}$,
  Rod Davies$^{4}$, \newauthor
  Richard Davis$^{4}$ and 
  Krzysztof Gorski$^{5,6,7}$
  \\
$^{1}$IUCAA, Post Bag 4, Ganeshkhind, Pune-411007, India\\  
$^{2}$ Universit\'e de Toulouse; UPS-OMP; IRAP;  Toulouse, France\\
$^{3}$ CNRS; IRAP; 9 Av. colonel Roche, BP 44346, F-31028 Toulouse cedex 4, France\\
$^{4}$ Jodrell Bank Centre for Astrophysics, Alan Turing Building, School of Physics \& Astronomy, The University of Manchester, \\
Oxford Road, Manchester, M13 9PL, UK\\
$^{5}$ Jet Propulsion Laboratory, 4800 Oak Grove Drive, Pasadena CA 91109, USA\\
$^{6}$ California Institute of Technology, Pasadena, CA 91125, USA\\
$^{7}$ Warsaw University Observatory, Aleje Ujazdowskie 4, 00-478 Warszawa, Poland\\
}

\date{\today}

\maketitle

\begin{abstract}
\emph{WMAP} data when combined with ancillary data on free-free, synchrotron and dust allow an improved understanding of the spectrum
of emission from each of these components. Here we examine the sky variation at intermediate and high latitudes using a cross-correlation 
technique. In particular, we compare the observed emission in several large partitions of the sky plus 33 selected sky regions to three 
``standard'' templates. The regions are selected using a criterion based on the morphology of these template maps.

The synchrotron emission shows evidence of steepening between GHz frequencies and the \emph{WMAP} bands. There are indications of 
spectral index variations across the sky but the current data are not precise enough to accurately quantify this from region-to-region.

The emission correlated with the \halpha template shows clear evidence of deviation from a free-free spectrum. The emission can be 
decomposed into a contribution from both free-free and spinning dust in the warm ionised medium of the Galaxy. The derived free-free 
emissivity corresponds to a  mean electron temperature of $\sim 6000$~K, although the value depends critically on the impact of dust
absorption on the  \halpha intensity. The WIM spinning dust emission has a peak emission in intensity in the range 40--50~GHz. 

The anomalous microwave emission associated with dust is detected at high significance in most of the 33 fields studied. The
anomalous emission correlates well with the Finkbeiner et al. (1999) model 8 predictions (FDS8) at 94 GHz, and is well described 
globally by a power-law emission model with an effective spectral index between 20 and 60~GHz of $\beta \approx -2.7$.  It is clear that
attempts to explain the emission by spinning dust models require multiple components, which presumably relates to a complex mix of 
emission regions along a given line-of-sight. An enhancement of the thermal dust contribution over the FDS8 predictions by a factor 
$\sim 1.2$ is required with such models. Furthermore, the emissivity varies by a factor of $\sim 50\%$ from cloud to cloud relative to 
the mean.

The significance of these results for the correction of CMB data for Galactic foreground emission is discussed.

\end{abstract}


\onecolumn

\section{Introduction}

A major goal of observational cosmology is to determine those parameters that describe the Universe. Observations of the Cosmic
Microwave Background (CMB) at frequencies in the range 20 -- 200 GHz provide unique data to achieve this by establishing the statistical
properties of temperature (and polarisation) measurements. However, an impediment to such studies arises due to foreground emission in our
own Galaxy from at least three sources -- synchrotron, free-free and thermal dust emission. As CMB studies move to ever higher precision
it is essential to determine the properties of these components to similarly high accuracy. Indeed, although the combined foreground
level reaches a minimum in this frequency range ($\nu \approx 70$\,GHz), it remains the dominant signal over large fractions of the sky. 
Of particular relevance to this discussion is the fact that each of the foreground components has a spectral index that varies from one 
line of sight to another so using a single spectral index can lead to significant uncertainties in the corrections required. It is 
therefore essential to study the nature of the Galactic signal at microwave wavelengths in its own right.

All-sky observations by the Wilkinson Microwave Anisotropy Probe \citep[\emph{WMAP},][]{Bennett_WMAP1:2003a} at the 5 frequencies of 
23, 33, 41, 61 and 94 GHz can provide the basis for improving our understanding of local foregrounds.  By comparing these maps with 
templates for synchrotron, free-free and dust emission, made at frequencies where the specific emission mechanisms dominate, it is
possible to clarify important properties of the emission. Indeed, new insights into the nature of the Galactic diffuse emission have 
arisen from studies of the \emph{WMAP} data, including the detection of several unexpected new contributions.

Anomalous dust-correlated emission \citep{Leitch:1997} was originally observed in the \emph{COBE}-DMR data \citep{Kogut_DMR:1996a} but 
was thought to be due to free-free emission. Draine \& Lazarian (1998a,b) shifted attention to the dust itself as the source of 
emission through a mechanism now referred to as ``spinning dust'', or dipole emission from very rapidly spinning grains. A reanalysis 
of the intermediate and high Galactic latitude data taken by \emph{COBE}-DMR and supplemented by 19~GHz observations 
\citep{Banday_DMR:2003} led to evidence for dust at intermediate Galactic latitudes emitting a spectrum consistent with the the form 
expected for spinning dust, specifically a hint of a turnover at frequencies below $\sim$ 20~GHz. However, study of the \emph{WMAP} data 
has allowed further refinement of our understanding of the emission. Cross-correlation of the K-band data with observations at 15~GHz 
\citep{dOC:2004} again indicated a plateau or downturn in foreground emission inconsistent with a free-free or synchrotron origin. 
\citet{Lagache_WMAP:2003} compared the \emph{WMAP} data to HI column density measurements,revealing an increase in emission with 
decreasing density, suggesting that the anomalous emission is connected to small, transient heated grains. In addition, a series of 
papers \citep{Finkbeiner_WMAP1:2004, Dobler_WMAP3:2008a, Dobler_WMAP3:2008b} have strongly confirmed the presence of anomalous dust 
emission in the \emph{WMAP} data, and claim to have found evidence of such a component from the diffuse warm ionised medium (WIM) of 
the Galaxy.  Specifically, the correlation with a H$_{\alpha}$ template, commonly utilised as a proxy for the free-free emission, is not 
consistent with the spectrum expected for ionised gas, and a broad bump is seen peaking towards $\sim$ 40~GHz. Subsequently, 
\citet{Dobler_WMAP5:2009} have attempted to constrain specific spinning dust parameters such as the density and typical electric dipole 
moment of the grains. Recently, \cite{Peel:2011} have shown that the K-band dust-correlated component is not strongly affected 
by the inclusion of a 2.3\,GHz synchrotron template, reducing the possibility of a flat-spectrum synchrotron component.

Finally, the so-called \emph{WMAP}-haze was identified by \citet{Finkbeiner_WMAP1:2004} although it was already clearly apparent
in the foreground residuals of \citet{Bennett_WMAP1:2003b} and subsequently in the SMICA analysis of \citet{Patanchon:2005}. The
initial physical interpretation of the haze was that it was associated with free-free emission from hot gas in excess of $10^5$\,K, 
but this was refuted on the basis of the lack of associated X-ray emission. The detection of the haze relies upon the use of standard 
templates to remove known foreground emission utilising spatially independent spectra over the entire (high-latitude) sky. It has been 
argued that such a crude approximation to the true behaviour of the foreground emission at microwave frequencies may well lead to 
unphysical results. Indeed, \citet{Gold_WMAP7:2011} find that a spatial variation of spectral index of order 0.25 between 408 MHz and 
K-band is sufficient to reproduce the haze amplitude. Furthermore, a corresponding polarised signal was found to be absent from the 
\emph{WMAP} data. However, the lack of polarised emission can be explained by the entanglement of the Galactic magnetic field towards 
the Galactic centre, leading to a lower polarisation fraction there as compared to the outer Galaxy. Nevertheless, the microwave haze 
remains an active area of research, particularly given its possible association with a gamma-ray counterpart observed in the 
\emph{Fermi} data \citep{Dobler_Fermi:2010}.

In this paper, we characterise the spatial variation in the foreground emission in terms of the various emission mechanisms noted above by
introducing a new partition of the sky into morphologically selected regions.  In previous work \citep[hereafter D06]{Davies_WMAP1:2006},
our approach was to identify regions away from the Galactic plane which were expected to be dominant in one of the three foreground
components, free-free, synchrotron or dust and to derive the spectrum for each component. Five regions covering angular scales of 
$3\deg$ to $20\deg$ were chosen for each component, based on foreground template maps, making 15 in all. Here, we generalise this 
approach and introduce an algorithm to define a set of 35 regions for further study. The selection is intended to minimise the potential 
cross-talk between the various physical components and to select regions over which the spectral behaviour is uniform thus supporting 
the use of a template-based comparison. We use a well-known and understood cross-correlation technique to undertake the analysis. 

The paper is organised as follows. Section~\ref{sec:data} describes the \emph{WMAP} data and foreground templates used in this analysis 
while Section~\ref{sec:masks} defines the regions of interest for investigation. The methodology of the cross-correlation analysis is 
outlined in Section~\ref{sec:methods} and the corresponding results are presented in Section~\ref{sec:results}.  Model-dependent spectral
fits for each component are considered in Section~\ref{sec:modelfits} and overall conclusions given in Section~\ref{sec:conclusions}.
Appendix~\ref{app:halpha} discusses in detail issues related to the \halpha template used in the analysis, Appendix~\ref{app:regions} 
defines the detailed method for partitioning the sky, and finally Appendix~\ref{app:complete_results} tabulates all of the template-fit
coefficients for all the regions analysed.

\section{Data used in the analysis}\label{sec:data}

\subsection{\emph{WMAP} data}
\label{sec:wmap_data}

We use the \emph{WMAP} seven-year data \citep{Jarosik_WMAP7:2011} provided in the  HEALPix\footnote{http://healpix.jpl.nasa.gov}
pixelisation scheme \citep{Gorski_HEALPix:2005} with associated resolution $N_{side}$=512 that can be obtained from the Legacy Archive 
for Microwave Background Data Analysis (LAMBDA) website\footnote{\texttt{http://lambda.gsfc.nasa.gov}}. The \emph{WMAP} satellite has 
10 so-called differencing assemblies (DAs) distributed over five frequencies from $\sim$23~GHz (K-band) to $\sim$ 94~GHz
(W-band), and with frequency-dependent resolutions ranging from approximately $0^{\circ}\!.93$ up to $0^{\circ}\!.23$. The K- and Ka-bands
have only one DA each, Q and V band have two each, while W-band has four. 

Multiple DAs at each frequency for the Q-, V- and W-bands are combined using simple averaging to generate a single map per frequency 
band. The data are then smoothed to a common resolution of $3^{\circ}$ (after deconvolving the effective azimuthally symmetric beam 
response for each map) and degraded to HEALPix $N_{side}$=64. We perform our analysis at this resolution, rather than the more typical 
$1^{\circ}$ studies, to account for the full covariance properties of the signal (see section~\ref{sec:methods}). We have also identified 
a problem with using the \halpha data at higher resolution, as discussed further in section~\ref{sec:template_data} and 
Appendix~\ref{app:halpha}. Finally, we convert the data to brightness (antenna) temperature units from thermodynamic temperature 
since this is more appropriate for studying the spectral dependences of foregrounds.

\subsection{Foreground templates}
\label{sec:template_data}
\defcitealias{Davies_WMAP1:2006}{D06}

Each of the templates used here has been discussed at length in many prior studies. Thus we provide only a brief review and refer the
reader to \citetalias{Davies_WMAP1:2006} and references therein.

\vspace{3mm}

\noindent {\it Synchrotron Template:}
The synchrotron emission arises mainly due to the acceleration of relativistic cosmic ray electrons in the Galactic magnetic field. 
Thus, the brightness temperature of the synchrotron spectrum depends on the energy spectrum of the cosmic ray electrons and the strength 
of the magnetic field. For an electron population with energy distribution $N(E)$ given by a power law, $N(E) dE \propto E^{-\delta} dE$ ,
the brightness temperature of the ensemble synchrotron spectrum is also a power law given by, $T(\nu) \propto \nu^{\beta_s}$,
where $\beta_s$ is related to spectral index of the energy spectrum $\delta$ by the relation $\beta_s =(\delta +3)/2$.
At very low frequency ($<$ 1~GHz), the observed sky signal is dominated by the synchrotron emission from our Galaxy, and is least 
contaminated by free-free emission, at least away from the Galactic plane. The spectral behaviour changes with frequency as a result of 
the details of cosmic-ray electron propagation. Indeed, the mean spectral index changes typically from $\beta_s=-2.55$ to $-2.8$ at 38 
and 800 MHz respectively \citep{Lawson:1987}. Moreover, $\beta_s$ varies across the sky and a range of values from $-2.3$ to $-3.0$ has 
been determined by \citet{Reich:1988b} between 408 and 1420~MHz data. The 408~MHz radio continuum survey of \citet{Haslam:1982} provides 
a full sky template for synchrotron studies at an angular resolution close to $1^{\circ}$. For our purpose, we use the version of the data 
provided by the LAMBDA site.

\vspace{3mm}

\defcitealias{DDD:2003}{DDD} 
\defcitealias{FDS:1999}{FDS}
\noindent {\it Free-Free Template:}
Free-free emission arises in regions of ionised hydrogen and is produced by free electrons scattering from ions without capture. 
The intensity spectrum of the free-free emission depends on the electron temperature and emission measure (EM) which is related to the 
number density of electrons ($n_e$) along a given line of sight as EM=$\int n_e^2 dl$. The optical \halpha  recombination line results 
from the capture of free electrons by a proton nucleus and is therefore also related to the EM. A high resolution full sky map of 
\halpha emission can then be used as a good tracer of the free-free continuum emission at radio wavelengths. There is a well-defined 
relationship between the \halpha intensity and radio-continuum brightness temperature with a strong dependence on frequency and modest 
dependence on electron temperature $T_{e}$. \citet[hereafter DDD]{DDD:2003} generated a full-sky \halpha map as a composite of 
\emph{WHAM} data in the northern sky \citep{Reynolds_WHAM:1998, Haffner_WHAM:1999} and the \emph{SHASSA} survey 
\citep{Gaustad_SHASSA:2001} in the south. We refer the reader to their paper for more complete details.

A significant uncertainty when using the \halpha template is the absorption of \halpha by foreground dust. The absorption can be
estimated by using the dust column density maps of \citet[hereafter SFD]{SFD:1998}, and the parameter $f_{d}$ corresponding to the 
fraction of dust in front of the \halpha that causes the absorption. \citetalias{DDD:2003} showed that for Galactic longitudes
in the range $l$ = 30$^{\circ}$ -- 60$^{\circ}$ and latitudes $\mid\, b\, \mid$ = 5$^{\circ}$ -- 15$^{\circ}$, $f_{d} \sim$ 0.3. However, for
local high latitude regions such as Orion and the Gum nebula there is little or no absorption by dust. This was also the interpretation
favoured for high latitude \halpha template fits to the \emph{COBE}-DMR data in \citet{Banday_DMR:2003}. We adopt a default
value $f_{d}\ =\ 0$ in this paper, but consider the impact of varying the absorption parameter on our results. We note that there is an
effective degeneracy between $f_{d}$ and electron temperature $T_{e}$ for the interpretation of the fits between the \emph{WMAP} data and
the \halpha template, and that the majority of such analyses assume that the dust is coextensive with the \halpha emission, i.e.  
$f_{d} \sim$ 0.5. 

An alternative \halpha template has been assembled by \citet{Finkbeiner:2003} using additional data that provides higher angular
resolution on limited areas of the sky.  We have examined and compared these two templates in detail and discuss the comparison in
Appendix~\ref{app:halpha}. In summary, we find that there are inconsistencies in the template fit results obtained with the two
templates that are resolved only by smoothing the data to a resolution lower than $1^{\circ}$. For our main results we choose a resolution
of $3^{\circ}$. Such differences were already visible in the results of \citetalias{Davies_WMAP1:2006}.  These may be connected with a 
subtle interplay between artefacts in the \halpha templates and the template fitting method we utilise that leads to unstable estimates
of the amplitude of the correlated emission at a given frequency. We use exclusively the \citetalias{DDD:2003} template fits in this work
since the template was explicitly constructed to have uniform resolution.

\vspace{3mm}

\noindent {\it Dust Template:}
Thermal dust emission is the dominant foreground at frequencies of 100 -- 1000 GHz. Its emissivity is generally modelled by a modified
blackbody spectrum, $I(\nu) \propto\ \nu^{\alpha_{d}}\, B_{\nu}(T_{d})$, where $\alpha_{d}$ is the emissivity index and $B_{\nu}(T_{d})$ is the
blackbody emissivity at a dust temperature $T_{d}$. \citet[hereafter FDS]{FDS:1999} predicted the thermal dust contribution at microwave 
frequencies from a series of models based on the \emph{COBE}-DIRBE 100 and 240 $\mu m$ maps tied to \emph{COBE}-FIRAS spectral data.
The preferred model 8 (FDS8) has an effective power-law spectral index in antenna temperature of $\beta_{d} \approx +1.55$ over the 
\emph{WMAP} frequencies. We use the FDS8 predicted emission at 94~GHz as our reference template for dust emission, and demonstrate that 
it also traces the anomalous component that predominates in the 10--100\,GHz frequency range. Fits of the template to the \emph{WMAP} 
data will help to constrain the spectral dependence of the dust-correlated foregrounds, and allow comparison to spinning dust models.
Nevertheless, we note that an alternative for the spinning dust component may ultimately be required. Indeed, 
\citet{Finkbeiner_WMAP1:2004} proposed that an anomalous dust template could be better constructed  using FDS8$\times {\rm T_{d}}^2$, 
however \citetalias{Davies_WMAP1:2006} find little evidence for this. More recently, \citet{Ysard_AME:2010} have found improved 
correlation between the AME and the 12~$\mu$m brightness divided by the intensity of the interstellar radiation field in 27 fields of 
area 5$^{\circ}$ squared. However, such a template is not available at present for the full-sky given the difficulty in adequately 
cleaning the 12~$\mu$m data from residual zodiacal emission.

\subsection{Masks for Global fits}
\label{sec:masks}
\defcitealias{SFD:1998}{SFD}

\begin{figure*}
\begin{center}
\begin{tabular}{cccc}
\epsfig{file=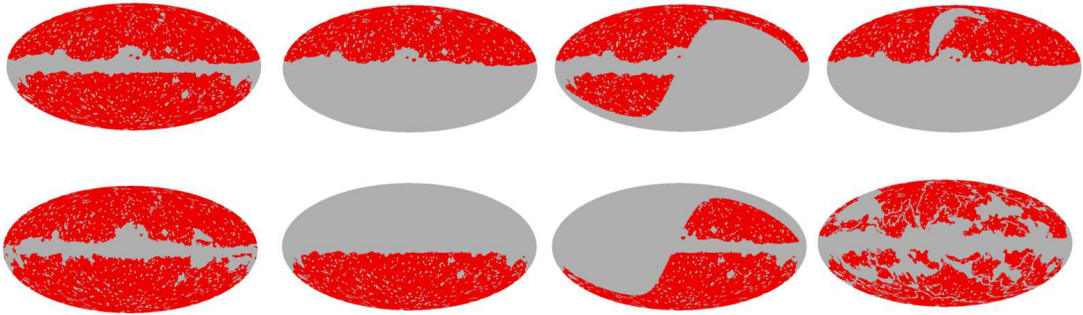,width=0.83\linewidth,angle=0,clip=} 
\end{tabular}
\caption{A representative set of global masks used on this analysis. Masked areas are in blue. From left to right and from top to bottom : 
  (1) EBV mask with point source mask included, 
  (2) Galactic North  Mask,
  (3) Ecliptic North Mask,
  (4) Galactic North Mask with North Polar Spur(NPS) removed
  (5) KQ85 mask with point source mask included,
  (6) Galactic South Mask
  (7) Ecliptic South Mask,
  (8) For reference, this shows the effective mask that would result if all of
  the 33 regions were combined and the EBV mask subsequently imposed.
  The individual regions are shown in Fig.~\ref{fig:regions}.
  Note that this mask has not been used for analysis.
}
\label{fig:globalmasks}
\end{center}
\end{figure*}

In order to examine the largest scale properties of the foregrounds, we have performed fits over large sky areas defined by two basic 
masks (EBV\footnote{The dust absorption correction is calibrated against an optical reddening law, or visual extinction, based on the 
(B - V) colours of a sample of galaxies, hence the name EBV for the mask.} and KQ85) and the intersection of the former with the Northern
and Southern hemispheres defined in various reference frames (Galactic and ecliptic). We have also tested the sensitivity of the Northern
hemisphere fits to the inclusion or otherwise of the North Polar Spur (NPS) region defined with reference to the 408 MHz radio survey. A 
representative subset of the effective masks are shown in Fig.~\ref{fig:globalmasks}. We collectively refer to analyses made using these 
masks as global fits.

The KQ85 mask was developed by the WMAP team for the primary purpose of cosmological analysis. Since we are interested in investigating
the properties of the foreground sky, our default EBV base mask is not as conservative as the KQ85 mask that may exclude some areas of
interest in our analysis.

We define the EBV mask by rejecting that fraction of the sky for which the dust absorption in the \halpha survey is unreliable. In 
practice, this corresponds to the sky area where the absorption, as related to the dust optical depth maps from \citetalias{SFD:1998} and
the conversion factors in \citetalias{DDD:2003}, exceeds 1 magnitude. This analysis is carried out at 3 degree FWHM resolution, then the 
extinction map is downgraded to a HEALPix pixel resolution $N_{side}=64$. This is then merged with the downgraded WMAP7 processing and 
point source masks in which any partially filled low resolution pixels are explicitly masked. Some additional pixels are also excluded 
corresponding to regions around the LMC, SMC and high-latitude clusters where some signal appears to leak outside the mask when applied 
to the WMAP7 K-band data smoothed to 3 degrees FWHM.

For reference, we have included some fits based on the KQ85 mask. The Galactic plane part of the mask was smoothed to 3 degree resolution,
then downgraded to $N_{side}=64$ and those pixels with a value less than 0.95 were then rejected (set to 0) and those above were accepted 
(set to 1). This was subsequently merged with the point source mask as used to create the EBV mask.

Finally, the NPS may impact the synchrotron spectral indices determined in the Northern hemisphere. To test this, we apply a crude
masking of the NPS. This is based on the observation that the 408 MHz map smoothed to 3 degrees resolution and divided by the cosecant of
the Galactic latitude (an approximation to remove the path-length dependence of the emission within the Galaxy), shows an obvious
enhancement towards the NPS compared to the rest of the sky. Thresholds are then applied to the map to eliminate data until
there are no pixels in the North with values exceeding the maximum in the South (for Galactic latitudes less than $-5$ degrees). The
resulting mask is certainly visually compelling as seen in Fig.~\ref{fig:globalmasks}.

\subsection{Regions Definition}\label{sec:region_def}

\begin{figure}
\begin{center}
\epsfig{file=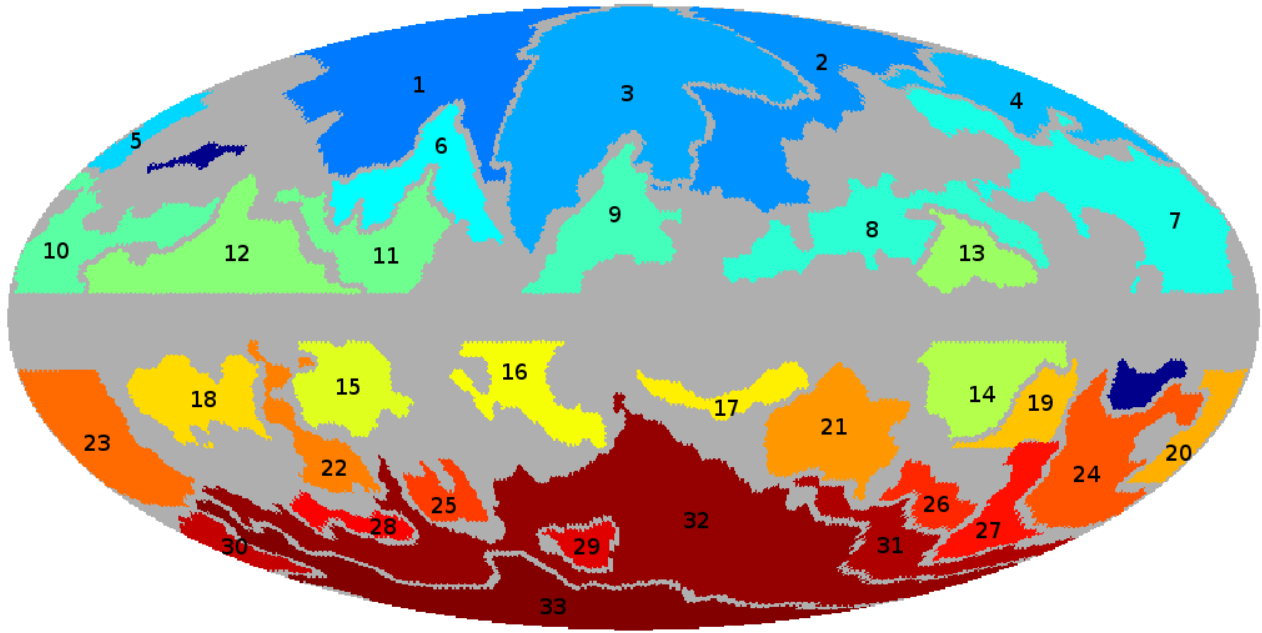,height=0.6\linewidth,angle=0,clip=0}
\caption{The set of regions defined in Appendix~\ref{app:regions}. 35 candidate regions have been identified, but two (here in dark 
blue and not enumerated) were considered to be too small after the EBV mask has also been applied for analysis purposes. 33 regions 
are analysed in this paper.}
\label{fig:regions}
\end{center}
\end{figure}

In this paper, we define a new set of regions generalising the approach from \cite{Davies_WMAP1:2006}. In this previous analysis, 
regions were selected on the basis that one of the three foregrounds (free-free, dust or synchrotron emission as traced by standard 
templates) was dominant in each region. We extend this morphological definition in Appendix~\ref{app:regions}, the result of which is 
the set of 35 candidate regions seen in Fig.~\ref{fig:regions}. For analysis purposes, the regional mask is combined with the EBV mask, 
and this renders two candidate regions sufficiently small that we elect to omit them, leaving 33 regions for further study. These two 
regions can be seen in the figure but are not enumerated.  Note that region 3 contains the NPS, but we do not attempt to suppress its 
impact by application of the NPS mask. In addition, our analysis is largely insensitive to the \emph{WMAP} haze given that the sky areas
where this is most prominent are not selected by our region definition scheme. This should be of little surprise given that the 
definition is based on the 408 MHz data in part, and haze emission is not seen therein. Various properties of the regions are
specified in Table~\ref{tab:summary_regions}.

We stress that we do not claim any absolute optimality of these regions for analysis. Indeed, they are by no means unique,
and many alternate methods for regional definition are plausible. In fact, we recognise that, although we have applied a cosecant
flattening to the data to allow their definition, such a flattening has not been applied in analysis and the fits may still be sensitive
to the largest-scale Galactic emission as a function of latitude. Given the morphology of our regions, and the large range of
latitudes that some cover (particularly towards the North Galactic pole) this could be a potential consequence of our approach.
However, if there were a strong effect, it would be seen by comparing intermediate latitudes with high latitudes.

\section{Methods}
\label{sec:methods}

The aim of this work is to cross-correlate datasets at different wavelengths in regions of the sky defined by the morphology of the
templates described in \S~\ref{sec:data}. The cross--correlation (C--C) method used here is a least--squares fit of one map to one or
more templates.  We perform the analysis at a resolution of 3$^{\circ}$ FWHM on sky maps degraded to a HEALPix pixel resolution
specified by $N_{side} = 64$ rather than the more typical $1^{\circ}$ analysis seen, for example, in \citetalias{Davies_WMAP1:2006}. This
allows us to take account of the full covariance properties of the signal, and also provides a solution to some issues related to the the
\halpha data at higher resolution, as discussed further in Appendix~\ref{app:halpha}.  In this way, we can include information
about the CMB through its signal covariance rather than having to correct for it.

It should be noted that in the high signal-to-noise regime probed by \emph{WMAP7} data, the CMB covariance
dominates the error budget in the fits. A different approach, as advocated for example by \citet{Finkbeiner_WMAP1:2004, Dobler_WMAP3:2008a, Dobler_WMAP3:2008b}, removes an estimate of the CMB signal from the data before cross-correlation.
Such an internal linear combination (ILC) map typically corresponds to a linear combination of the data at the 5 \emph{WMAP} frequencies.
However, in principle this also correlates the fits between frequencies due to the common noise properties of the ILC map.
More seriously, since the ILC contains foreground residuals, the subtraction changes the relative levels of foreground emissions at 
each frequency depending on the spectral characteristics of a given component. An approach to rectify this has been specified in
\citet{Dobler_WMAP3:2008a}. However, we prefer to retain the approach from \citetalias{Davies_WMAP1:2006} in this analysis.

\subsection{Cross - Correlation Analysis}

The cross--correlation measure, $\alpha$, between a data vector, ${\bf d}$ and a template vector ${\bf t}$ can be measured by
minimising:

\begin{equation}
\chi^2 = ({\bf d}-\alpha {\bf t})^T \cdot {\bf M}^{-1}_{SN}\cdot ({\bf d}-\alpha {\bf t}) = {\bf \tilde{d}}^T \cdot {\bf M}^{-1}_{SN}\cdot {\bf \tilde{d}}
\end{equation}

where ${\bf M}_{SN}$ is the covariance matrix including both signal and noise for the template--corrected data vector ${\bf \tilde{d}}
\equiv {\bf d} - \alpha {\bf t}$.  Solving for $\alpha$ then becomes:

\begin{equation}
\alpha = \frac{ {\bf t}^T\cdot{\bf M}^{-1}_{SN}\cdot {\bf d} }{ {\bf t}^T\cdot{\bf M}^{-1}_{SN}\cdot {\bf t} } \label{eq:cc_basic}
\end{equation}

To compare multiple template components ${\mathbf t}_j$, {\it e.g.}, different foregrounds, to a given dataset, the problem becomes a
matrix equation. In the case where we have $N$ different foreground components, we end up with the simple system of linear equations 
${\bf Ax}={\bf b}$,where
\begin{gather}
A_{kj}=\mathbf{t}^T_k \cdot \mathbf{M}_{\textrm{SN}}^{-1} \cdot \mathbf{t}_j, \notag \\
b_k = \mathbf{t}^T_k \cdot \mathbf{M}_{\textrm{SN}}^{-1} \cdot \mathbf{d} \notag, \\
x_k =\alpha_k.
\end{gather}
When only one template is present, this reduces to equation~\ref{eq:cc_basic} above. 

The signal covariance is that for theoretical CMB anisotropies, $M^S_{ij} = \frac{1}{4\pi} \sum^\infty_{\ell=0} (2\ell+1) C_{\ell}
B^2_{\ell} B^2_{\ell, pix} P_{\ell}(\hat{n}_i\cdot\hat{n}_j) $, where $B_{\ell}$ is the beam window function for a Gaussian beam of $3\degr$ 
FWHM, and $B_{\ell, pix}$ is the HEALPix window function. The power spectrum, $C_{\ell}$, is taken from the \emph{WMAP} best fit 
$\Lambda$CDM power law spectrum \citep{Jarosik_WMAP7:2011}. The noise covariance is determined from the uncorrelated pixel noise
as specified for each pixel in the \emph{WMAP} data, and subsequently convolved and degraded appropriately to match the processing of the
signal. The properties of the covariance matrix were also verified against noise simulations processed in the same way as the 
temperature maps.

In this paper, we utilise three templates to describe the Galactic emission mechanisms, but also add a further template to describe
monopole terms. Significant residual monopoles and dipoles may be present in the \emph{WMAP} data, and can impact foreground studies, 
particularly when based on parametric approaches \citep[see for example][]{Eriksen_WMAP3:2008,Dickinson_WMAP5:2009}. Here, template 
fitting relies on the morphological content of the data, and the effective mean emission within a given region provides
no relevant information. The inclusion of a single monopole term can account for offset contributions in all templates and the \emph{WMAP}
data in a way that does not bias the results \citep{Macellari:2011}. We follow this approach here, as indeed we did in 
\citetalias{Davies_WMAP1:2006}. Dipole terms can be similarly treated, but are more important for the global fit studies.

\section{Basic Results}
\label{sec:results}

At each \emph{WMAP} band the emissivity of each of the 3 foreground components (free-free, dust and synchrotron) has been estimated as a 
ratio of template brightness; $\mu$K~R$^{-1}$, $\mu$K/$\mu$K$_{\rm FDS8}$ and $\mu$K~K$^{-1}$ respectively.  The analysis was a joint solution
derived for all 3 components simultaneously, together with monopole and dipole terms. Results for the three templates can be found in 
Tables~\ref{tab:results_synch_33regions},\ref{tab:results_freefree_33regions} and \ref{tab:results_dust_33regions}.

\begin{figure*}
\begin{tabular}{ccc}
\epsfig{file=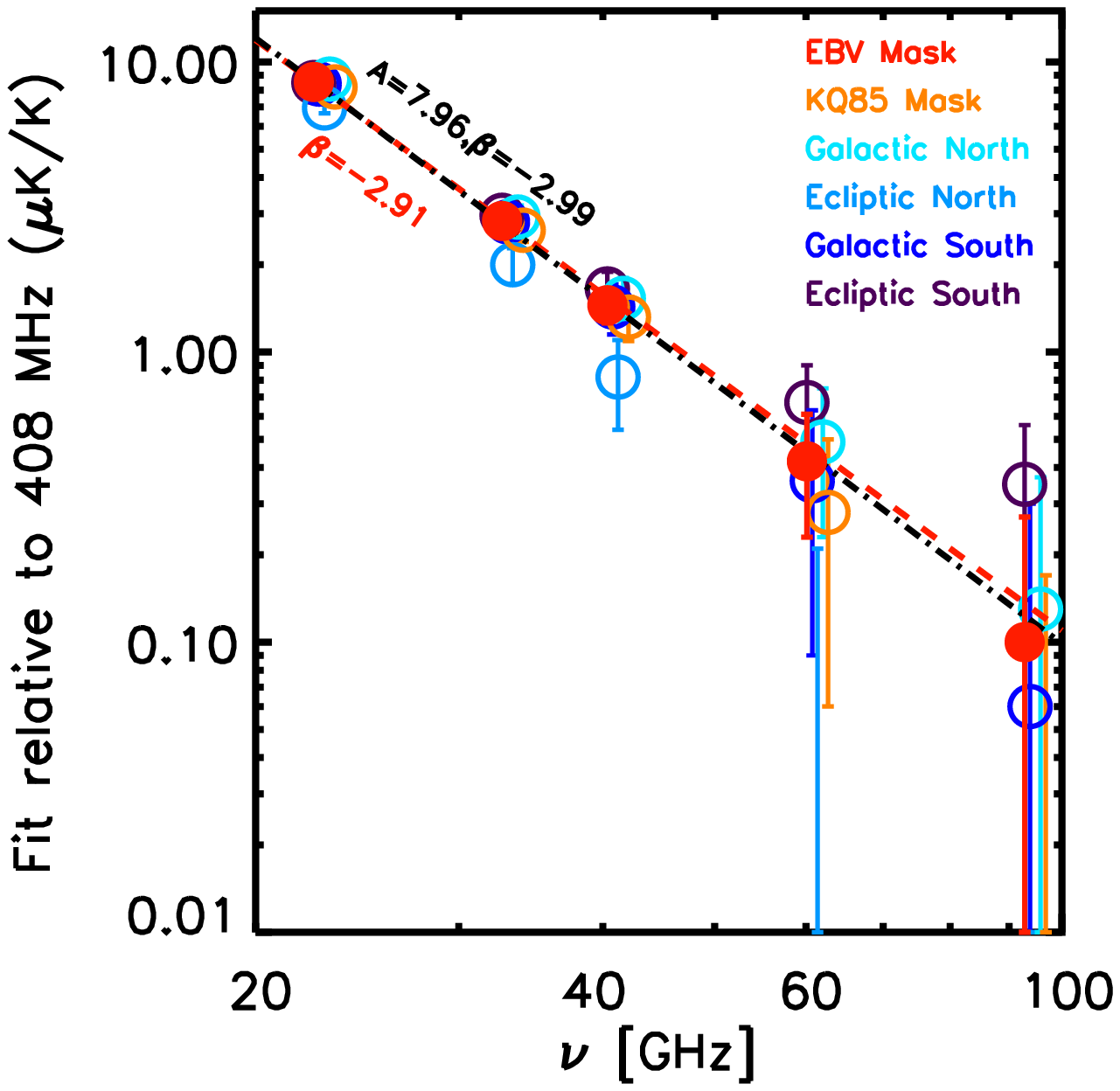,width=0.3\linewidth} &
\epsfig{file=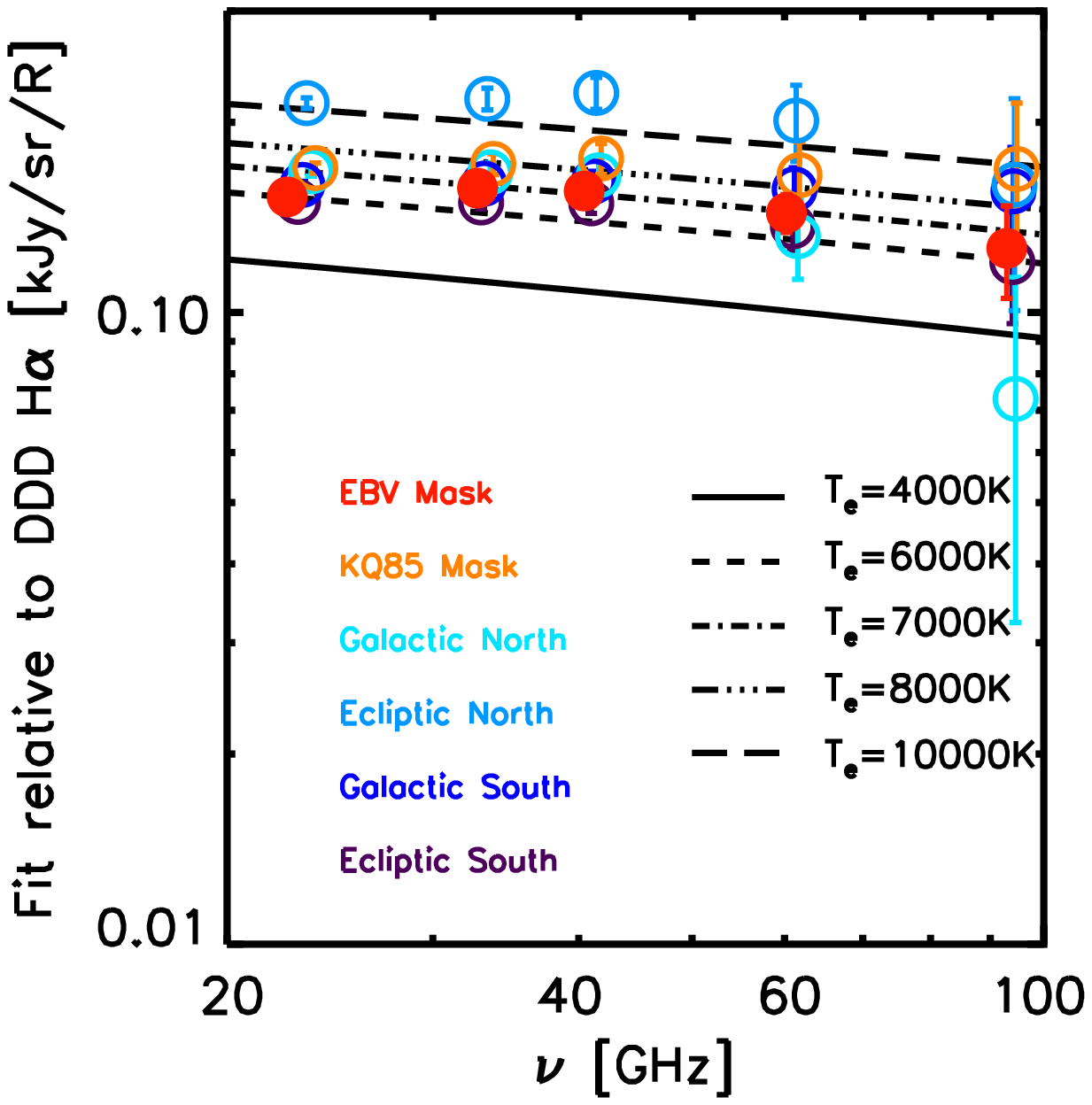,width=0.3\linewidth} &
\epsfig{file=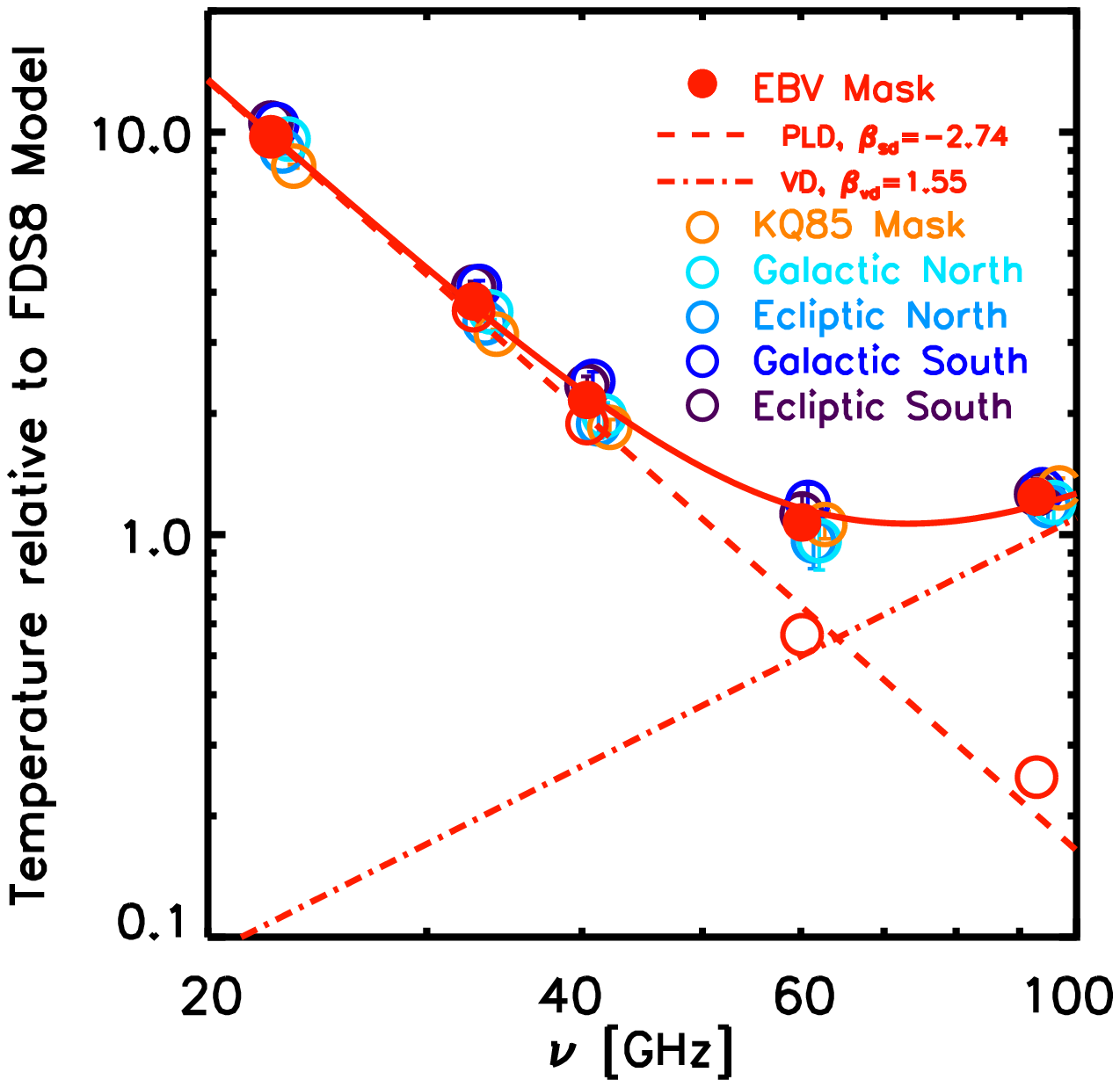,width=0.3\linewidth} \\
\epsfig{file=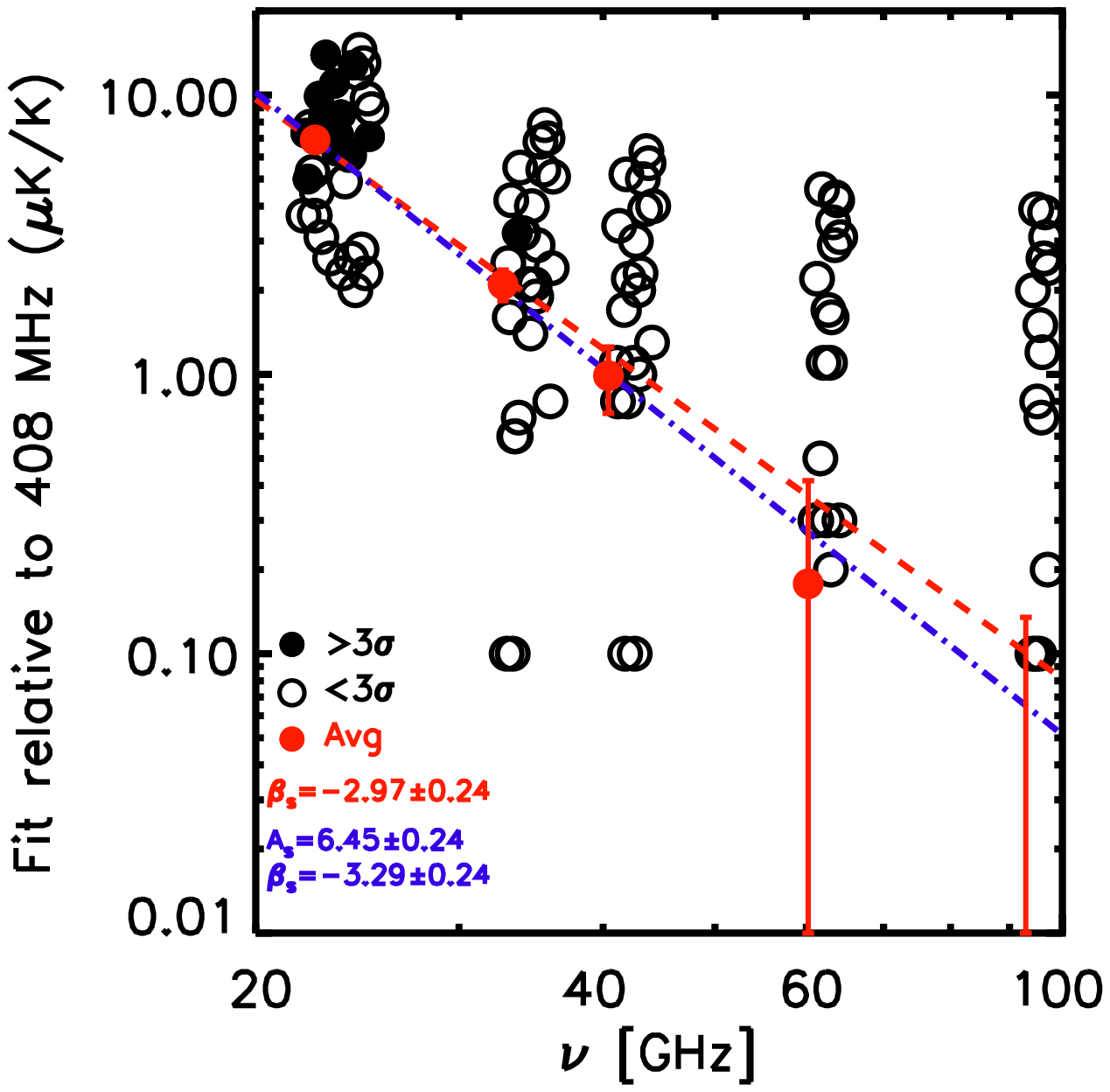,width=0.3\linewidth,clip=} &
\epsfig{file=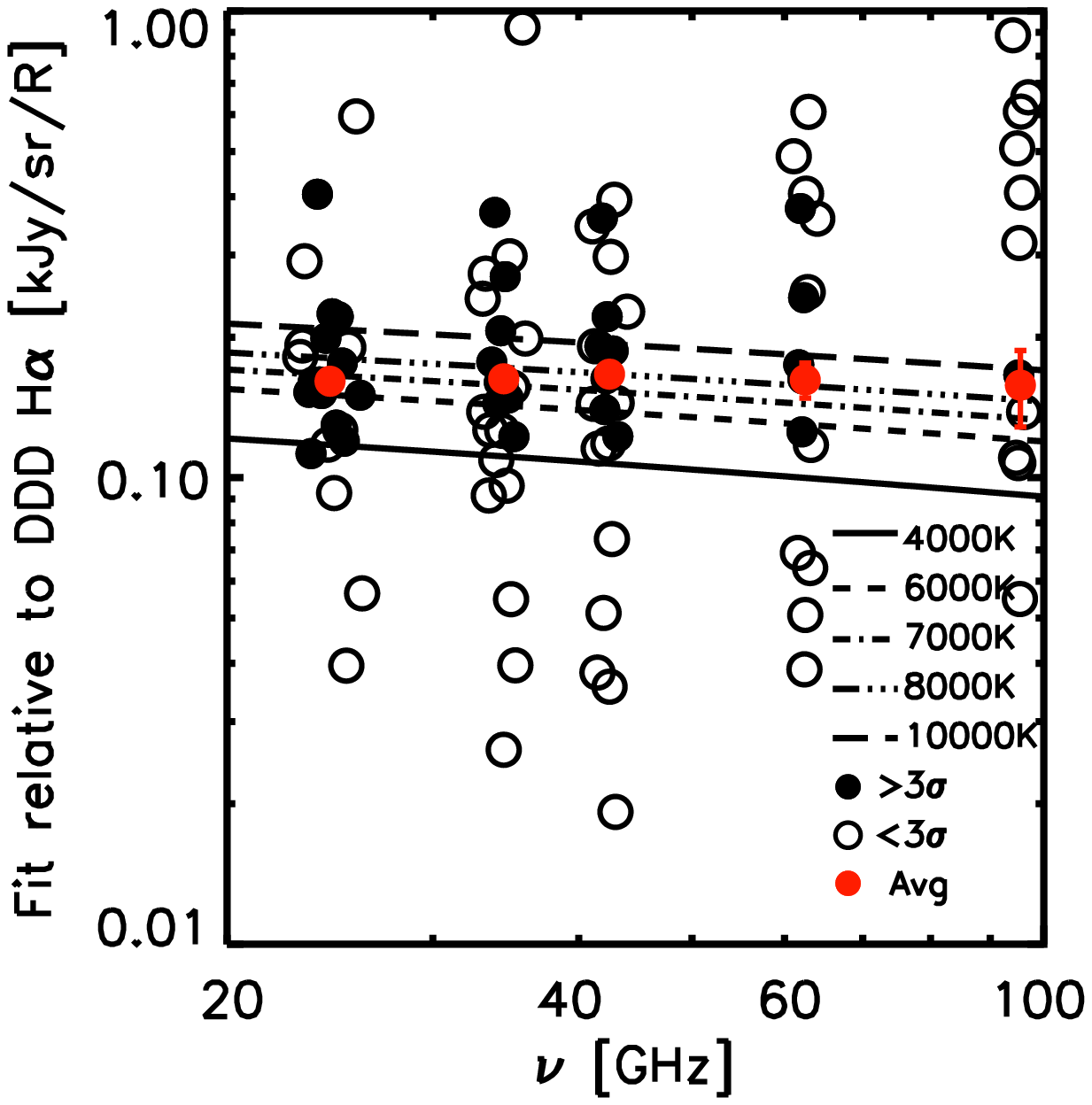,width=0.3\linewidth,clip=} &
\epsfig{file=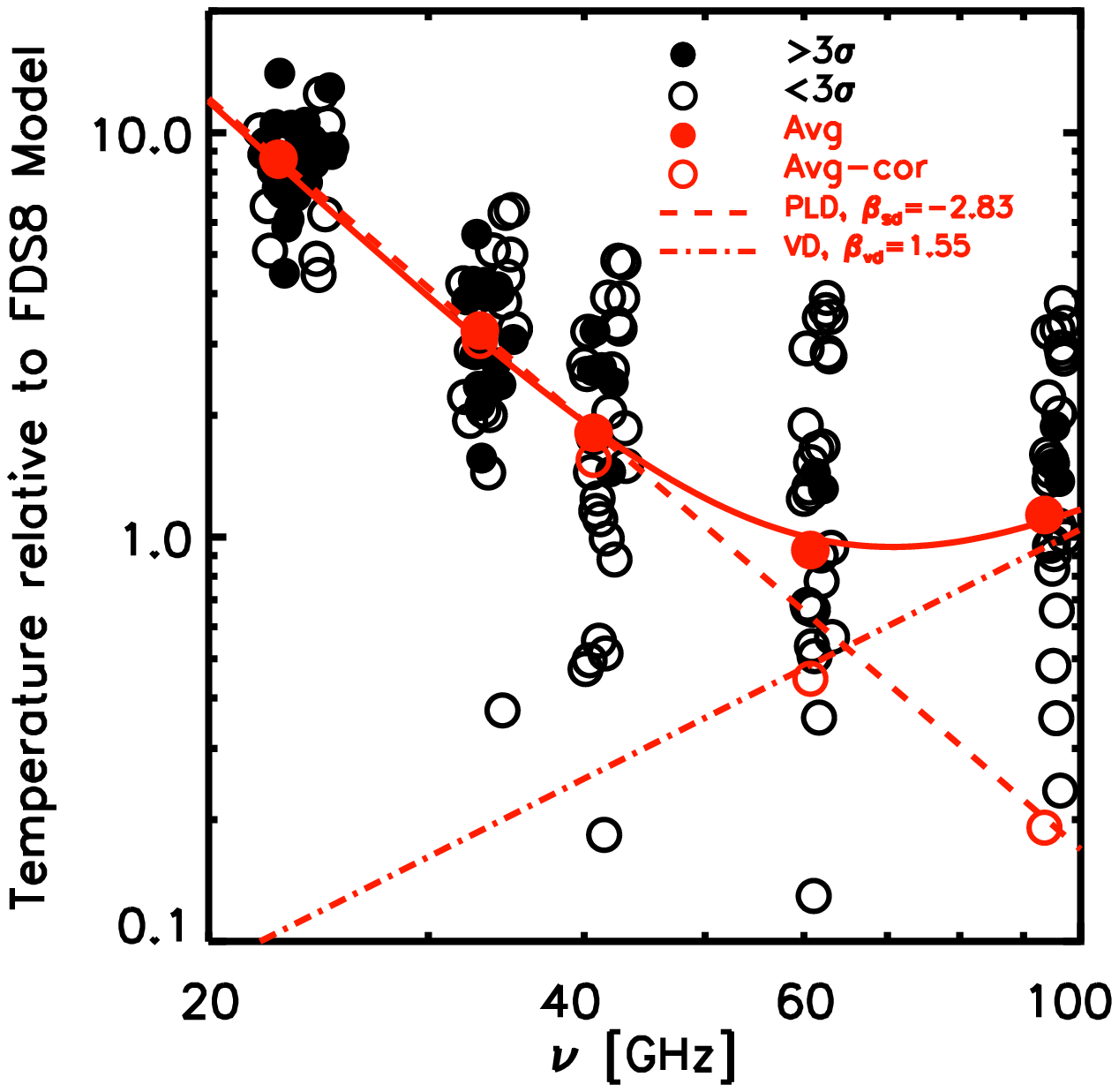,width=0.3\linewidth,angle=0,clip=} \\
\end{tabular}
\caption{Summary of template fit coefficients. The {\it top row} presents results for the full-sky and individual hemispheres, the 
{\it bottom row} shows the values for the 33 regions (after excluding any pixels rejected by the EBV mask). For the regional plots,  
greater than 3-$\sigma$ detections are shown with filled black  circles whilst all other values are shown by open circles. The weighted 
averages at each frequency of the coefficients for the 33 regions are shown the filled red circles. {\it Left column}: Synchrotron - two 
fits for power-law spectral models are shown; in red the fit includes the 408 MHz reference point, in blue the fit only includes the 5 
\emph{WMAP} frequencies. Neither model includes spectral curvature. {\it Middle column}: Free-free - the predicted coefficients (in 
intensity units) for different electron temperatures as given by \citet{DDD:2003} are shown. {\it Right column}: Dust - Straight lines 
show power spectra fit to the data: in black is a single power law fit to K through Q; in red, two power laws are fit, one with a fixed 
index of $\beta=1.55$.}
\label{fig:results_summary}
\end{figure*}

In Fig.~\ref{fig:results_summary} we provide a graphical summary of the results for both the global analyses and fits to individual 
regions. We present general inferences about the physical emission mechanisms below, and detailed comparisons of the fits to models in 
Section~\ref{sec:modelfits}.

\subsection{Synchrotron}\label{sec:results_sync}

\begin{table}
\scriptsize 
\begin{center}
\begin{tabular}{lccc}
\hline
            & \multicolumn{3}{c}{\bf Synchrotron spectral indices}\\
\hline
Region & K/408 &  Ka/408 & Q/408\\
\hline
                      EBV      &$\bf   -2.90         ^{+    0.01         }_{   -0.01     }$&$\bf   -2.91         ^{+    0.02         }_{   -0.02     }$&$\bf   -2.92         ^{+    0.03         }_{   -0.03        }$\\
                      KQ85       &$\bf   -2.91         ^{+    0.01         }_{   -0.01     }$&$\bf   -2.93         ^{+    0.02         }_{   -0.02     }$&$\bf   -2.94         ^{+    0.03         }_{   -0.04        }$\\
                      NPS      &$\bf -3.03     ^{+ 0.03}_{-0.03}$&$                      -3.06 ^{+0.08}_{-0.12} $&$ < -2.90 $\\
                       GN       &$\bf   -2.90         ^{+    0.01         }_{   -0.01     }$&$\bf   -2.90         ^{+    0.02         }_{   -0.02     }$&$\bf   -2.91         ^{+    0.04         }_{   -0.04        }$\\
                      GN$_{\rm reduced}$       &$\bf   -2.88         ^{+    0.01         }_{   -0.01     }$&$\bf   -2.88         ^{+    0.02         }_{   -0.02     }$&$\bf   -2.88         ^{+    0.03         }_{   -0.04        }$\\
                      EN      &$\bf   -2.95         ^{+    0.01         }_{   -0.01     }$&$\bf   -2.99         ^{+    0.03         }_{   -0.03        }$&$   -3.04         ^{+    0.06         }_{   -0.09        }$\\
                      EN$_{\rm  reduced }$       &$\bf   -2.94         ^{+    0.01         }_{   -0.01     }$&$\bf   -2.97         ^{+    0.03         }_{   -0.04     }$&$  -3.02         ^{+    0.06         }_{   -0.09        }$\\
                      GS       &$\bf   -2.91         ^{+    0.01         }_{   -0.01     }$&$\bf   -2.91         ^{+    0.02         }_{   -0.02     }$&$\bf   -2.92         ^{+    0.04         }_{   -0.05        }$\\
                      ES       &$\bf   -2.90         ^{+    0.01         }_{   -0.01     }$&$\bf   -2.90         ^{+    0.02         }_{   -0.02     }$&$\bf   -2.89         ^{+    0.03         }_{   -0.03        }$\\
\hline
1       &$ <    -2.96      $&$ <    -2.88      $&$ <    -2.80         $\\
 2          &$ \bf  -2.94         ^{+    0.04         }_{   -0.05     }$&$ <    -2.79      $&$ <    -2.74         $\\
 3          &$ \bf   -3.03         ^{+    0.03         }_{   -0.03        }$&$   -3.04         ^{+    0.07         }_{   -0.10     }$&$ <    -2.89         $\\
 4       &$ <    -2.77      $&$ <    -2.60      $&$ <    -2.50         $\\
 5       &$ <    -2.75      $&$ <    -2.62      $&$ <    -2.50         $\\
 6          &$   -3.11         ^{+    0.10         }_{   -0.16     }$&$ <    -2.87      $&$ <    -2.79         $\\
 7          &$   -3.06         ^{+    0.09         }_{   -0.15     }$&$ <    -2.84      $&$ <    -2.78         $\\
 8          &$  \bf  -2.86         ^{+    0.01         }_{   -0.01        }$&$ \bf   -2.88         ^{+    0.04         }_{   -0.04        }$&$ \bf   -2.89         ^{+    0.06         }_{   -0.09        }$\\
 9          &$   -3.15         ^{+    0.09         }_{   -0.13     }$&$ <    -2.91      $&$ <    -2.84         $\\
10          &$ \bf   -2.90         ^{+    0.06         }_{   -0.09        }$&$   -2.76         ^{+    0.08         }_{   -0.14        }$&$   -2.64         ^{+    0.08         }_{   -0.14        }$\\
11          &$ \bf  -2.78         ^{+    0.02         }_{   -0.03        }$&$   -2.88         ^{+    0.08         }_{   -0.14     }$&$ <    -2.76         $\\
12          &$  \bf  -2.92         ^{+    0.06         }_{   -0.08     }$&$ <    -2.72      $&$ <    -2.63         $\\
13       &$ <    -3.00      $&$ <    -2.93      $&$ <    -2.85         $\\
15          &$ \bf   -2.84         ^{+    0.04         }_{   -0.04     }$&$ <    -2.78      $&$ <    -2.80         $\\
16          &$ \bf   -2.98         ^{+    0.03         }_{   -0.03        }$&$   -2.97         ^{+    0.07         }_{   -0.09     }$&$ <    -2.82         $\\
17          &$ \bf   -2.94         ^{+    0.04         }_{   -0.05     }$&$ <    -2.85      $&$ <    -2.82         $\\
18          &$ \bf   -2.90         ^{+    0.06         }_{   -0.07     }$&$ <    -2.68      $&$ <    -2.59         $\\
19       &$ <    -2.85      $&$ <    -2.67      $&$ <    -2.58         $\\
20       &$ <    -2.82      $&$ <    -2.65      $&$ <    -2.53         $\\
21          &$ \bf   -2.98         ^{+    0.07         }_{   -0.09     }$&$ <    -2.78      $&$ <    -2.70         $\\
22          &$   -2.98         ^{+    0.09         }_{   -0.15     }$&$ <    -2.69      $&$ <    -2.59         $\\
23       &$ <    -2.94      $&$ <    -2.87      $&$ <    -2.81         $\\
24          &$ \bf   -2.80         ^{+    0.04         }_{   -0.05        }$&$   -2.71         ^{+    0.07         }_{   -0.10        }$&$   -2.65         ^{+    0.08         }_{   -0.14        }$\\
25       &$ <    -2.92      $&$ <    -2.76      $&$ <    -2.65         $\\
26          &$   -2.82         ^{+    0.10         }_{   -0.16     }$&$ <    -2.53      $&$ <    -2.45         $\\
27          &$   -2.77         ^{+    0.08         }_{   -0.11     }$&$ <    -2.51      $&$ <    -2.42         $\\
28       &$ <    -2.79      $&$ <    -2.66      $&$ <    -2.57         $\\
29          &$   -2.80         ^{+    0.10         }_{   -0.17     }$&$ <    -2.50      $&$ <    -2.40         $\\
30       &$ <    -2.82      $&$ <    -2.67      $&$ <    -2.57         $\\
31          &$   -2.87         ^{+    0.08         }_{   -0.13     }$&$ <    -2.70      $&$ <    -2.67         $\\
32          &$ \bf   -2.95         ^{+    0.04         }_{   -0.04        }$&$   -2.95         ^{+    0.08         }_{   -0.14     }$&$ <    -2.76         $\\
33          &$   -2.89         ^{+    0.08         }_{   -0.11     }$&$ <    -2.62      $&$ <    -2.52         $\\
\hline
Average  &$ \bf  -2.95         ^{+    0.01         }_{   -0.01 }$&$ \bf   -2.97   ^{+    0.03         }_{   -0.03     }   $&$ \bf   -3.00 ^{+    0.05         }_{   -0.06     }         $\\
\hline
\end{tabular}
\end{center}
\caption{Synchrotron fits between 408 MHz and \emph{WMAP }K-, Ka- and Q-bands for various regions of the sky. For those regions that are 
detected at 2-$\sigma$ significance, the fit values from Table~\ref{tab:results_synch_33regions} are converted into a spectral index 
using the usual power-law relation $\beta$ ($T_b \propto \nu^{\beta}$). Associated errors are determined by using the fit values plus and 
minus the 1$\sigma$ error bars.  Otherwise, one-sided 95\% confidence level upper limits on the index are quoted based on the fit value 
plus  1.64$\sigma$. Bold text denotes detections at 3-$\sigma$ significance or more.  For the global fits, NPS -- North Polar Spur, 
GN -- Galactic North, GN$_{\rm reduced}$ -- Galactic North with the NPS removed, EN -- Ecliptic North, EN$_{\rm reduced}$ -- Ecliptic North 
with the NPS removed, GS -- Galactic South, ES -- Ecliptic South.}
\label{tab:simple_synch_spectra}
\end{table}

As expected, the template fit coefficients between the Haslam 408 MHz data and the \emph{WMAP} sky maps all fall with frequency in a 
manner consistent with power-law emission.

In Table~\ref{tab:simple_synch_spectra} we define simple pairwise spectral indices between 408 MHz and the K-, Ka- and Q-bands. The 
global fits exhibit a typical index of order $-2.90$ at K-band, with the results from all masks slightly flatter than the
value of $-3.01$ found in the \citetalias{Davies_WMAP1:2006} analysis of the \emph{WMAP} first-year data for the then-preferred Kp2
sky-coverage. Analyses of the lower-frequency surveys at 408, 1420 and 2326\,MHz by \citet{Giardino_synch_pol:2002} and 
\citet{Platania_indices:2003} suggest a spectral index over this lower frequency range of approximately $-2.7$, thus our results support 
the idea of spectral steepening, continuing beyond K-band. We will consider this further in Section~\ref{sec:modelfits}.

However, there are differences in the coefficients depending on the exact sky coverage that must reflect genuine spectral
variations on the sky. The NPS is recognised to be an arc of steep spectrum emission at lower frequencies, thus it is not
surprising that it is notably steeper than the rest of the high latitude sky.  The spectral index value of $-3.03$ at K-band is quite
consistent with the value adopted in \citet{Finkbeiner_WMAP1:2004} to remove the prominent emission from the \emph{WMAP} data.
The presence of the NPS  also impacts measures of the spectral index in both the Galactic and Ecliptic northern sky, resulting in a modest
steepening of the index. Interestingly, the northern Ecliptic hemisphere is notably steeper than the other hemispheres, and exhibits an 
increasingly steep index with frequency. Conversely, the corresponding southern hemisphere hints at spectral flattening, whereas both 
the north and south Galactic hemispheres are consistent with simple  power-law behaviour.

From the 33 regions of interest, there are 13 regions where the synchrotron fit coefficients are detected at greater than $3 \sigma$
confidence at K-band. Most of these regions include contributions close to the Galactic plane, although regions 2, 3 and 32 are mostly 
at high Galactic latitude. The inferred spectral indices span the range $-2.78$ to $-3.03$, inconsistent with statistical variation 
alone, and likely representing genuine spectral variations on the sky. We note that region 3  contains the NPS and its behaviour seems 
to be dominated by that component. Significant emission is detected for region 8 at K-, Ka- and Q-band. Regions 13 and 14 correspond to 
an area of the sky containing the Gum nebula, and show no evidence for detection of synchrotron emission, with upper limits consistent 
with a steep spectrum, particularly in the southern region. Interestingly, region 9 has one of the steepest spectral indices on the sky, 
despite the putative presence of the \emph{WMAP} haze. However, the overlap between the brightest regions of the haze emission as seen 
in \citet{Dobler_WMAP3:2008a} and regions 9, 16 and 17 is very small, and unlikely to affect any studies here.

It is interesting that the spectral indices inferred from the mean of the regional fit coefficients are steeper than the typical global 
fit values. This trend was also seen in our previous work \citepalias{Davies_WMAP1:2006}, and may be due to a selection effect in that 
the regional subdivisions partly favour stronger synchrotron emission regions which may exhibit steeper spectra than normal due to 
synchrotron losses. In addition, evidence of spectral steepening with frequency is again generally seen although regions 10 and 24 
show inconsistent behaviour with the other regions in that a flattening of the spectrum is indicated.

\subsection{Free-free}\label{sec:results_freefree}

In this paper, we use an H$_{\alpha}$ template as a proxy for the free-free emission. We see significant correlation between the
\emph{WMAP} data and the template for the global fits at all frequencies (Table~\ref{tab:results_freefree_33regions}). As with the 
synchrotron results, there are interesting variations depending on the exact sky coverage, with the northern Ecliptic hemisphere 
showing significantly enhanced amplitude, whilst the south indicates a lower emissivity. 14 individual regions are detected at 
3$\sigma$ significance at K-band. Most of these regions lie close to the Galactic plane ($|b|\lesssim 20^{\circ}$).

\citetalias{DDD:2003} detail the relationship between the expected free-free brightness temperature and the related H$_{\alpha}$
intensity, and its dependence on both frequency and electron temperature ($T_e$) in the ionised medium. As can be seen from their
Fig.~5, the spectral dependence of the emission shows weak curvature, but over the range of frequencies covered by \emph{WMAP} a
reasonable approximation is a power-law with index $-2.15$.

Inspection of the coefficients in Table~\ref{tab:results_freefree_33regions} indicates that there are
departures from this behaviour. Following \citet{Dobler_WMAP3:2008b} we plot these results in intensity units in
Fig.~\ref{fig:results_summary}. It should be apparent that a bump in emission is seen around 40--50~GHz for the global fits, a feature 
that \citet{Dobler_WMAP3:2008b} argue is indicative of contributions from both classical free-free emission plus a spinning dust 
component in the WIM. The mean spectrum of the regional fits perhaps indicates a slightly broader bump in the spectrum. However, it is 
also the case that there is a range of behaviour seen amongst the individual regions, some of which are consistent with emission from a 
single physical component only - either free-free emission or a more steeply falling spectrum as expected from spinning dust.
In Section~\ref{sec:modelfits}, we will undertake a more detailed modelling of the emission in terms of these components.

Such a feature in the spectrum of course has implications for the determination of physical parameters such as $T_e$. Nevertheless, we
can make some general inferences, particularly by examining the K- and W-band amplitudes that are least affected by a putative WIM 
spinning dust component. The global fits seem to be consistent with values in the range 6000 -- 8000\,K, with some dependency on the 
exact mask used. The average free-free electron temperature inferred from the 33 regions is also in this range, although for individual 
regions there is a spread of values between 4000 and more than 10000\,K. These values are somewhat higher than seen previously in 
\citet{Davies_WMAP1:2006}, and this is due to the use of 3$^{\circ}$ smoothed data here, rather than the 1$^{\circ}$ resolution data used 
earlier for reasons provided in Appendix~\ref{app:halpha}. It is also interesting to note that the 7000\,K temperature inferred from the 
31.5 GHz $COBE$-DMR data at  7$^{\circ}$ resolution in \citet{Banday_DMR:2003} is quite consistent with the Ka-band values determined here.
Moreover, the higher values in this paper are in better agreement with the electron temperatures derived from radio recombination line 
studies of extended HII regions \citep{Shaver:1983,Paladini_HII:2004,Alves:2011} which derive an average $T_e$ value of $\approx 7000$\,K 
in the vicinity of the solar neighbourhood. There is therefore no need to invoke a large fraction of scattered \halpha light to account 
for this discrepancy (e.g., \citealt{Witt_Scattered_Halpha:2010}).

\subsection{Impact of Dust Extinction}\label{sec:dust_extinct}

\begin{figure*}
\begin{tabular}{ccc}
\epsfig{file=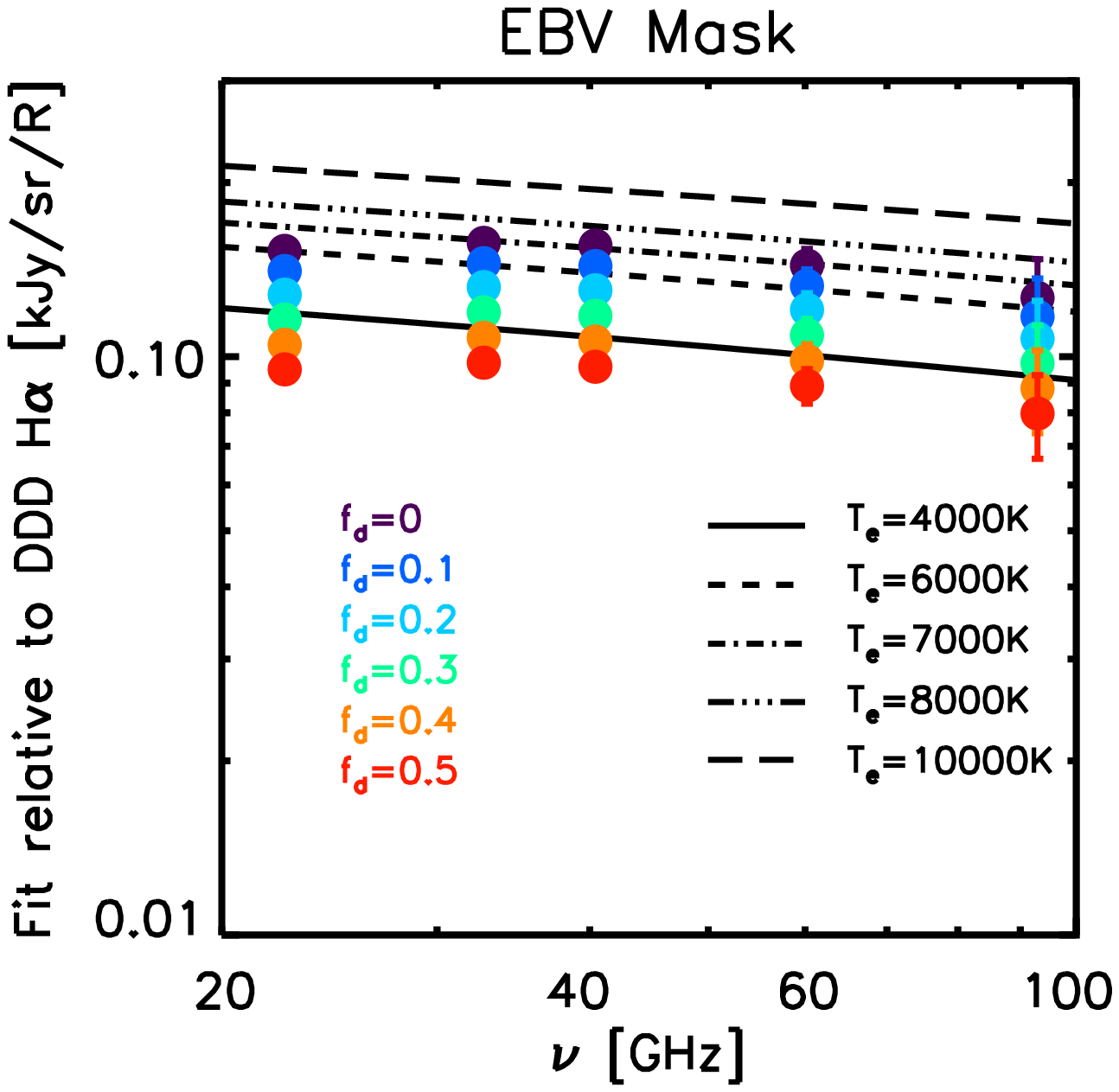,width=0.3\linewidth,angle=0,clip=}&
\epsfig{file=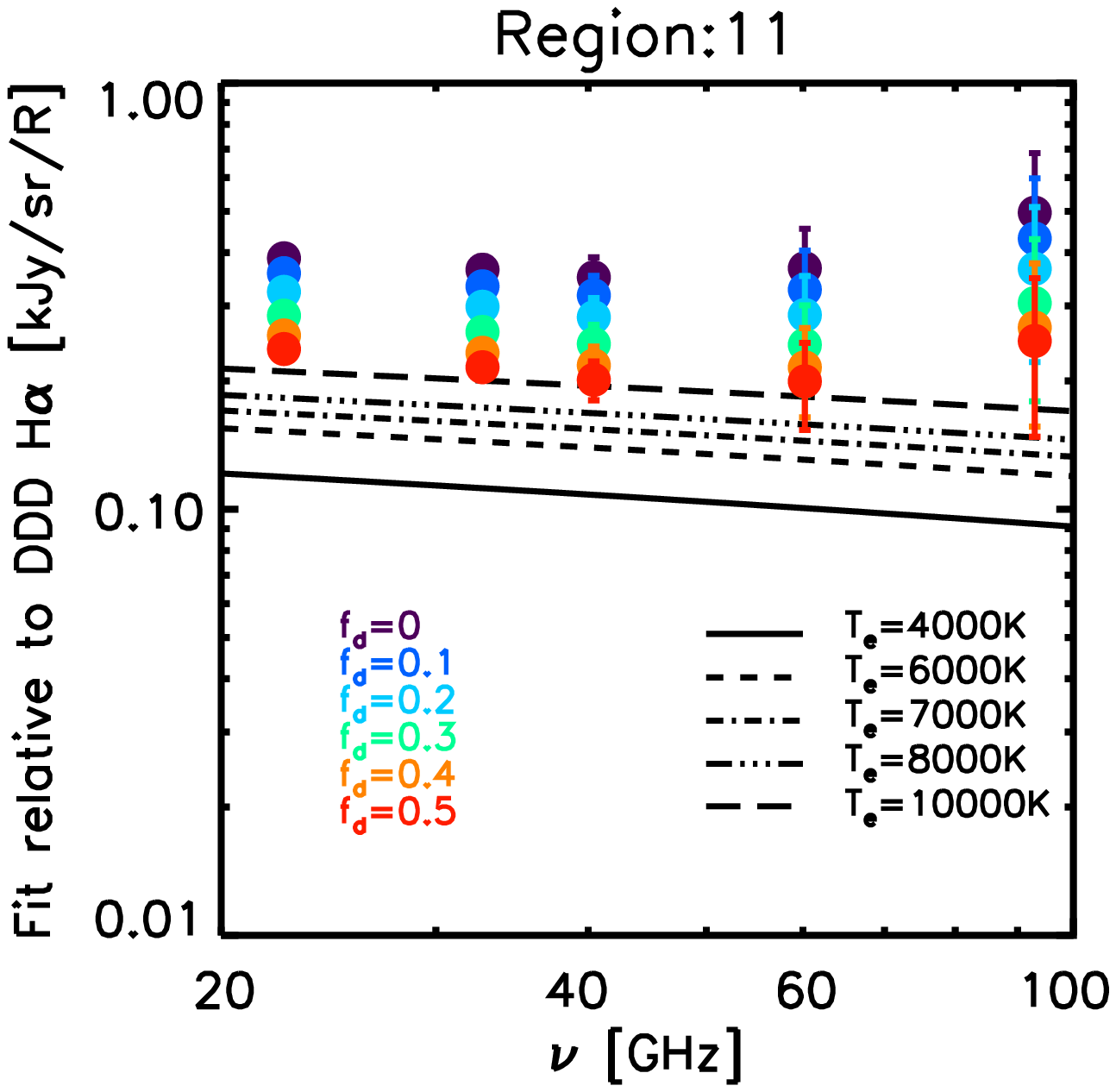,width=0.3\linewidth,angle=0,clip=}&
\epsfig{file=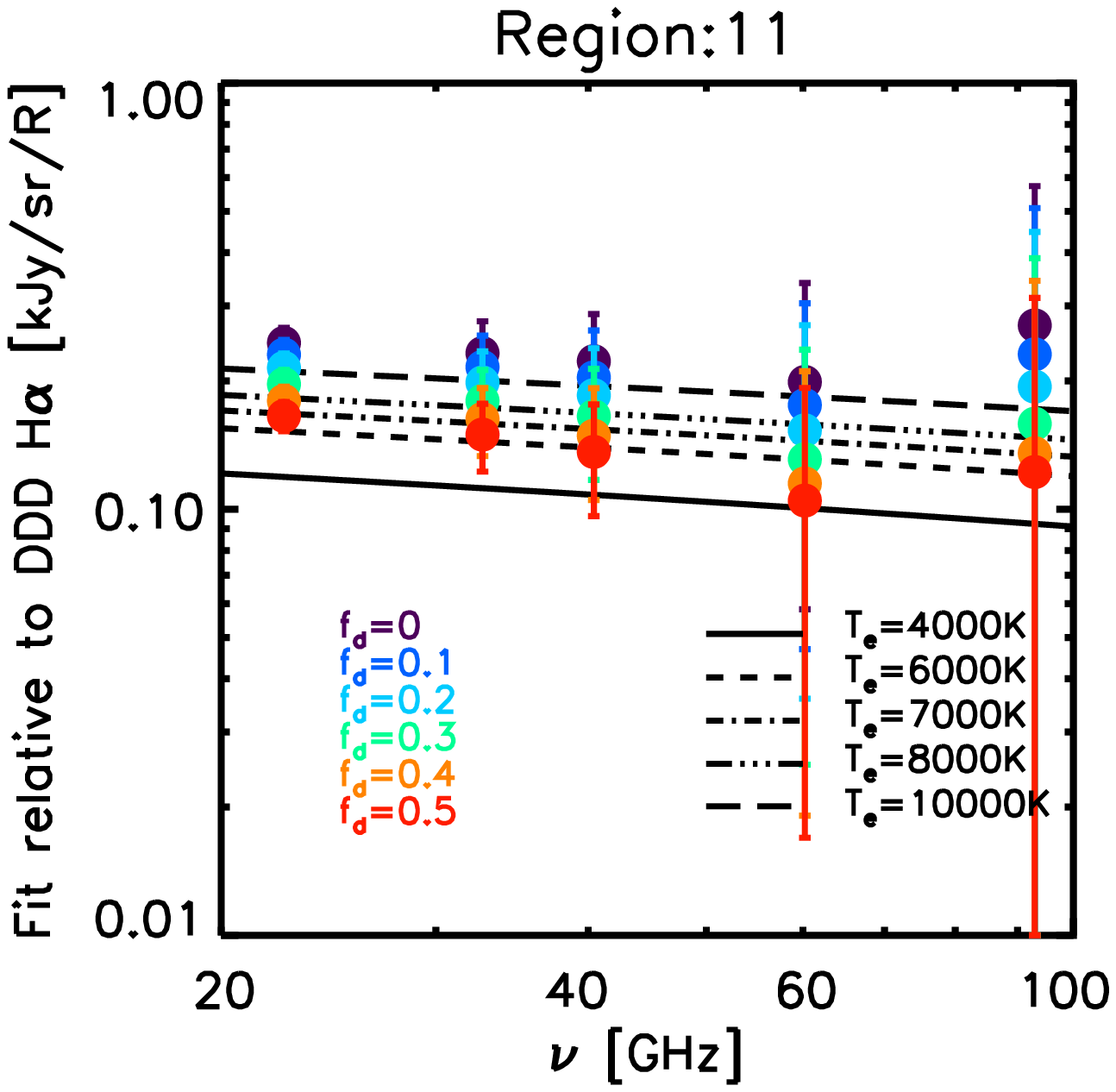,width=0.3\linewidth,angle=0,clip=}\\
\epsfig{file=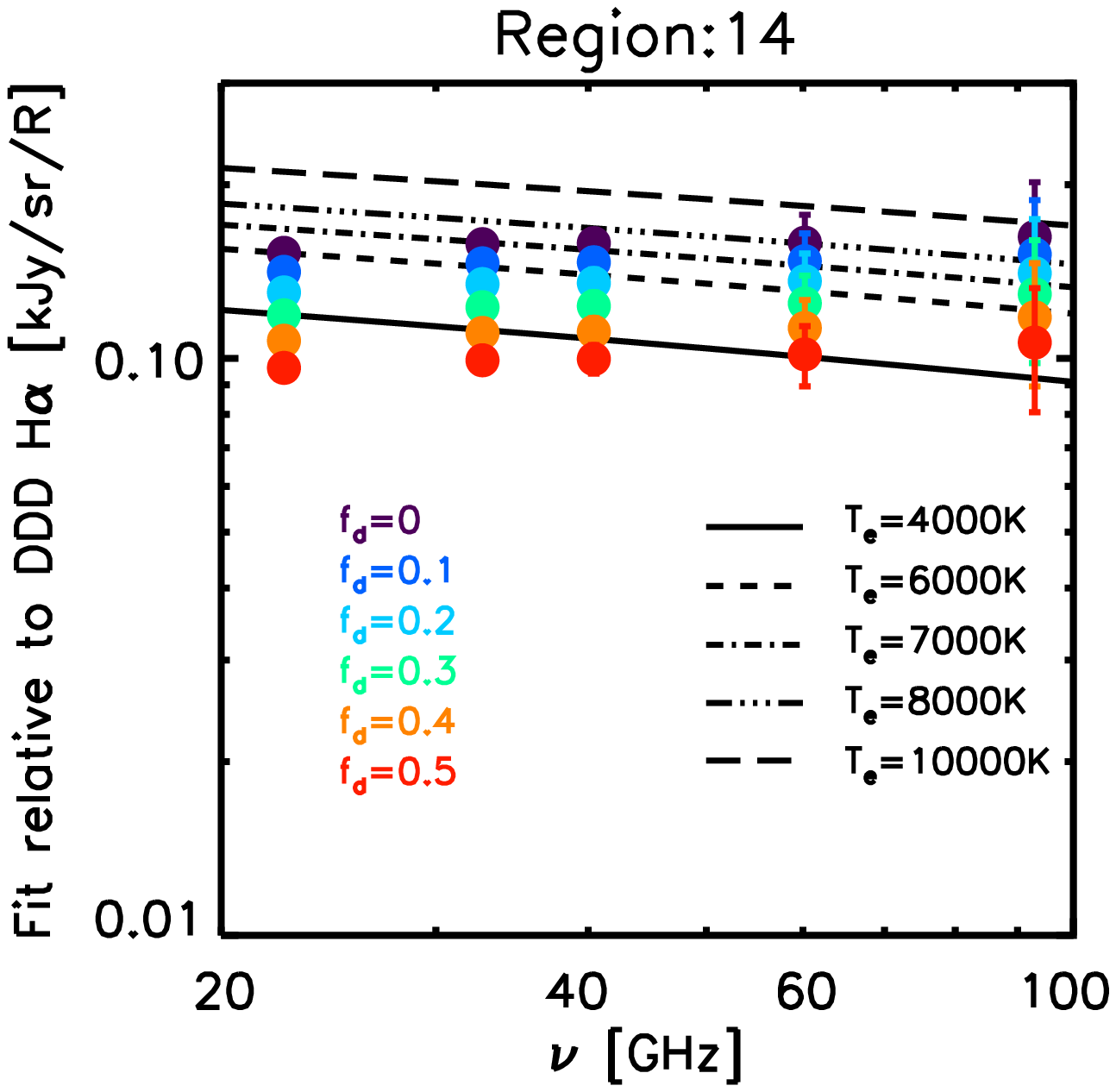,width=0.3\linewidth,angle=0,clip=}&
\epsfig{file=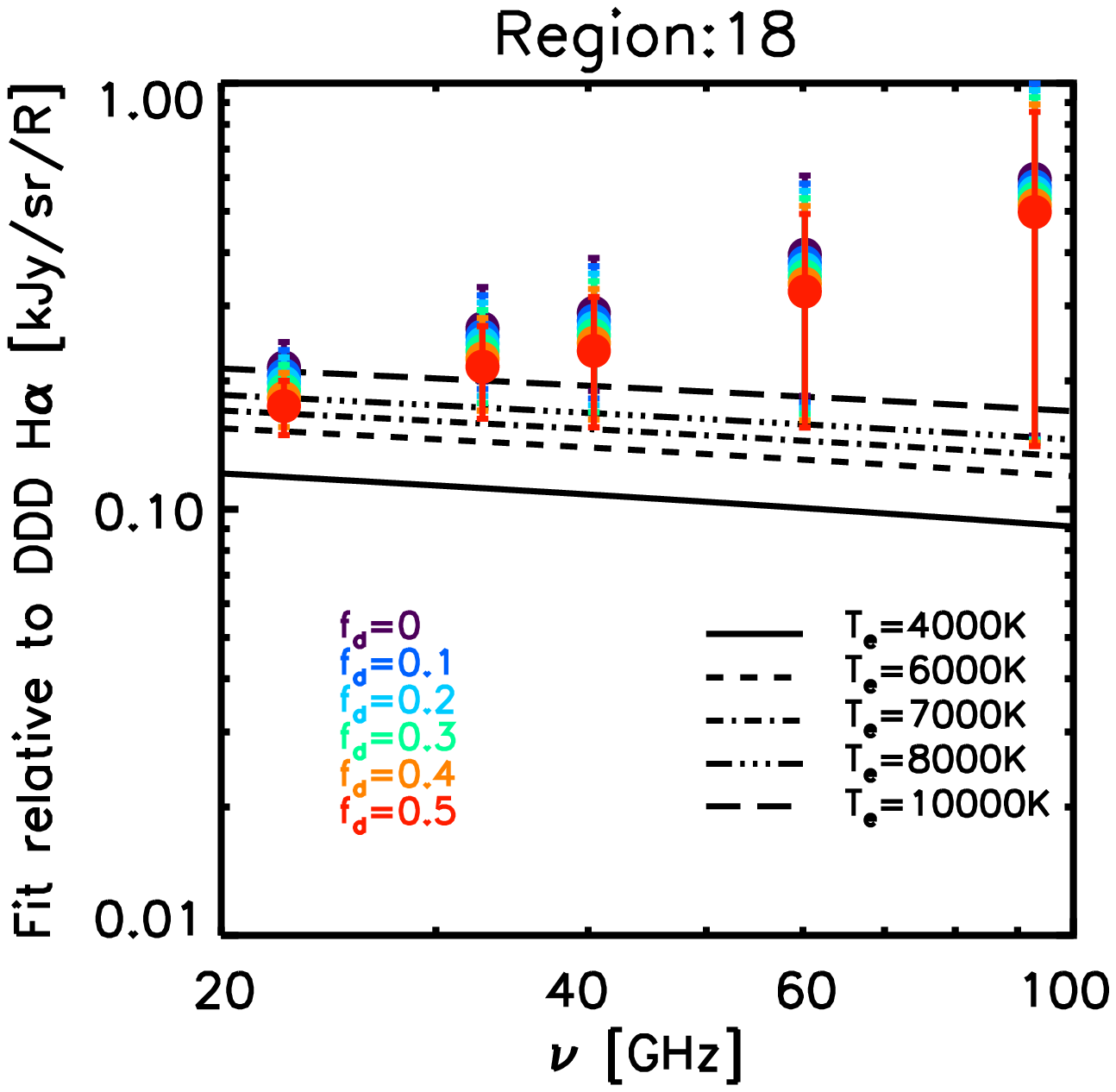,width=0.3\linewidth,angle=0,clip=}&
\epsfig{file=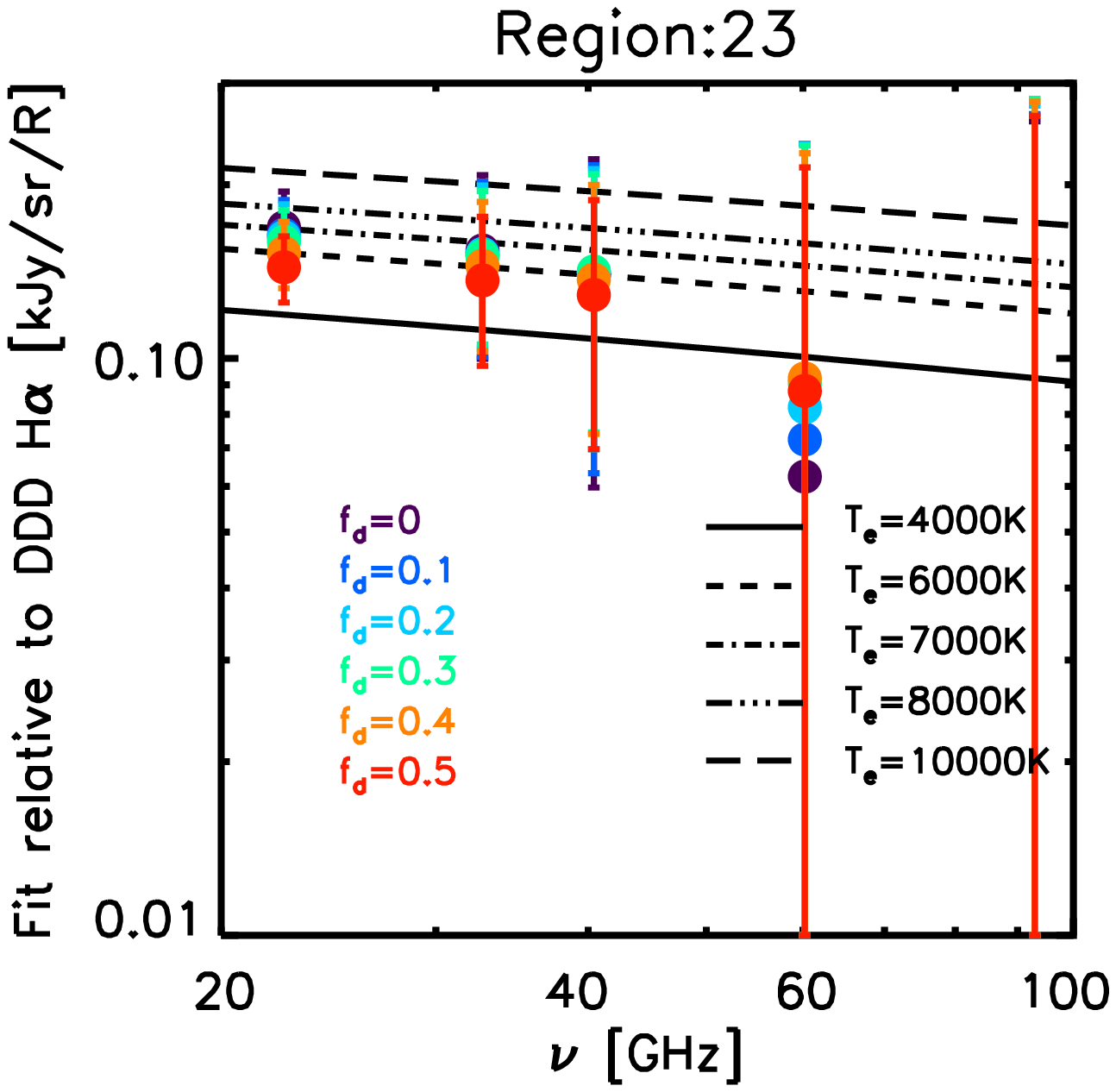,width=0.3\linewidth,angle=0,clip=}\\
\end{tabular}
\caption{Dependence of the correlation coefficients between the \emph{WMAP} data and the \halpha template as a function of the absorption
correction $f_{d}$. The union mask refers to the combination of all 33 small regions defined in this paper. Thus the results for 
``Union + Galactic North'' refers to the inverse-noise-weighted average coefficients for the subset of regions in the northern Galactic hemisphere. The observed dependencies for regions 13, 14 and 23 are representative of that seen for all regions with a significant detection amplitude at K-band.}
\label{fig:coefficients_fd}
\end{figure*}

\citet{Bennett_WMAP1:2003b} summarise various uncertainties in the use of the H$_{\alpha}$ template to trace free-free emission in the
Galaxy. These include uncertainties in the value of the electron temperature $T_e$, and in the value of the dust absorption correction,
specified here by the $f_d$ parameter. In principle, there will be variations in both throughout the Galaxy. As noted previously,
\citetalias{DDD:2003} have determined that for local regions such as Barnard's Arc and the Gum nebula, there is little absorption by dust,
and this is the default situation that we have assumed in our analyses. However, for mid-Galactic latitudes a value $f_{d} \sim$ 0.3 is 
preferred, and \citet{Finkbeiner:2003} adopted the assumption that the H$_{\alpha}$ emission is co-extensive with dust emission along the 
line-of-sight, ie. $f_{d} \sim$ 0.5.

In this section we discuss the impact of varying $f_d$ on the template fit coefficients. In Fig.~\ref{fig:coefficients_fd} we show this 
variation as a function of $f_d$, and compare against the behaviour expected for a range of values of $T_e$. In general, we expect that 
as the amplitude of the H$_{\alpha}$ template is corrected upwards by increasing $f_d$, then the fit coefficients will decrease, implying a
lower value for the electron temperature. This is indeed what is seen, but the extent of the correction depends on the region of sky 
under consideration.

For the largest sky area that we consider, the EBV mask, the use of a template corrected absorption specified by $f_{d} = 0.5$ results in
coefficients consistent with values of $T_e$ of order 3000\,K. However, assuming no dust absorption correction yields values closer to 
6000\,K.  A similar result is seen for the southern extension of the Gum nebula (region 14). If it is indeed the case that there is no 
evidence for dust absorption, then a higher temperature of 6000\,K is determined. Conversely, regions 15, 18 and 21 yield coefficients 
at K-band consistent with temperatures closer to 10000\,K if no absorption correction is applied, whereas values of order 8000\,K are
found for $f_d = 0.5$. Given that these regions exhibit rising spectra that may require the presence of significant emission from a 
spinning dust component, then the latter value would be more consistent. However, there are also regions of the sky typified by region 23,
where apparently acceptable values of the electron temperature are inferred over a range of values of $f_d$.
Indeed, it may be that the large spread in coefficients seen for the different regions reflects changes in the fraction of dust mixed with
the WIM as much as variations in $T_e$. 

It does appear, therefore, that there are a range of values for $T_e$ and $f_d$ throughout the Galaxy, and reaching conclusions about 
their values solely from studies of H$_{\alpha}$ correlations suffers from degeneracies between the parameters. 
The adoption of a value for the dust absorption correction of 0.5 in, for example,  \citet{Dobler_WMAP3:2008b}, thus
seems to be associated, at least in part, with the low electron temperatures inferred. This uncertainty then has
implications for modelling the emission of the diffuse component.

A  more serious complication would arise if the validity of using  H$_{\alpha}$ as a tracer of free-free emission were questioned.
\citet{Mattila_Scattered_Halpha:2007} have argued that the H$_{\alpha}$ excess towards the high-latitude interstellar cloud LDN~1780 
is the result of scattering of  H$_{\alpha}$ photons produced elsewhere in the Galaxy by dust in the cloud. Earlier computations by
\citet{Wood_Scattered_Halpha:1999} suggested that this contribution would typically be 5--20\% at high-latitudes, although their model 
has been criticised due to the assumption of a smooth distribution of material in the ISM. \citet{Witt_Scattered_Halpha:2010} have used 
an empirical relation to relate the scattered H$_{\alpha}$ intensity in the translucent cloud LDN1780 to the \emph{IRAS} 100~$\mu$m diffuse
background intensity and conclude that this estimate is reasonable for 50\% of the high-latitude sky, but that the scattered 
contribution can be highly structured and result in contributions of between 25 and 50\% of the observed intensity for a
further 40\% of this region of the sky. Such a result would clearly have implications for using an H$_{\alpha}$ template to trace
free-free emission, and a complex relationship between the template and dust would arise both due to the scattering contribution and due
to any dust absorption correction applied to the data. Applying a correction for the former would effectively result in an
increase of the  H$_{\alpha}$-\emph{WMAP} correlation coefficients and therefore of the inferred electron temperature, whereas 
application of the latter results in the opposite behaviour. More recently, \citet{Brandt_Scattered_Halpha:2011} have estimated the 
fraction of high-latitude H$_{\alpha}$ that is scattered to be $19\pm 4\%$, a value consistent with that proposed by 
\citet{Dong_WIM_Halpha:2011} to reconcile the low ratio of radio free-free to  H$_{\alpha}$.

Ultimately, unravelling the degeneracy between the electron temperature, H$_{\alpha}$ scattering and dust absorption requires additional 
observations. Detailed RRL surveys in the Galactic plane together with radio-continuum surveys at frequencies of $\sim$5~GHz, as expected
from the C-BASS \citep{King_CBASS:2010} will be important in this respect. 

Finally, we would like to make some remarks about region 11, the coefficients of which suggest an exceptionally high temperature of 
more than 25000\,K. If this result were considered unphysical, then naively a value of $f_d = 1$ would be required in order to lower the
inferred temperature to the 6000\,K seen in the EBV fit. In fact, a more realistic assessment of the situation is that the EBV mask is not
large enough to eliminate some parts of region 11 close to the Galactic plane where the simple dust absorption correction is
untrustworthy. If instead we apply the KQ85 mask before analysing the region, then for a more plausible value of $f_{d} = 0.5$ the 
inferred electron temperature is again consistent with 6000\,K.

\subsection{Dust}\label{sec:results_dust}

As can be seen in the right-hand panels of Fig.~\ref{fig:results_summary}, the template fit coefficients determined between the 
\emph{WMAP} data and the FDS8 dust template prediction at W-band are consistent with a rising thermal dust contribution at frequencies
higher than V-band, and a rising spectrum to lower frequencies below it. The latter corresponds to the now widely identified anomalous
microwave emission (AME).

It appears that the FDS8 template underpredicts the W-band amplitude by approximately 30\% for the global fits, consistent with the 
results of \citetalias{Davies_WMAP1:2006}. A broad range of values for the individual regions is seen, but only three detect emission at 
a statistically significant level. The mean emissivity at W-band of all regions shows a more modest enhancement, but is nevertheless 
consistent with the FDS8 predictions, as indeed are the three significant regions.

At K-band,  there are variations in the global fit amplitudes depending on  sky coverage. The KQ85 mask indicates a lower emissivity
compared to the EBV as might be expected. The Southern Galactic and Ecliptic hemispheres have the highest amplitudes. All coefficients 
are higher than those determined for the Kp2 sky coverage in \citetalias{Davies_WMAP1:2006}. 26 of the individual regions detect emission
at 3$\sigma$ significance or higher. The mean emission amplitude lies between that for EBV and KQ85,  and again somewhat higher than in 
\citetalias{Davies_WMAP1:2006}.  The regions indicate a variation around the mean of approximately 50\% of its amplitude, inconsistent 
with statistical errors alone and indicating genuine spatial variations in the AME emissivity.

The overall spectrum would appear to be well-described by a superposition of two power-law emissivities. The thermal dust emission
described by the FDS8 model is adequately represented by an emissivity index $\beta_d = 1.55$ over the \emph{WMAP} range of frequencies. 
We fit the power-law AME spectrum with the thermal dust index fixed to this value, and find AME spectral indices of order $-2.7$ for both
the global fits and regional mean. 

\citet{Draine_spinning:1998} first proposed that the AME could be explained by electric dipole radiation from rotationally excited small
interstellar grains, or spinning dust. Our results have ramifications for such models of the emission. In particular, given that the 
spinning dust spectra typically fall off steeply with frequency beyond their peak, then it is unlikely that a single such spectrum could 
account adequately for the effective power-law emission. Indeed, the observed spectrum is presumably formed from a superposition of 
components with varying spectra as a consequence of their differing physical environments along a given line-of-sight. We will discuss 
detailed fits of the observed emission in Section~\ref{sec:modelfits}, and their implications for more refined models of the AME. 

\section{Model Fits}
\label{sec:modelfits}

In order to compare the derived template fit coefficients of a given foreground component with various theoretically motivated foreground
models, we adopt the simple procedure used previously in, for example,  \citet{Dobler_WMAP3:2008b}. The model parameters of the 
foreground model are extracted using the least-square minimization defined as,
\[
 \chi^2 = \sum_i \left [ \frac{F(\nu_i)-F^M(\nu_i)}{\sigma (\nu_i)} \right ]^2 \,
\]
where $F(\nu_i)$ and $\sigma(\nu_i)$ are the observed fit coefficients and standard deviation at WMAP frequency bands and $F^M(\nu)$ are 
the fit coefficients given the foreground model. We use a Levenberg-Marquardt method to determine the coefficients, and quote the reduced
chi-square for a given model fit. The degree of freedom (dof) is defined as $n-m$ where $n$ represent the number of data points
to be fitted and $m$ represents the number of free parameters for a given foreground model. We note that the interpretation of such
values can be problematic when the number of data points is small, as is the case here. This is compounded by the correlated errorbars that are dominated by a CMB common to all frequencies. Indeed, the model fits are only to be considered 
as indicative rather than definitive. A more robust approach would apply a multi-frequency analysis to account for the correlated errors 
at each frequency due to the dominant CMB term in the covariance matrix. This will form the basis of a future publication where models
are fitted directly to the data rather than to previously derived template fit coefficients.

\subsection{Spinning Dust preamble}\label{sec:spdust}

\citet{Erickson:1957} first proposed the basic mechanism of spinning dust emission from the rotation of small dust grains with electric
dipole moments. \citet{Ferrara:1994} later suggested that such grains in the diffuse ionised medium should exhibit significant radio
emission peaking at a frequency between 10 and 100~GHz. However, it was \citet{Draine_spinning:1998} that suggested such emission could
explain the AME, and provided detailed computations of its spectral shape in \citet{Draine_electric:1998}.

Indeed, testing for the presence of such emission requires detailed predictions of the emission spectra to compare with
observations. However, given the large number of parameters in the model \citep{SpDust:2009}, and the fact that most lines-of-sight
likely average over many emission regions, this is difficult unless specific objects are considered. Such an analysis was performed 
recently with early data from the \emph{Planck} mission \citep{Tauber_PlanckMission:2010}, specifically for the
Perseus and $\rho$-Ophiuchus regions \citep{Planck_ERXX_AME:2011}. Thus, it is usual to adopt spectra for spinning dust in a variety of
phases of the ISM, computed for \lq typical' values of the parameters for those physical conditions.  Alternatively, 
\citet{Gold_WMAP5:2009} proposed a generalised analytic form for the spinning dust emission for their analysis of the \emph{WMAP} 5-year 
data (see also \citealt{Bonaldi:2007}). However, a fit to the exact CNM form from DL98 underestimates the emission at
frequencies beyond the peak. This is problematic given that more recent calculations by \citet{Hoang:2010} indicate that the emission
in this region may be broadened by the inclusion of additional physical processes.

We will adopt spinning dust emission templates based on the {\tt{SPDUST2}} code \citep{SpDust2:2011} for emission in the cold
neutral medium (CNM), warm neutral medium (WNM), and warm ionised medium (WIM).  We then allow the spectra to be shifted in both
amplitude and peak frequency to fit the data, considering that this approximately mimics the effect of varying the physical parameters as
actually required. 

\subsection{Synchrotron}\label{sec:discussion_synch}

The following analytic forms are used to fit the synchrotron coefficients.
\begin{itemize}

\item Model SI : 
Given the power law energy distribution of cosmic ray electrons, we assumed a power-law emissivity in terms of brightness temperature
over the WMAP frequencies as
\begin{equation}
T_A (\nu) = A_{s} \times \left( \frac{\nu}{23} \right )^{\beta_{s}}_{\text{GHz}}
\end{equation}
where $\beta_s$ is the spectral index and $A_s$ is the normalised amplitude expressed in $\mu \text{K}$ with respect to the frequency 
$\nu_0 = 23$\,GHz.

\item Model SII : 
A power-law emissivity is assumed to extend from 408~MHz up to and through the WMAP frequencies. Since we use the 408~MHz survey as a
template for the synchrotron emission, the amplitude at the low frequency must be reproduced perfectly. This results in an effective
constraint to be applied to the analytical form above, and we then fit for the spectral index $\beta_{s}$ only.

\begin{equation}
T_A (\nu) = 10^6 \times \left( \frac{\nu}{0.408} \right )^{\beta_{s}}_{\text{GHz}}
\end{equation}

\item Model SIII : 
The cosmic ray electron energy spectrum is expected to steepen with time due to radiation energy loss. A review of cosmic-ray propagation
including electrons can be found in \citet{Strong_etal:2007}, whilst \citet{Strong_etal:2011} directly test propagation models based on 
cosmic-ray and gamma-ray data against synchrotron data from 22~MHz to 94~GHz as averaged over mid-latitude regions 
($10^{\circ} < \mid b \mid < 45^{\circ}$). The latter analysis confirms the need for a low-energy break in the cosmic-ray electron
injection spectrum to account for the steepening synchrotron spectrum. Since we do not include synchrotron information at frequencies
intermediate to 408~MHz and the \emph{WMAP} data, we follow the treatment of \citet{Gold_WMAP5:2009}. Specifically, 
the emissivity is assumed to follow a power-law from 408~MHz until K-band and then to exhibit spectral curvature as follows,
\begin{align*}
T_A (\nu) &= A_{s} \times \left( \frac{\nu}{23} \right )_{\text{GHz}}^{\beta_{s}}  &\nu < \nu_\text{K} \\
               &= A_{s} \times \left( \frac{\nu}{23} \right )_{\text{GHz}}^{\beta_{s} + c_s ln(\frac{\nu}{\nu_\text{K}})}  & \nu > \nu_\text{K} 
\end{align*}
As above, the 408~MHz point is fixed, thus $A_s$ can be written in terms of $\beta_s$, and we are left to fit this spectral
index and the curvature $c_s$. For a WMAP frequency point, the above equation reduces to the form:
\begin{equation}
T_A (\nu) = 10^6 \times \left( \frac{\nu}{0.408} \right )^{\beta_s}_{\text{GHz}} \times \left( \frac{\nu}{23} \right )^{ c_s ln(\frac{\nu}{\nu_K})}_{\text{GHz}}
\end{equation}

\end{itemize}

\begin{table}
\scriptsize 
\begin{center}
\begin{tabular}{lcccccccc}
\hline
            &\multicolumn{3}{c}{SI} &\multicolumn{2}{c}{SII} &\multicolumn{3}{c}{SIII}\\
\hline
Region & $A_{s}$ & $\beta_{s}$ & $\chi^2$   & $\beta_{s}$  & $\chi^2$ & $\beta_{s}$ & $c_s$ & $\chi^2$  \\
\hline
                                   EBV    &    7.96 $\pm$    0.18    &   -2.99 $\pm$    0.14    &   0.047    &   -2.91 $\pm$    0.01    &   0.106    &   -2.91 $\pm$    0.01    &   -0.17 $\pm$    0.28    &   0.011   \\
                                   KQ85    &    7.64 $\pm$    0.20    &   -3.11 $\pm$    0.19    &   0.246    &   -2.92 $\pm$    0.01    &   0.470    &   -2.92 $\pm$    0.01    &   -0.44 $\pm$    0.39    &   0.071   \\
                                    NPS     &    4.68 $\pm$    0.55    &   -3.43 $\pm$    0.92    &   0.021   &   -3.05 $\pm$    0.03    &   0.069  &   -3.04 $\pm$    0.03    &   -0.91 $\pm$    2.18    &   0.003   \\                                                                 GN    &    8.16 $\pm$    0.25    &   -2.94 $\pm$    0.19    &   0.007    &   -2.91 $\pm$    0.01    &   0.013    &   -2.91 $\pm$    0.01    &   -0.06 $\pm$    0.35    &   0.006   \\
                                    EN    &    6.41 $\pm$    0.26    &   -3.55 $\pm$    0.33    &   1.077    &   -2.97 $\pm$    0.01    &   1.914    &   -2.96 $\pm$    0.01    &   -1.38 $\pm$    0.81    &   0.628   \\
                                   GS    &    7.84 $\pm$    0.26    &   -2.99 $\pm$    0.21    &   0.082    &   -2.92 $\pm$    0.01    &   0.093    &   -2.92 $\pm$    0.01    &   -0.20 $\pm$    0.42    &   0.031   \\
                                   ES    &    7.95 $\pm$    0.22    &   -2.74 $\pm$    0.16    &   0.337    &   -2.91 $\pm$    0.01    &   0.504    &   -2.91 $\pm$    0.01    &    0.37 $\pm$    0.23    &   0.075   \\
\hline
     2    &    6.82 $\pm$    1.25    &   -3.05 $\pm$    1.23    &   0.018    &   -2.95 $\pm$    0.04    &   0.015    &   -2.95 $\pm$    0.05    &   -0.34 $\pm$    2.59    &   0.012   \\
     3    &    4.66 $\pm$    0.53    &   -3.03 $\pm$    0.76    &   0.006    &   -3.04 $\pm$    0.03    &   0.005    &   -3.05 $\pm$    0.03    &    0.08 $\pm$    1.30    &   0.005   \\
     8    &    9.23 $\pm$    0.53    &   -3.01 $\pm$    0.38    &   0.002    &   -2.88 $\pm$    0.01    &   0.037    &   -2.87 $\pm$    0.01    &   -0.25 $\pm$    0.72    &   0.005   \\
    10    &    7.67 $\pm$    2.04    &   -0.56 $\pm$    0.42    &   0.076    &   -2.86 $\pm$    0.05    &   2.305    &   -2.87 $\pm$    0.05    &    1.69 $\pm$    0.28    &   0.700   \\
    11    &   12.61 $\pm$    1.24    &   -4.33 $\pm$    1.16    &   0.560    &   -2.81 $\pm$    0.02    &   1.403    &   -2.79 $\pm$    0.03    &   -3.67 $\pm$    3.07    &   0.441   \\
    12    &    7.33 $\pm$    1.88    &   -2.09 $\pm$    1.03    &   0.028    &   -2.92 $\pm$    0.06    &   0.121    &   -2.93 $\pm$    0.06    &    1.01 $\pm$    0.88    &   0.010   \\
    15    &    9.74 $\pm$    1.62    &   -5.38 $\pm$    2.78    &   1.284    &   -2.88 $\pm$    0.04    &   1.801    &   -2.85 $\pm$    0.04    &   -5.99 $\pm$    7.08    &   1.194   \\
    16    &    5.88 $\pm$    0.62    &   -2.96 $\pm$    0.69    &   0.002    &   -2.99 $\pm$    0.03    &   0.002    &   -2.99 $\pm$    0.03    &    0.01 $\pm$    1.32    &   0.002   \\
    17    &    6.42 $\pm$    1.26    &   -5.08 $\pm$    3.08    &   0.574    &   -2.98 $\pm$    0.05    &   0.822    &   -2.95 $\pm$    0.05    &   -4.99 $\pm$    7.95    &   0.527   \\
    18    &    7.95 $\pm$    1.96    &   -1.62 $\pm$    0.77    &   0.070    &   -2.89 $\pm$    0.05    &   0.373    &   -2.90 $\pm$    0.06    &    1.26 $\pm$    0.55    &   0.055   \\
    21    &    5.76 $\pm$    1.69    &   -3.18 $\pm$    1.99    &   0.001    &   -2.99 $\pm$    0.07    &   0.003    &   -2.99 $\pm$    0.07    &   -0.40 $\pm$    4.20    &   0.001   \\
    22    &    5.65 $\pm$    2.39    &   -1.44 $\pm$    1.17    &   0.031    &   -2.97 $\pm$    0.09    &   0.195    &   -2.98 $\pm$    0.09    &    1.40 $\pm$    0.83    &   0.038   \\
    24    &   11.98 $\pm$    2.20    &   -1.36 $\pm$    0.48    &   0.104    &   -2.78 $\pm$    0.04    &   1.056    &   -2.79 $\pm$    0.04    &    1.27 $\pm$    0.35    &   0.231   \\
    32    &    6.64 $\pm$    0.98    &   -2.91 $\pm$    0.93    &   0.007    &   -2.96 $\pm$    0.04    &   0.006    &   -2.96 $\pm$    0.04    &    0.05 $\pm$    1.63    &   0.007   \\
\hline
\end{tabular}
\end{center}
\caption{Model fits to the synchrotron coefficients determined between the 5 WMAP frequencies and the Haslam 408 MHz template for large 
sky areas (upper part of table) and for those regions that indicate a 3$\sigma$ significant amplitude at K-band (lower part). The models 
SI, SII and SIII are described in Section~\ref{sec:discussion_synch}. $A_{s}$ represents the normalisation amplitude at K-band, 
$\beta_{s}$ the synchrotron spectral index and  $c_{s}$ the spectral curvature. The key for the global fits is as for Table~\ref{tab:simple_synch_spectra}. \label{tab:synch_model_results}}
\end{table}

\begin{center}
\begin{figure}
\begin{tabular}{ccc}
\epsfig{file=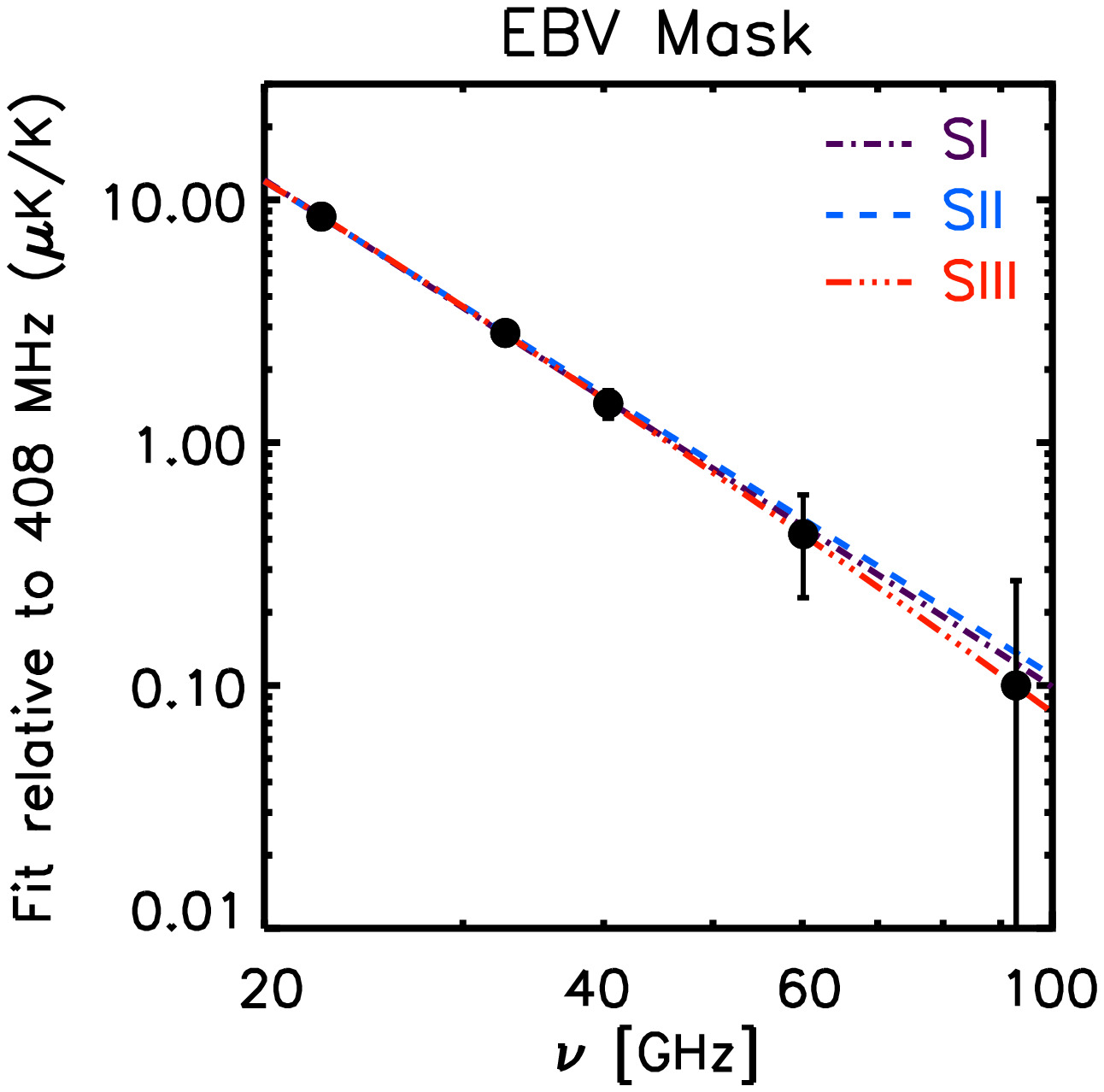,width=0.3\linewidth,angle=0,clip=} &
\epsfig{file=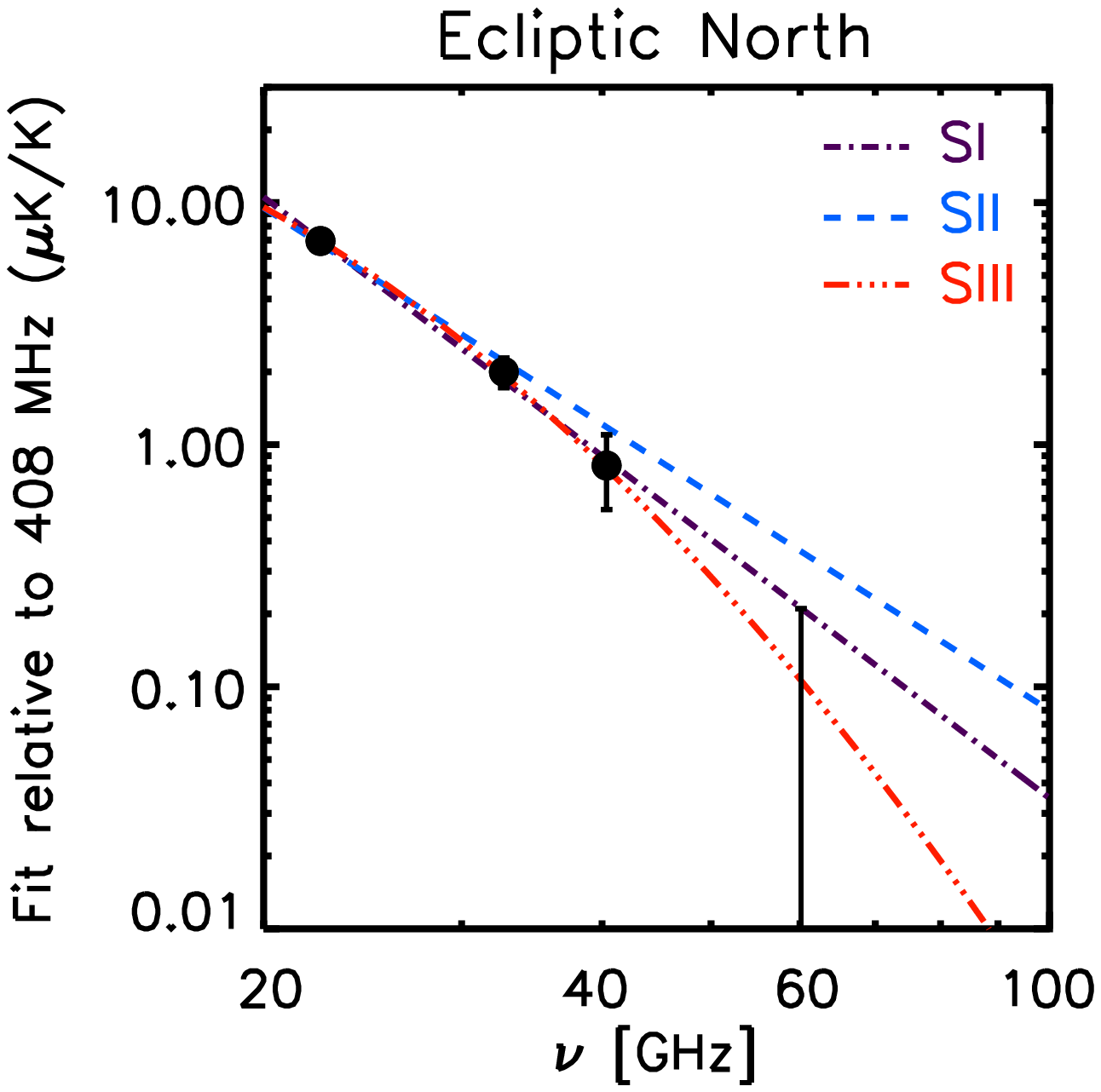,width=0.3\linewidth,angle=0,clip=} &
\epsfig{file=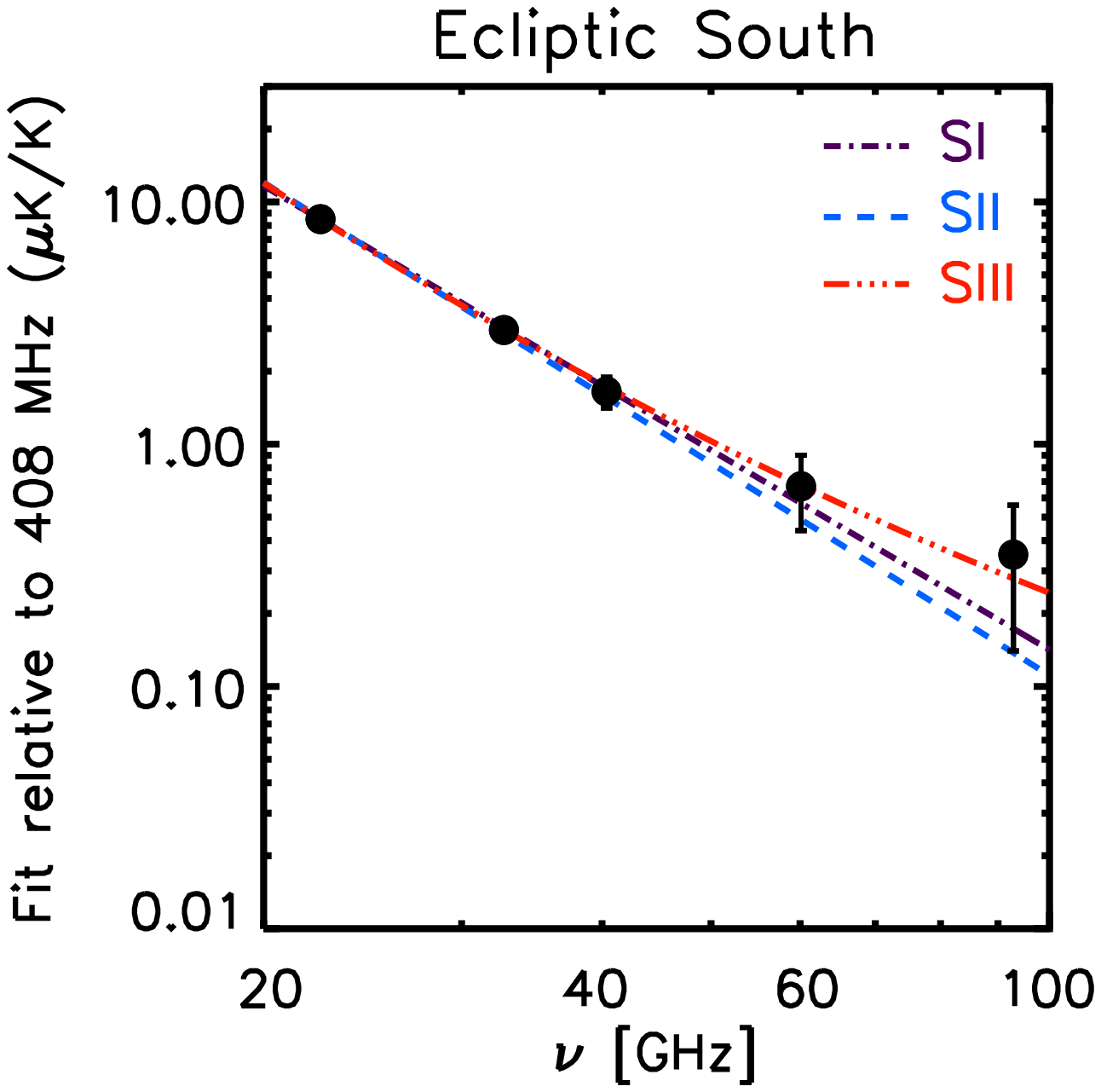,width=0.3\linewidth,angle=0,clip=} \\
\epsfig{file=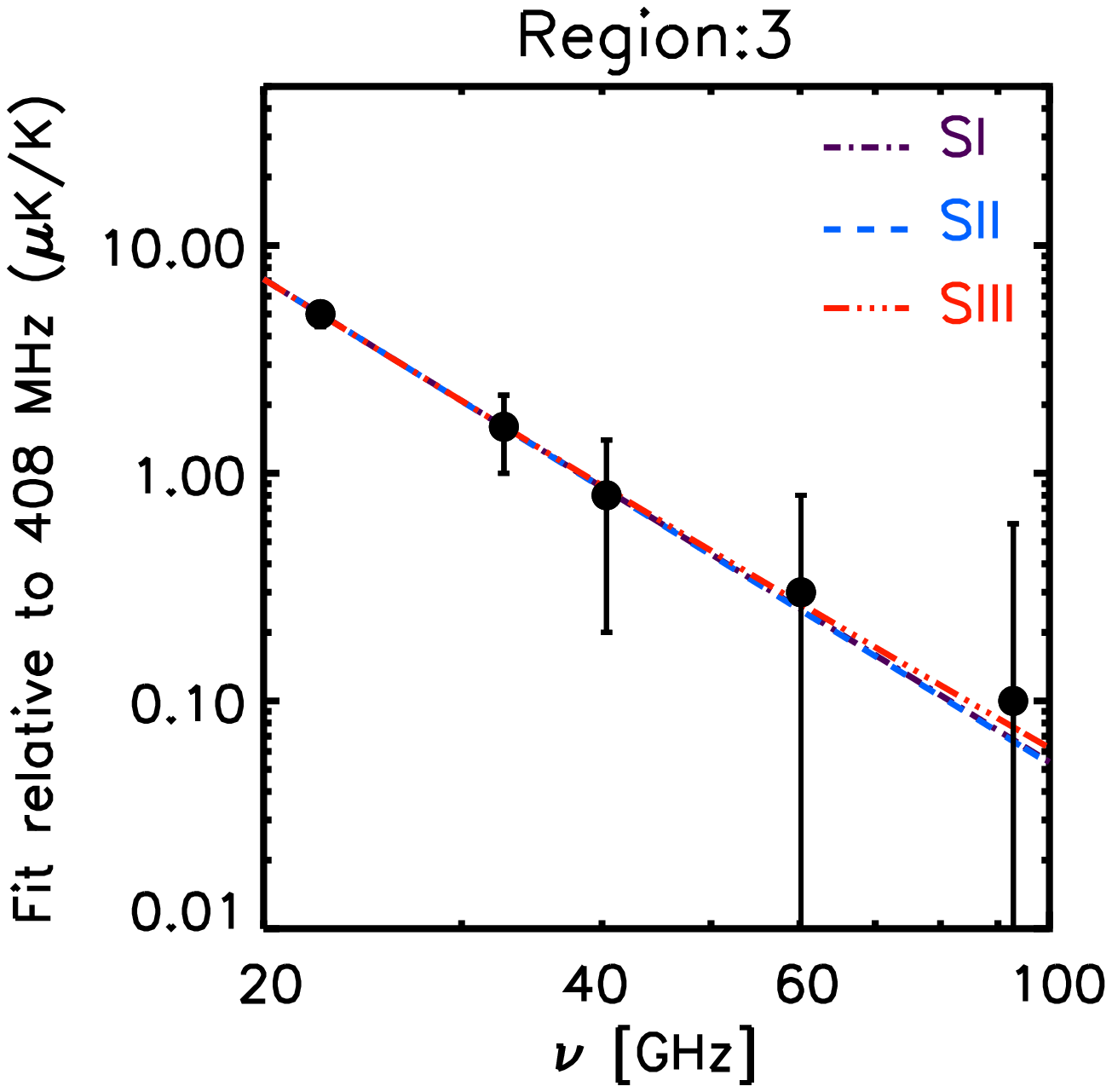,width=0.3\linewidth,angle=0,clip=}  &
\epsfig{file=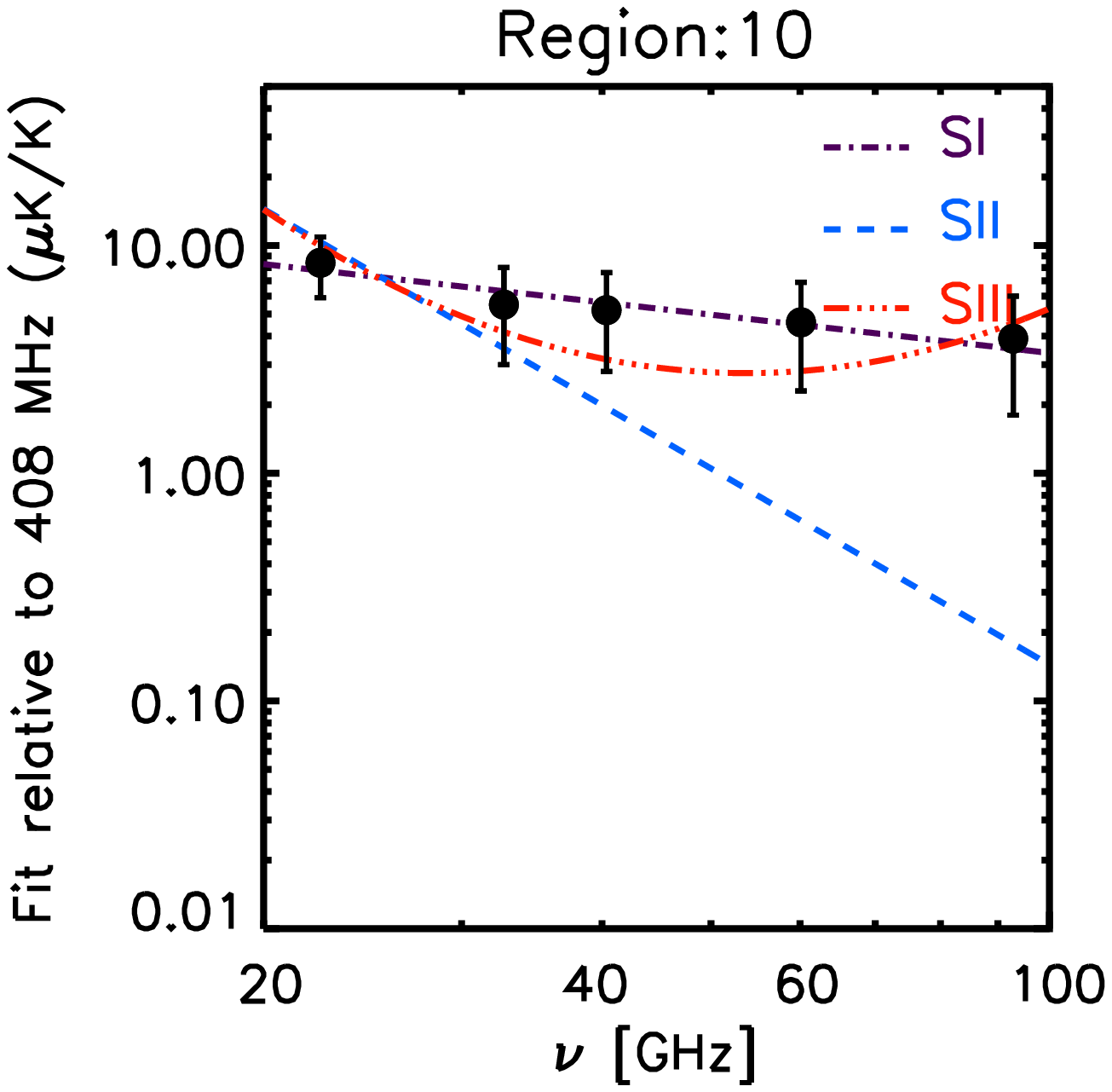,width=0.3\linewidth,angle=0,clip=}  &
\epsfig{file=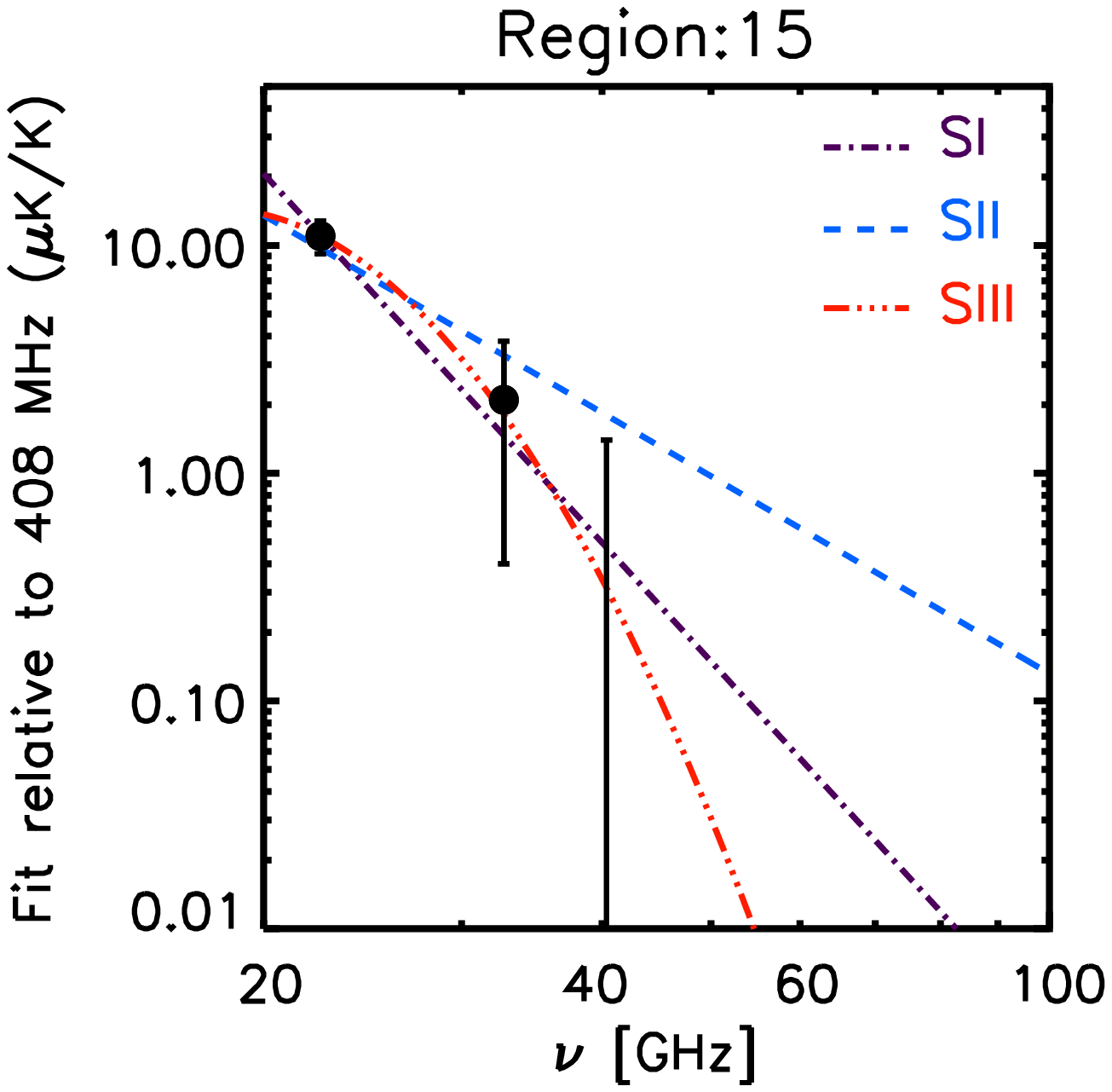,width=0.3\linewidth,angle=0,clip=} \\
\end{tabular}
\caption{Synchrotron spectral fits for various regions of the sky. A comparison is made of the observed template fits amplitudes 
derived from the 5 \emph{WMAP} frequency bands and the 408~MHz survey with the 3 models SI, SII and SIII as defined in 
Section~\ref{sec:discussion_synch}. The observed spectral dependencies of regions 3, 10 and 15 are representative of those seen for 
all regions with a significant detection at K-band.\label{fig:synch_model_plots}}
\end{figure}

\end{center}

The results are summarised in Table~\ref{tab:synch_model_results}. Fig.~\ref{fig:synch_model_plots} presents a comparison of the
model fits for three of the global masks, plus three of the regions that can be considered representative of the general results seen.
The $\chi^2$ results generally should be considered indicative rather than definitive, as would be expected given the small number of
data points and the comparable number of model parameters. 

The synchrotron fit coefficients obtained for the global masks that include the 408~MHz datum as a reference point (model SII) indicate a
typical power-law spectral index of $\beta_s=-2.9\pm0.1$. Model SI fits to the 5 \emph{WMAP} frequencies are generally steeper, however,
only the Ecliptic North region shows evidence of such behaviour at a significant level (a 2$\sigma$ shift in the spectral index). Indeed,
a model with negative spectral curvature is the best fit to the data, indicating steepening due to cosmic ray energy loss
mechanisms. Curiously, the Ecliptic South indicates spectral flattening, although at much lower significance. The SII fit to the
NPS region finds a spectral index $\beta_s=-3.05\pm0.03$ which is steeper than the average spectral index variation over the remaining
sky, as expected \citep{Lawson:1987}.

For the individual regions, the results are generally similar to the global fits. The spectral behaviour is consistent with emission from
power-law cosmic ray electron spectra with only hints of a steepening between 408~MHz and K-band. Region 11 might be considered to show 
weak evidence for negative spectral curvature. More interestingly, three regions (10, 12 and 24) indicate positive curvature that might be
understood as due to the presence of multiple emission regions with varying spectral behaviour along the line-of-sight.  However, region
10 has a very flat spectrum between K- and W-band that is strongly favoured over a steeper spectrum extending to 408~MHz, and perhaps
even mildly inconsistent with the curvature model adopted here. Whether this indicates some problem with the analysis in this
region, due to problems with the templates or cross-talk between components, is difficult to determine. It is apparent that the region
lies close to the North Celestial Pole where the 408~MHz template still retains considerable striations from the original survey, 
whereas we certainly consider that the template fitting methodology has by now been extensively tested. However, the spectral flattening 
seen in other analyses is less dramatic. For example, \citet{Gold_WMAP7:2011} find a range of values of $c_{s}$ between $\pm 0.7$ with a 
variance of $\sim 0.4$ for their MCMC analysis of the \emph{WMAP} data combined with the 408~MHz data.

\subsection{Free-free}\label{sec:discussion_freefree}

The free-free spectral index is almost constant over the \emph{WMAP} frequency range and changes only slightly with the the electron
temperature. Over this frequency range, the spectral dependence of the corresponding \halpha  emission can then be approximated with a
power-law model of fixed spectral index $-0.15$. However, \citet{Dobler_WMAP3:2008a} observed deviations from such
power-law behaviour, which they attributed to the presence of spinning dust emission in the WIM also traced by 
\halpha.  We investigate our template fit coefficients in terms of the following three models.

\begin{itemize}
\item Model FI :
We consider that the emission is due entirely to the free-free mechanism, ie. a power-law model emissivity with a fixed spectral index 
of $-0.15$ is assumed, 
\begin{equation}
I(\nu) = A_{f}\times \left( \frac{\nu}{23} \right )^{-0.15}_{\text{GHz}} 
\end{equation}
Thus only the free-free amplitude ($A_{f}$)  needs to be estimated from the template fit coefficients. \\

\item Model FII : 
The coefficients are fitted with an empirically motivated power-law model emissivity. This is particularly illustrative as to the 
extent that the standard free-free emissivity index is a poor fit to the data.
\begin{equation}
I(\nu) = A_{PL} \times \left( \frac{\nu}{23} \right )^{\beta_{PL}}_{\text{GHz}} 
\end{equation}
In this case, both the foreground amplitude ($A_{PL}$) and spectral index  ($\beta_{PL}$) are to be estimated.\\

\item Model FIII : 
The coefficients are fitted with a combination of free-free emission and a spinning dust model computed by the 
{\tt{SPDUST2}} code \citep{SpDust2:2011} for typical WIM conditions. This model peaks around 30~GHz, but we allow a simple
shift $\Delta \nu_{\text{WIM}}$ to be applied to the spectrum, as an approximation that represents the effect of varying the WIM physical 
parameters to match the model spectrum to the data.
\begin{equation}
I(\nu) =   A_{f} \times\left( \frac{\nu}{23} \right)^{-0.15}_{\text{GHz}}  + A_{\text{WIM}}  \times D_{\text{WIM}}(\nu - \Delta \nu_{\text{WIM}})  
\end{equation}
Here, $D_{\text{WIM}}(\nu)$ represents the normalised spinning dust spectral model for the WIM at a given frequency.
Clearly, we must now fit 3 parameters -- the free-free amplitude ($A_{f}$),  WIM amplitude ($A_{\text{WIM}}$ ) and WIM frequency shift 
($\Delta \nu_{\text{WIM}}$). We have also considered fits of this model to the coefficients derived using the \halpha after correction for 
dust absorption ($f_{d}=0.5$).\\

\end{itemize}

\begin{table}
\scriptsize 
\begin{center}
\begin{tabular}{lccccccccc}
\hline
            & \multicolumn{2}{c}{FI} & \multicolumn{3}{c}{FII} & \multicolumn{4}{c}{FIII} \\
\hline
Region & $A_{f}$ & $\chi^2$ & $A_{PL}$ & $\beta_{PL}$ & $\chi^2$ &  $A_{f}$ & $A_{WIM}$ & $\Delta \nu_{\text{WIM}}$ & $\chi^2$ \\ 
\hline
                                     EBV    &   0.156 $\pm$   0.001    &   6.374    &   0.153 $\pm$   0.002    &   0.010 $\pm$   0.036    &   1.633    &   0.150 $\pm$   0.004    &   2.094 $\pm$   0.429    &  13.575 $\pm$   4.727    &   0.340   \\
                                    KQ85    &   0.166 $\pm$   0.003    &   2.449    &   0.163 $\pm$   0.003    &   0.061 $\pm$   0.067    &   0.205    &   0.161 $\pm$   0.004    &   2.822 $\pm$   1.204    &  15.930 $\pm$   6.463    &   0.439   \\
                                      GN    &   0.163 $\pm$   0.002    &   2.057    &   0.161 $\pm$   0.003    &  -0.073 $\pm$   0.065    &   2.264    &   0.141 $\pm$   0.017    &   3.483 $\pm$   1.979    &   5.343 $\pm$   3.317    &   0.704   \\
                                      EN    &   0.210 $\pm$   0.004    &   2.177    &   0.207 $\pm$   0.004    &   0.031 $\pm$   0.070    &   0.730    &   0.203 $\pm$   0.008    &   3.442 $\pm$   1.245    &  14.290 $\pm$   7.196    &   0.062   \\
                                      GS    &   0.156 $\pm$   0.002    &   4.622    &   0.153 $\pm$   0.002    &   0.039 $\pm$   0.044    &   0.296    &   0.152 $\pm$   0.002    &   2.386 $\pm$   0.806    &  16.354 $\pm$   4.503    &   1.163   \\
                                      ES    &   0.146 $\pm$   0.001    &   3.499    &   0.144 $\pm$   0.002    &  -0.003 $\pm$   0.041    &   1.031    &   0.141 $\pm$   0.005    &   1.775 $\pm$   0.491    &  12.932 $\pm$   6.538    &   0.202   \\
\hline
     7    &   0.126 $\pm$   0.029    &   1.003    &   0.141 $\pm$   0.030    &  -2.043 $\pm$   2.257    &   0.601    &   0.000 $\pm$   0.000    &  15.936 $\pm$   3.512    &   0.000 $\pm$   0.000    &   1.381   \\
     8    &   0.111 $\pm$   0.021    &   0.086    &   0.110 $\pm$   0.024    &  -0.043 $\pm$   0.831    &   0.109    &   0.053 $\pm$   0.152    &   8.168 $\pm$  18.511    &   3.214 $\pm$  10.241    &   0.024   \\
     9    &   0.162 $\pm$   0.004    &   3.619    &   0.156 $\pm$   0.004    &   0.205 $\pm$   0.101    &   1.046    &   0.152 $\pm$   0.006    &   5.470 $\pm$   1.890    &  15.818 $\pm$   5.131    &   0.115   \\
    11    &   0.388 $\pm$   0.011    &   0.298    &   0.386 $\pm$   0.012    &  -0.105 $\pm$   0.128    &   0.358    &   0.388 $\pm$   0.011    &   0.000 $\pm$   0.000    &   0.000 $\pm$   0.000    &   0.596   \\
    12    &   0.135 $\pm$   0.034    &   0.731    &   0.150 $\pm$   0.036    &  -1.693 $\pm$   2.087    &   0.481    &   0.000 $\pm$   0.000    &  17.099 $\pm$   4.139    &   0.000 $\pm$   0.000    &   0.909   \\
    13    &   0.145 $\pm$   0.005    &   0.037    &   0.145 $\pm$   0.006    &  -0.119 $\pm$   0.169    &   0.039    &   0.137 $\pm$   0.038    &   1.238 $\pm$   4.593    &   4.601 $\pm$  20.905    &   0.003   \\
    14    &   0.156 $\pm$   0.003    &   2.336    &   0.153 $\pm$   0.003    &   0.060 $\pm$   0.066    &   0.096    &   0.152 $\pm$   0.003    &   2.858 $\pm$   1.513    &  17.556 $\pm$   5.663    &   1.060   \\
    15    &   0.200 $\pm$   0.011    &   0.836    &   0.191 $\pm$   0.012    &   0.227 $\pm$   0.192    &   0.051    &   0.192 $\pm$   0.012    &   6.819 $\pm$   6.145    &  20.334 $\pm$   7.279    &   0.627   \\
    18    &   0.235 $\pm$   0.027    &   0.772    &   0.216 $\pm$   0.029    &   0.620 $\pm$   0.358    &   0.018    &   0.219 $\pm$   0.031    &  19.115 $\pm$  17.644    &  21.162 $\pm$   7.240    &   0.612   \\
    20    &   0.135 $\pm$   0.012    &   1.495    &   0.123 $\pm$   0.013    &   0.685 $\pm$   0.276    &   0.116    &   0.127 $\pm$   0.014    &  14.890 $\pm$   9.172    &  23.658 $\pm$   4.864    &   1.101   \\
    21    &   0.250 $\pm$   0.043    &   1.159    &   0.212 $\pm$   0.044    &   1.062 $\pm$   0.411    &   0.014    &   0.222 $\pm$   0.048    &  42.774 $\pm$  29.924    &  23.582 $\pm$   5.653    &   0.893   \\
    23    &   0.164 $\pm$   0.023    &   0.157    &   0.168 $\pm$   0.025    &  -0.447 $\pm$   0.708    &   0.134    &   0.066 $\pm$   0.170    &  12.174 $\pm$  20.912    &   0.000 $\pm$   0.000    &   0.144   \\
    24    &   0.118 $\pm$   0.005    &   0.340    &   0.116 $\pm$   0.006    &   0.074 $\pm$   0.195    &   0.048    &   0.114 $\pm$   0.015    &   1.977 $\pm$   1.888    &  13.646 $\pm$  21.049    &   0.093   \\
    32    &   0.164 $\pm$   0.041    &   0.374    &   0.144 $\pm$   0.042    &   0.914 $\pm$   0.657    &   0.019    &   0.150 $\pm$   0.046    &  19.687 $\pm$  27.751    &  22.391 $\pm$  10.970    &   0.368   \\
\hline
\end{tabular}
\end{center}
\caption{Model fits to the free-free coefficients determined between the 5 WMAP frequencies and the DDD \halpha template for large sky 
areas (upper part of table) and for those regions that indicate a 3$\sigma$ significant amplitude at K-band (lower part). The models FI, 
FII and FIII are fully defined in Section~\ref{sec:discussion_freefree}. $A_{f}$ represents the normalisation amplitude of the free-free
emission at K-band, $A_{PL}$ represents the normalisation amplitude for the power-law model emission at K-band, $\beta_{PL}$ the corresponding power-law spectral index, $A_{WIM}$ is the amplitude of the WIM spinning dust model in units of 10$^{20}$ R cm$^{-2}$, and 
$\Delta \nu_{\text{WIM}}$ the shift in frequency of the peak of the dust model to better the fit the data. The key for the global fits as 
for Table~\ref{tab:simple_synch_spectra}. \label{tab:freefree_model_hemispheres}}
\end{table}

\begin{center}
\begin{figure}
\begin{tabular}{ccc}
\epsfig{file=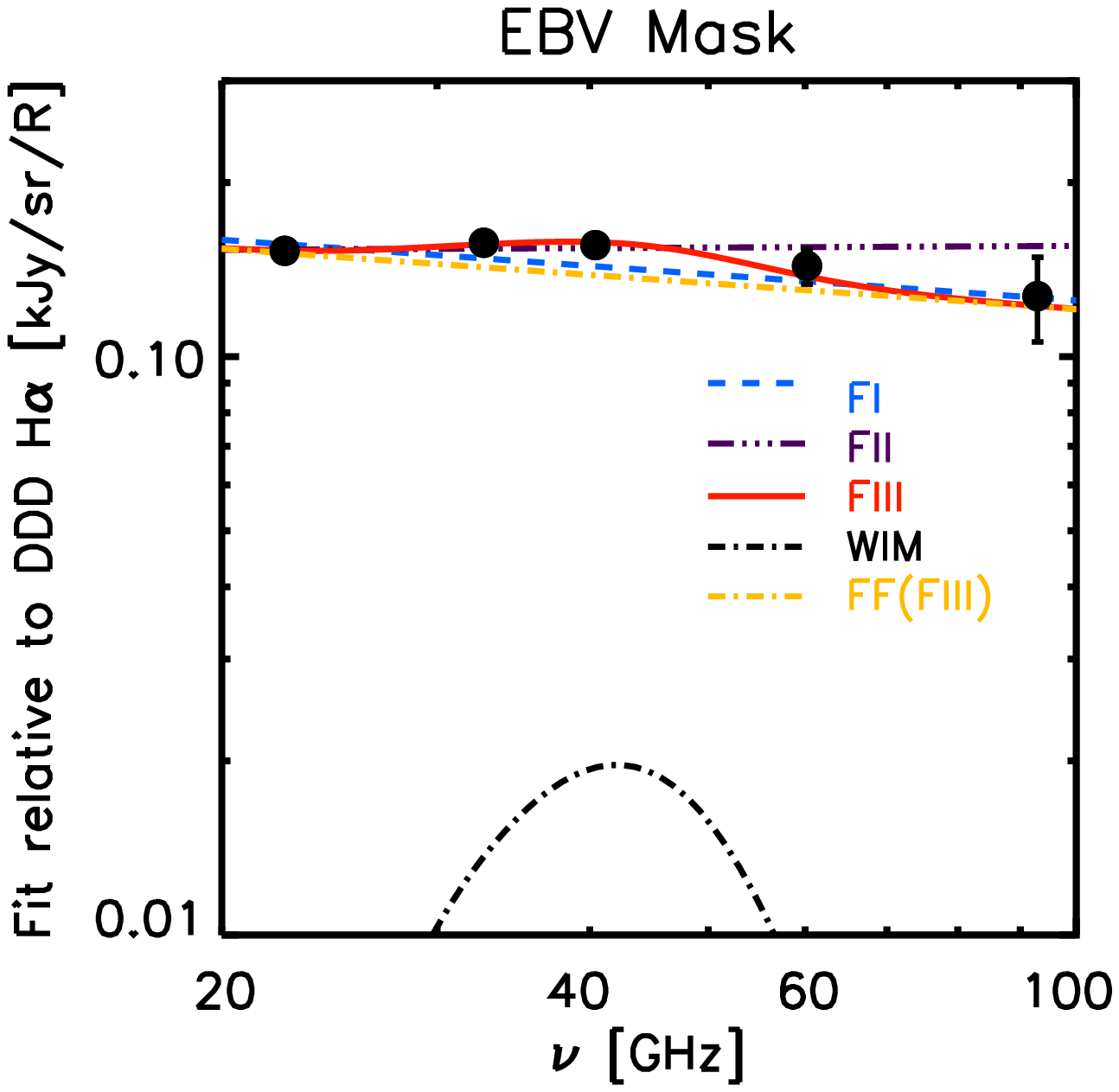,width=0.3\linewidth,angle=0,clip=} &
\epsfig{file=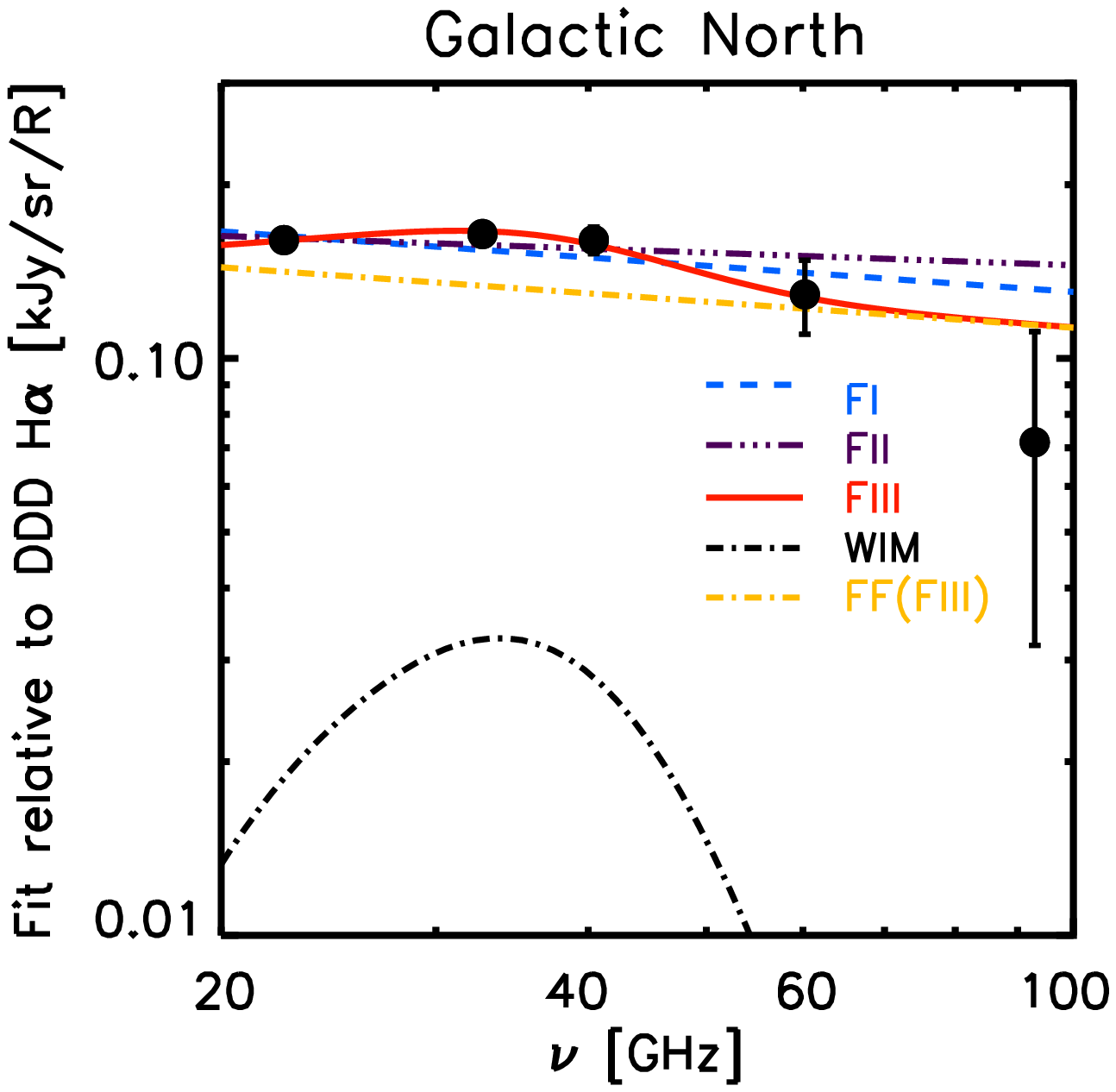,width=0.3\linewidth,angle=0,clip=} &
\epsfig{file=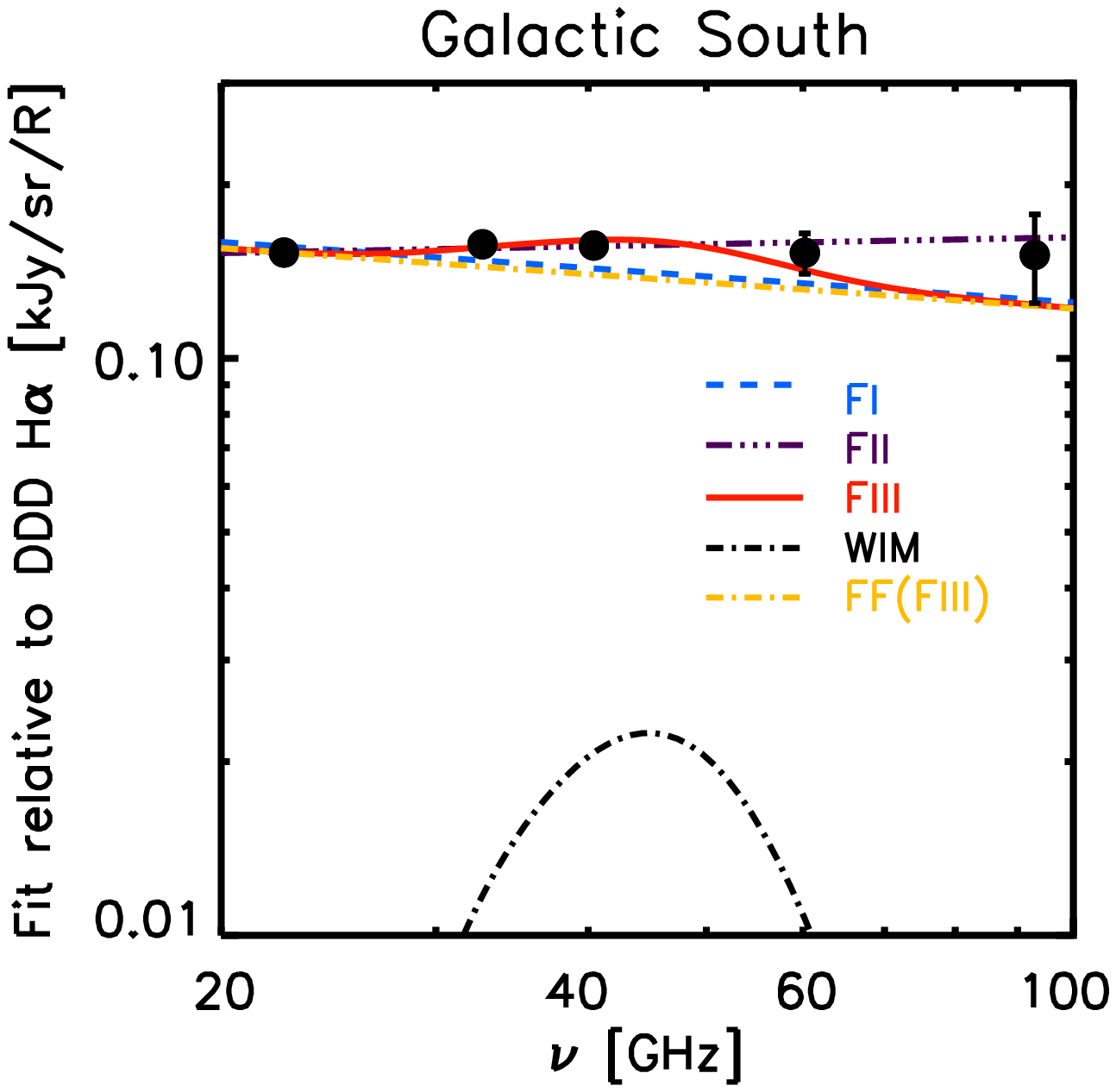,width=0.3\linewidth,angle=0,clip=} \\
\epsfig{file=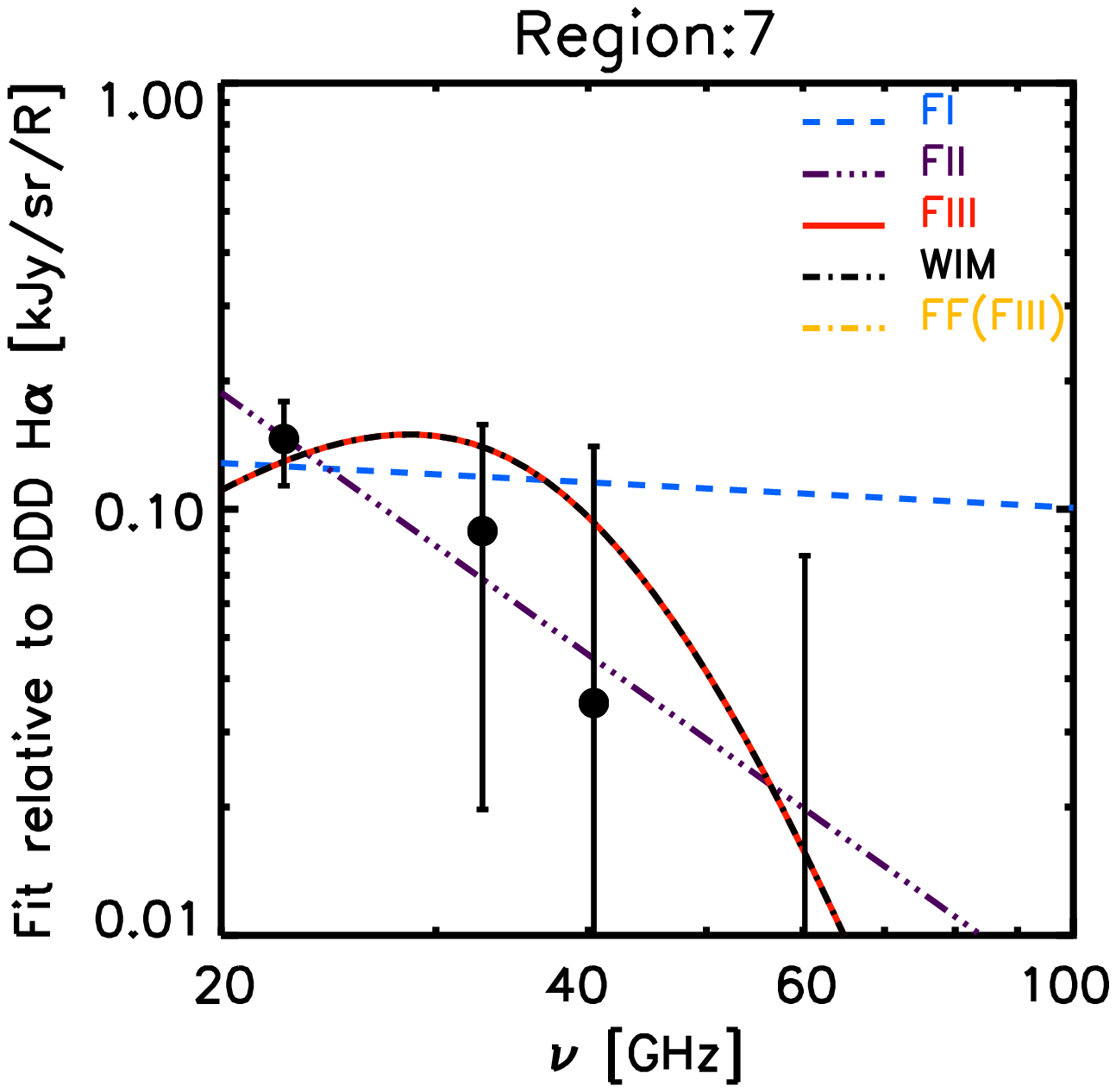,width=0.3\linewidth,angle=0,clip=} &
\epsfig{file=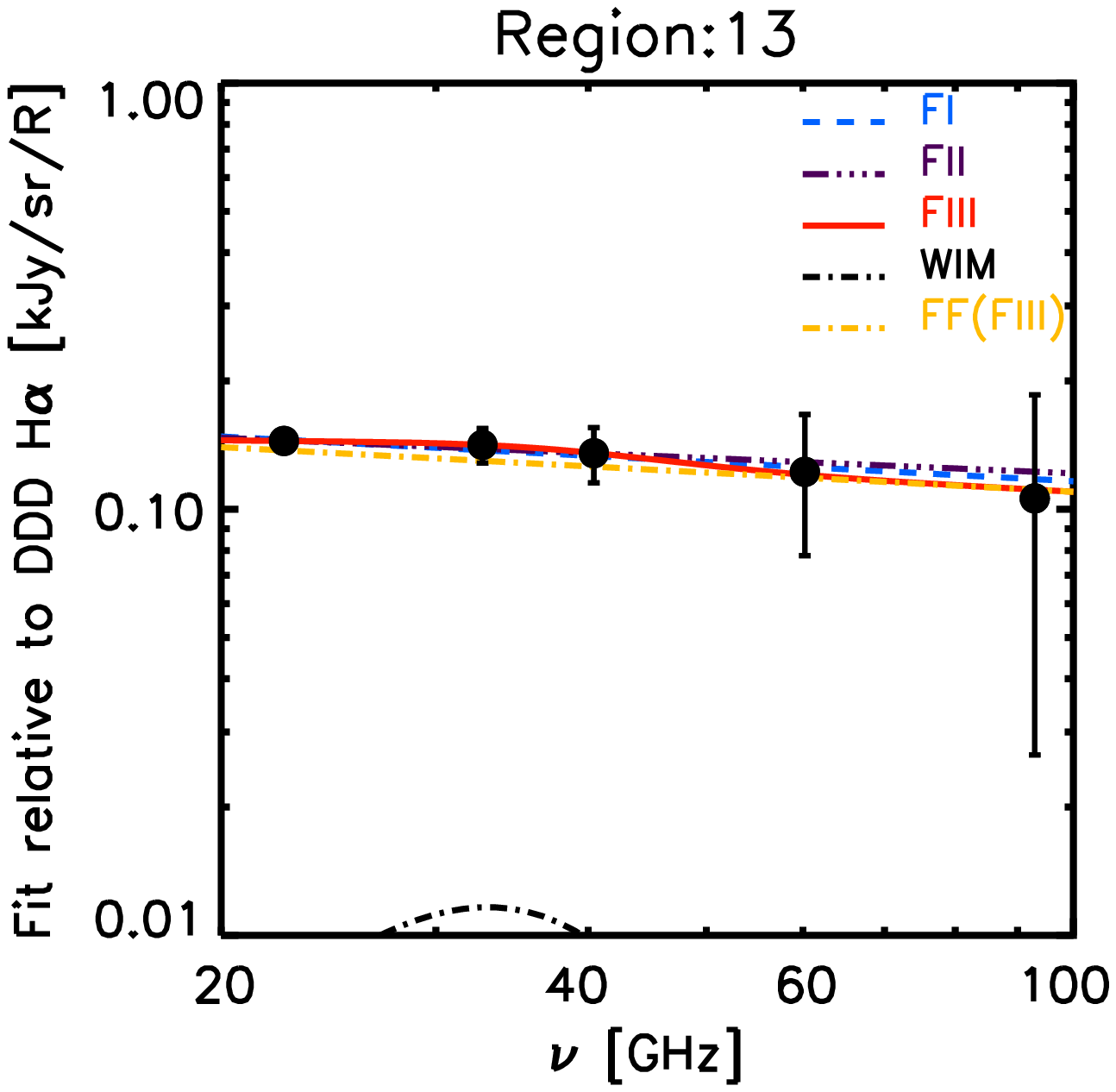,width=0.3\linewidth,angle=0,clip=} &
\epsfig{file=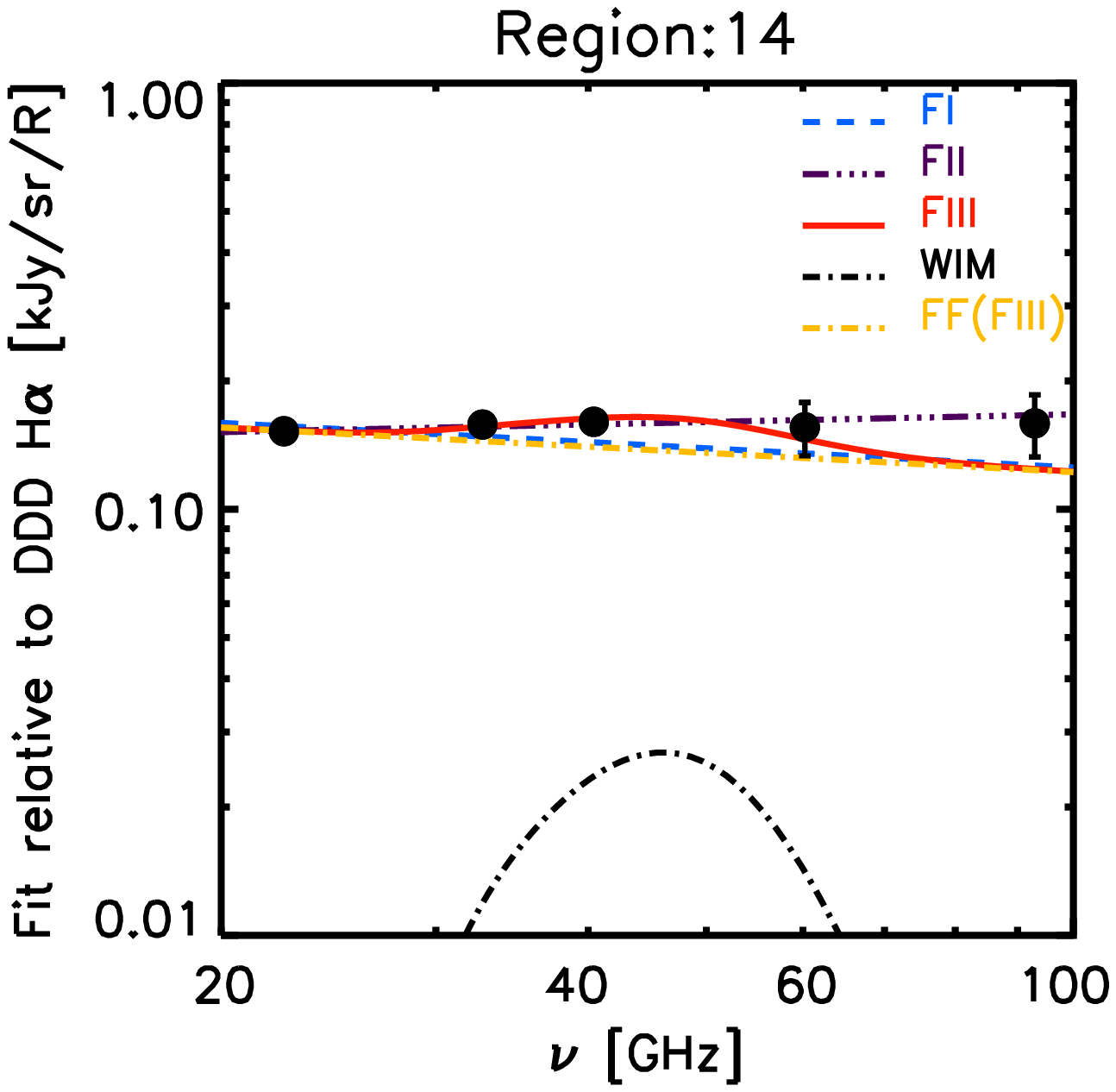,width=0.3\linewidth,angle=0,clip=} \\
\end{tabular}
\caption{Free-free spectral fits for various regions of the sky. A comparison is made of the observed template fits amplitudes derived 
from the 5 \emph{WMAP} frequency bands and the  DDD \halpha template with the 3 models FI, FII and FIII as defined in 
Section~\ref{sec:discussion_freefree}. Also shown in the plot are the separate free-free (FF) and WIM spinning dust components that 
constitute model FIII. The observed spectral dependencies of regions 7, 13 and 14 are representative of those seen for all regions with 
a significant detection at K-band. \label{fig:freefree_model_hemispheres}}
\end{figure}
\end{center}

The results are summarised in Table~\ref{tab:freefree_model_hemispheres}. Fig.~\ref{fig:freefree_model_hemispheres} presents a comparison
of the model fits for three of the global masks, plus three of the regions that can be considered representative of the general results 
seen.

The global fits all indicate a deviation from the free-free emission law (FI) at high significance. Unconstrained power-law fits (FII) 
are generally flatter, and models containing both free-free and spinning dust emission (FII) are typically preferred. For the EBV mask, 
the ratio of free-free to spinning dust emission (at the shifted peak frequency) is of order 7,  but this drops to 4 for the Galactic 
North. All regional fits are consistent with a 15\,GHz shift to higher frequencies of the peak intensity for the WIM spinning dust 
component with the exception of the Galactic North region which favours a smaller value of 5\,GHz. 

The fits to individual regions are generally consistent with the global mask results, and only region 9 shows a clear rejection of
model FI. The typical frequency shift is again of order 15~GHz, with a larger dispersion with several regions preferring no shift at all.
Of those, regions 7 and 12 can be explained by a single component - either free-free, power-law emission or spinning dust alone with no
frequency shift. However, it should be noted that the \halpha fit coefficients for these regions are only significant at K-band.
Region 11 is consistent with free-free only emission, and yields no evidence for spinning dust, but the region itself has been
flagged as anomalous as discussed previously in section \ref{sec:dust_extinct}.

Region 13 corresponds to the Northern Gum nebula, and is dominated by free-free emission with a small contribution from spinning dust at a
lower peak frequency, $\sim 35$~GHz, than is typical. The Southern part of the nebula is contained in region 14. This is the most clearly
detected structure on the sky, significant at all frequencies, as traced by the \halpha template, and shows significant evidence
for a spinning dust contribution with a free-free to spinning dust ratio of $\sim$ 5. Otherwise, this ratio varies considerably from
region to region.

Some regions that have a rising spectrum in terms of the \halpha coefficients and naturally favour models with a spinning dust
contribution over pure free-free are better fitted still by a flatter/rising power-law emission model. This might be alleviated with
more detailed spinning dust modelling (beyond the scope of our paper), or by including physical effects that increase the spinning dust
amplitude at frequencies higher than the peak. Such modifications have been investigated by \citet{Hoang:2010}.

The $A_{f}$ fit coefficients can, of course, be converted to estimates of the thermal electron temperature in the ionised medium. Since 
the global mask fits require the presence of a spinning dust component, we only consider the $T_e$ results from model FIII, as presented 
in Table~\ref{tab:freefree_electron_temperatures}. The table also includes results derived from fits to an \halpha template
corrected for dust absorption assuming $f_d=0.5$. We do not include the detailed coefficient results here since the interpretation
presented above remains essentially unchanged, and only conclusions about $T_{e}$ are affected. The global masks are consistent with
values of the electron temperature of order 6000\,K without any dust absorption correction, falling to 3000\,K when $f_d=0.5$ is
assumed. These values are for guidance only -- since the regions are not independent, an average is meaningless. It is interesting to note
that the Ecliptic North shows an enhanced temperature some 50\% higher than these typical values. Whether this reflects some property of 
the \halpha template is difficult to say, but the coverage is dominated by measurements from \emph{WHAM} data. Conversely, the Ecliptic 
South global fit gives a lower temperature that may also suggest issues with the template, given that it is largely comprised of the 
\emph{SHASSA} fields. New observations from the southern extension of the \emph{WHAM} survey \citep{Haffner_WHAM-S:2011} should help 
resolve this issue in the near future. Nevertheless, problems may still remain near the ecliptic poles given difficulties with removing 
the geocoronal \halpha contribution. For the individual regions (ignoring region 11 which is considered to be anomalous), we
find a weighted average of $6300 \pm  200$\,K (dropping to 5900\,K if region 11 is excluded) without a correction for
dust absorption, and $2900\pm 100$\,K otherwise. These values are in good agreement with the global averages as might be expected. 
However, note that the dispersion of values is considerably larger than the quoted uncertainty, implying true variations in temperature 
on the sky. Moreover, as discussed previously, it is likely that some of the dispersion seen reflects the existence of a range of values 
for both $T_{e}$ and $f_{d}$ throughout the Galaxy.

\begin{table}
\scriptsize
\begin{center}
\begin{tabular}{lcc}
\hline
                           &  \multicolumn{2}{c}{Inferred $T_{e}$  (K) for model FIII} \\
               Region &  $f_{d} = 0.0$ & $f_{d} = 0.5$ \\
\hline
         EBV  &   5900 $\pm$    300   &   2600 $\pm$    200   \\
        KQ85  &   6600 $\pm$    300   &   3200 $\pm$    200   \\
          GN  &   5300 $\pm$   1000   &   2500 $\pm$    500   \\
          EN  &   9500 $\pm$    600   &   5300 $\pm$    600   \\
          GS  &   6000 $\pm$    100   &   2600 $\pm$    100   \\
          ES  &   5300 $\pm$    300   &   2200 $\pm$    200   \\
\hline
    7$^{(a)}$  &   4400 $\pm$   1700   &   5600 $\pm$   1100   \\
    8$^{(a)}$  &   3600 $\pm$   1200   &   2300 $\pm$    900   \\
           9  &   6000 $\pm$    400   &   3200 $\pm$    300   \\
          11  &  25500 $\pm$   1100   &  11800 $\pm$    500   \\
   12$^{(a)}$  &   5000 $\pm$   2100   &   3300 $\pm$   1300   \\
          13  &   5100 $\pm$   2200   &   3500 $\pm$   1500   \\
          14  &   6000 $\pm$    200   &   2700 $\pm$    100   \\
          15  &   8600 $\pm$    800   &   5900 $\pm$    600   \\
          18  &  10700 $\pm$   2300   &   7700 $\pm$   1700   \\
          20  &   4500 $\pm$    800   &   2400 $\pm$    500   \\
          21  &  10900 $\pm$   3600   &   4000 $\pm$   1800   \\
   23$^{(a)}$  &   6800 $\pm$   1400   &   5400 $\pm$   1000   \\
          24  &   3700 $\pm$    600   &   2300 $\pm$    700   \\
          32  &   5900 $\pm$   2900   &   5600 $\pm$   2500   \\
\hline
\end{tabular}
\end{center}
\caption{Inferred free-free electron temperature $T_{e}$ in Kelvins corresponding to model fit FIII with dust corrections $f_d=0.0$ and 
$0.5$. $^(a)$ These regions show an effective degeneracy between free-free only or WIM spinning dust only solutions, thus FIII solutions 
have no contribution from free-free emission. We have therefore used the FI results to compute $T_{e}$ in these cases.}
\label{tab:freefree_electron_temperatures}
\end{table}

\subsection{Dust}\label{sec:discussion_dust}

The total dust emission is modelled as a combination of the relatively well-understood thermal dust emission and the AME. The former is 
assumed to have a fixed spectral index relative to the FDS8 94~GHz template over the \emph{WMAP} frequency range as determined
directly from the FDS8 model. We consider the following two models in order to fit the dust coefficients.

\begin{itemize}
\item Model DI : 
The dust coefficients are fitted with a combination of thermal (vibrational) dust and a power law dust-correlated AME. \\
\begin{equation}
T_A (\nu)  =  A_{PLD} \times \left( \frac{\nu}{23.} \right)^{\beta_{PLD}}_{\text{GHz}} + A_{TD}\times \left( \frac{\nu}{94} \right)^{1.55}_{\text{GHz}} 
\end{equation}
Three parameters -- the thermal dust amplitude ($A_{TD}$), the power law dust amplitude ($A_{PLD}$) and power law dust spectral index ($\beta_{PLD}$) -- are fitted to the coefficients.

\item Model DII : 
The dust coefficients are fitted with a combination of thermal dust and two spinning dust components (CNM and WNM).
The two spectra are generated using the {\tt{SPDUST2}} code assuming typical CNM and WNM conditions \citep{Draine_spinning:1998}. Given 
that both spectra peak at approximately the same frequency ($\approx 30$\,GHz), and the limited number of degrees of freedom available 
in the fit, it is only possible to apply a frequency shift to one component in order to match observations. \citet{Hoang:2011} found 
that modifying the CNM properties to increase its peak frequency yields a closer match to the \emph{WMAP} observations, and therefore we 
elect to allow a frequency shift of this component. \\
\begin{equation}
T_A (\nu)  =  A_{\text{WNM}}  \times D_{\text{WNM}}(\nu) + A_{\text{CNM}}  \times D_{\text{CNM}}(\nu - \Delta \nu_{\text{CNM}}) + A_{TD} \times  \left( \frac{\nu}{94} \right )^{1.55}_{\text{GHz}}  
\end{equation}
We fit four parameters to the template fit coefficients: the thermal dust amplitude ($A_{TD}$), the WNM amplitude ($A_{\text{WNM}}$) 
normalised at 23~GHz, the CNM amplitude ($A_{\text{CNM}}$) normalised at 41~GHz, and the CNM peak frequency shift ($\Delta \nu_{\text{CNM}}$).

\end{itemize}

\begin{table}
\scriptsize 
\begin{center}
\begin{tabular}{lccccccccc}
\hline
Region &\multicolumn{4}{c}{Model DI} &\multicolumn{5}{c}{Model DII}\\
\hline
 & $A_{PLD}$ &$\beta_{PLD}$ & $A_{TD}$ & $\chi^2$ & $A_{WNM}$ & $A_{CNM}$ & $\Delta {\nu_{\text{CNM}}}$ & $A_{TD}$ & $\chi^2$  \\
\hline
            EBV    &    9.15 $\pm$    0.09    &   -2.74 $\pm$    0.06    &    1.01 $\pm$    0.08    &   1.710    &    9.35 $\pm$    0.10    &    0.67 $\pm$    0.09    &   22.42 $\pm$    1.22    &    1.22 $\pm$    0.08    &   0.825   \\
           KQ85    &    7.61 $\pm$    0.08    &   -2.78 $\pm$    0.07    &    1.14 $\pm$    0.08    &   0.667    &    7.78 $\pm$    0.09    &    0.53 $\pm$    0.08    &   23.34 $\pm$    1.33    &    1.29 $\pm$    0.08    &   1.208   \\
      GN   &    8.91 $\pm$    0.14    &   -2.86 $\pm$    0.11    &    1.00 $\pm$    0.13    &   0.769    &    9.12 $\pm$    0.15    &    0.50 $\pm$    0.14    &   23.35 $\pm$    2.51    &    1.18 $\pm$    0.13    &   0.829   \\
      EN    &    8.32 $\pm$    0.13    &   -2.83 $\pm$    0.11    &    0.99 $\pm$    0.13    &   0.617    &    8.51 $\pm$    0.14    &    0.49 $\pm$    0.14    &   23.35 $\pm$    2.42    &    1.16 $\pm$    0.13    &   0.679   \\
      GS  &    9.73 $\pm$    0.13    &   -2.64 $\pm$    0.08    &    1.02 $\pm$    0.13    &   0.699    &    9.93 $\pm$    0.13    &    0.85 $\pm$    0.13    &   21.99 $\pm$    1.41    &    1.28 $\pm$    0.12    &   0.114   \\
      ES   &    9.82 $\pm$    0.13    &   -2.69 $\pm$    0.08    &    0.99 $\pm$    0.12    &   1.162    &   10.01 $\pm$    0.13    &    0.78 $\pm$    0.12    &   22.07 $\pm$    1.47    &    1.24 $\pm$    0.11    &   0.285   \\
\hline
       2    &    8.03 $\pm$    1.60    &   -4.14 $\pm$    2.16    &    0.00 $\pm$    0.00    &   0.093    &    8.17 $\pm$    1.63    &    0.00 $\pm$    0.00    &    0.00 $\pm$    0.00    &    0.00 $\pm$    0.00    &   0.378   \\
       3    &    8.73 $\pm$    0.99    &   -2.60 $\pm$    0.66    &    2.03 $\pm$    0.95    &   0.016    &    8.94 $\pm$    1.04    &    0.88 $\pm$    0.97    &   23.97 $\pm$    8.94    &    2.20 $\pm$    0.91    &   0.001   \\
       6    &    7.49 $\pm$    2.33    &   -3.06 $\pm$    2.31    &    0.84 $\pm$    2.20    &   0.007    &    7.68 $\pm$    2.44    &    0.17 $\pm$    2.50    &   25.27 $\pm$  100.72    &    0.99 $\pm$    2.23    &   0.002   \\
       7    &    9.79 $\pm$    0.63    &   -2.56 $\pm$    0.40    &    1.19 $\pm$    0.64    &   0.034    &    9.96 $\pm$    0.67    &    0.98 $\pm$    0.65    &   21.14 $\pm$    6.79    &    1.49 $\pm$    0.60    &   0.008   \\
       8    &    6.70 $\pm$    0.63    &   -2.73 $\pm$    0.62    &    1.41 $\pm$    0.62    &   0.019    &    6.86 $\pm$    0.66    &    0.56 $\pm$    0.68    &   24.81 $\pm$    8.69    &    1.48 $\pm$    0.61    &   0.018   \\
       9    &   13.09 $\pm$    0.45    &   -2.61 $\pm$    0.22    &    1.09 $\pm$    0.52    &   0.121    &   13.34 $\pm$    0.48    &    1.20 $\pm$    0.48    &   21.58 $\pm$    3.83    &    1.49 $\pm$    0.50    &   0.003   \\
      10    &    6.55 $\pm$    0.71    &   -3.16 $\pm$    0.76    &    0.67 $\pm$    0.60    &   0.051    &    6.69 $\pm$    0.74    &    0.11 $\pm$    0.70    &   25.45 $\pm$   44.79    &    0.78 $\pm$    0.61    &   0.110   \\
      11    &    4.20 $\pm$    0.63    &   -2.45 $\pm$    0.84    &    0.81 $\pm$    0.64    &   0.061    &    4.21 $\pm$    0.86    &    0.44 $\pm$    0.59    &   16.51 $\pm$   25.45    &    1.02 $\pm$    0.59    &   0.006   \\
      12    &    5.29 $\pm$    0.36    &   -3.79 $\pm$    0.68    &    0.35 $\pm$    0.38    &   0.304    &    5.40 $\pm$    0.37    &    0.00 $\pm$    0.00    &    0.00 $\pm$    0.00    &    0.34 $\pm$    0.35    &   1.691   \\
      13    &    5.61 $\pm$    0.71    &   -3.16 $\pm$    0.88    &    0.78 $\pm$    0.60    &   0.048    &    5.73 $\pm$    0.75    &    0.09 $\pm$    0.69    &   24.66 $\pm$   62.57    &    0.88 $\pm$    0.61    &   0.091   \\
      14    &    9.70 $\pm$    0.45    &   -2.53 $\pm$    0.28    &    1.20 $\pm$    0.43    &   0.033    &    9.87 $\pm$    0.48    &    1.03 $\pm$    0.46    &   21.51 $\pm$    4.49    &    1.50 $\pm$    0.40    &   0.000   \\
      15    &    8.25 $\pm$    0.54    &   -2.41 $\pm$    0.38    &    1.64 $\pm$    0.55    &   0.015    &    8.40 $\pm$    0.58    &    1.06 $\pm$    0.56    &   21.79 $\pm$    5.14    &    1.92 $\pm$    0.50    &   0.037   \\
      16    &    6.29 $\pm$    0.62    &   -4.72 $\pm$    1.36    &    0.00 $\pm$    0.00    &   0.327    &    6.38 $\pm$    0.63    &    0.00 $\pm$    0.00    &    0.00 $\pm$    0.00    &    0.00 $\pm$    0.00    &   3.088   \\
      17    &    6.59 $\pm$    1.79    &   -3.90 $\pm$    2.82    &    0.17 $\pm$    1.62    &   0.055    &    6.73 $\pm$    1.83    &    0.00 $\pm$    0.00    &    0.00 $\pm$    0.00    &    0.15 $\pm$    1.52    &   0.148   \\
      18    &    6.99 $\pm$    0.90    &   -3.41 $\pm$    1.10    &    0.57 $\pm$    0.78    &   0.053    &    7.14 $\pm$    0.91    &    0.00 $\pm$    0.00    &    0.00 $\pm$    0.00    &    0.67 $\pm$    0.72    &   0.127   \\
      19    &    9.29 $\pm$    1.88    &   -1.87 $\pm$    0.99    &    2.81 $\pm$    2.23    &   0.050    &    9.48 $\pm$    2.01    &    2.19 $\pm$    1.97    &   22.96 $\pm$    7.81    &    3.36 $\pm$    1.81    &   0.062   \\
      20    &    9.85 $\pm$    0.72    &   -2.90 $\pm$    0.52    &    0.87 $\pm$    0.71    &   0.070    &   10.09 $\pm$    0.76    &    0.45 $\pm$    0.76    &   23.50 $\pm$   14.84    &    1.08 $\pm$    0.71    &   0.024   \\
      21    &    6.92 $\pm$    0.45    &   -2.98 $\pm$    0.47    &    0.96 $\pm$    0.40    &   0.049    &    7.08 $\pm$    0.47    &    0.28 $\pm$    0.47    &   24.34 $\pm$   12.08    &    1.08 $\pm$    0.41    &   0.124   \\
      22    &    8.73 $\pm$    0.81    &   -2.40 $\pm$    0.54    &    1.69 $\pm$    0.87    &   0.019    &    8.87 $\pm$    0.87    &    1.08 $\pm$    0.85    &   20.69 $\pm$    8.30    &    2.03 $\pm$    0.80    &   0.023   \\
      23    &    8.96 $\pm$    0.36    &   -2.54 $\pm$    0.22    &    1.12 $\pm$    0.32    &   0.079    &    9.12 $\pm$    0.38    &    0.95 $\pm$    0.30    &   21.38 $\pm$    3.44    &    1.40 $\pm$    0.30    &   0.005   \\
      24    &    7.75 $\pm$    0.54    &   -3.56 $\pm$    0.63    &    0.04 $\pm$    0.49    &   0.207    &    7.92 $\pm$    0.55    &    0.00 $\pm$    0.00    &    0.00 $\pm$    0.00    &    0.10 $\pm$    0.45    &   0.867   \\
      28    &    8.46 $\pm$    1.80    &   -2.03 $\pm$    1.09    &    2.47 $\pm$    2.01    &   0.038    &    8.64 $\pm$    1.91    &    1.73 $\pm$    1.89    &   23.34 $\pm$    9.27    &    2.83 $\pm$    1.71    &   0.031   \\
      30    &    9.73 $\pm$    3.14    &   -2.00 $\pm$    1.62    &    2.30 $\pm$    3.47    &   0.006    &    9.85 $\pm$    3.36    &    2.09 $\pm$    3.18    &   21.92 $\pm$   14.67    &    2.81 $\pm$    2.91    &   0.002   \\
      31    &   12.00 $\pm$    3.23    &   -1.93 $\pm$    1.32    &    2.70 $\pm$    3.66    &   0.005    &   12.13 $\pm$    3.46    &    2.67 $\pm$    3.27    &   21.47 $\pm$   12.17    &    3.45 $\pm$    3.00    &   0.011   \\
      32    &    8.15 $\pm$    0.81    &   -2.87 $\pm$    0.70    &    0.85 $\pm$    0.81    &   0.003    &    8.33 $\pm$    0.85    &    0.53 $\pm$    0.87    &   24.26 $\pm$   12.86    &    0.95 $\pm$    0.81    &   0.098   \\
      33    &    8.53 $\pm$    1.88    &   -3.08 $\pm$    1.60    &    0.78 $\pm$    1.70    &   0.025    &    8.74 $\pm$    1.99    &    0.12 $\pm$    1.91    &   23.77 $\pm$  135.62    &    0.99 $\pm$    1.72    &   0.004   \\
\hline
\end{tabular}
\end{center}
\caption{Model fits to the dust coefficients determined between the 5 WMAP frequencies and the FDS8  template  for large sky areas 
(upper part of table) and for those regions that indicate a 3$\sigma$ significant amplitude at K-band (lower part). The models DI and
DII are fully defined in Section~\ref{sec:discussion_dust}. $A_{PLD}$ and $\beta_{PLD}$ represent the normalisation amplitude  at K-band
and spectral index respectively of a power-law anomalous dust emission component. $A_{WNM}$ is the amplitude of the WNM spinning dust 
model at K-band, $A_{CNM}$ is the amplitude of the CNM spinning dust model normalised at 41~GHz, and $\Delta \nu_{CNM}$ is the shift
in frequency of the peak of the CNM dust model to better the fit the data. $A_{TD}$ is the amplitude of the thermal dust emission with 
an assumed spectral index  $\beta_{TD} = 1.55$ as determined directly from the FDS8 dust model. The key for the global fits as for 
Table~\ref{tab:simple_synch_spectra}.}
\label{tab:dust_model_hemispheres}
\end{table}

\begin{center}
\begin{figure}
\begin{tabular}{ccc}
\epsfig{file=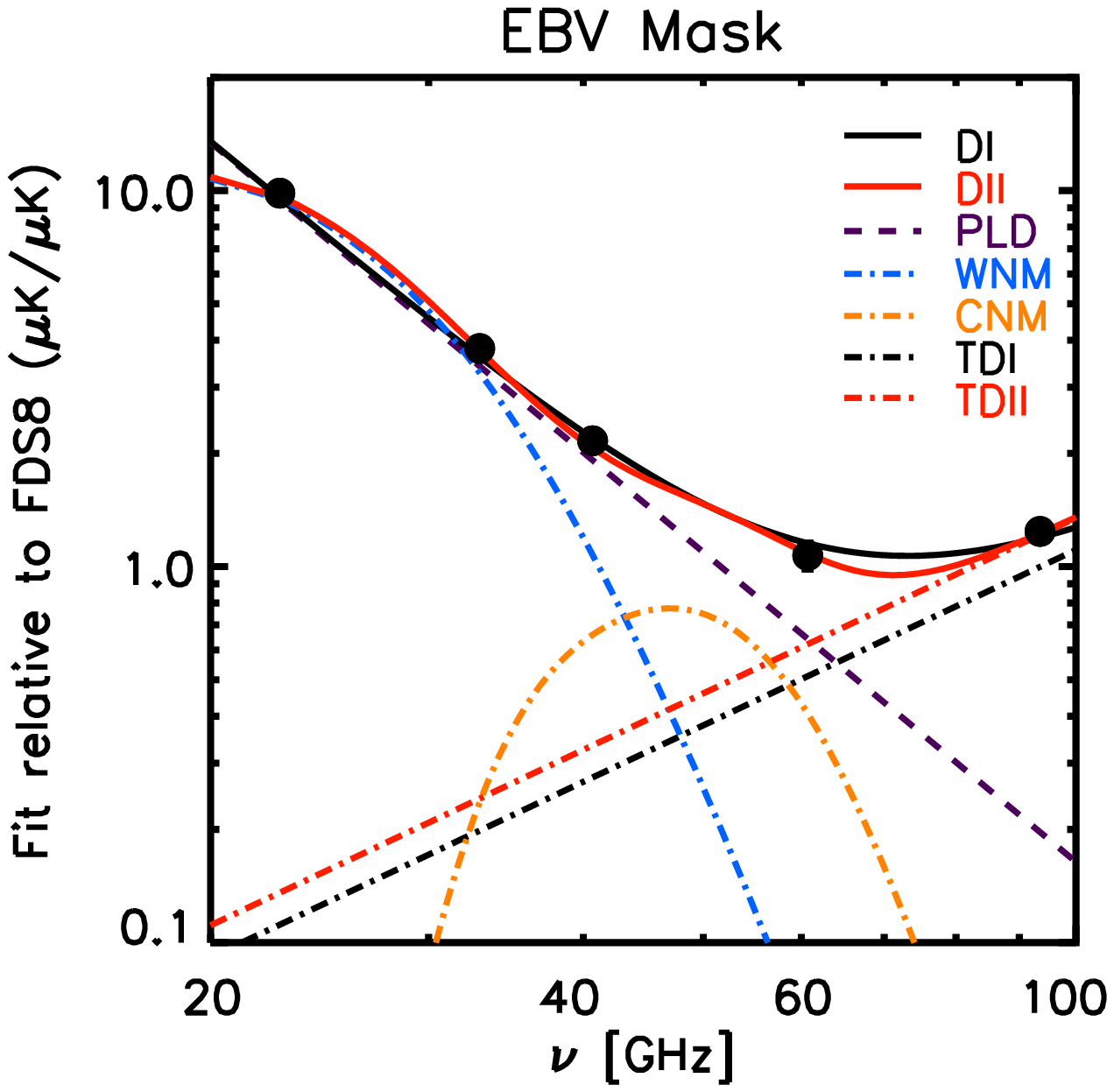,width=0.3\linewidth,angle=0,clip=} &
\epsfig{file=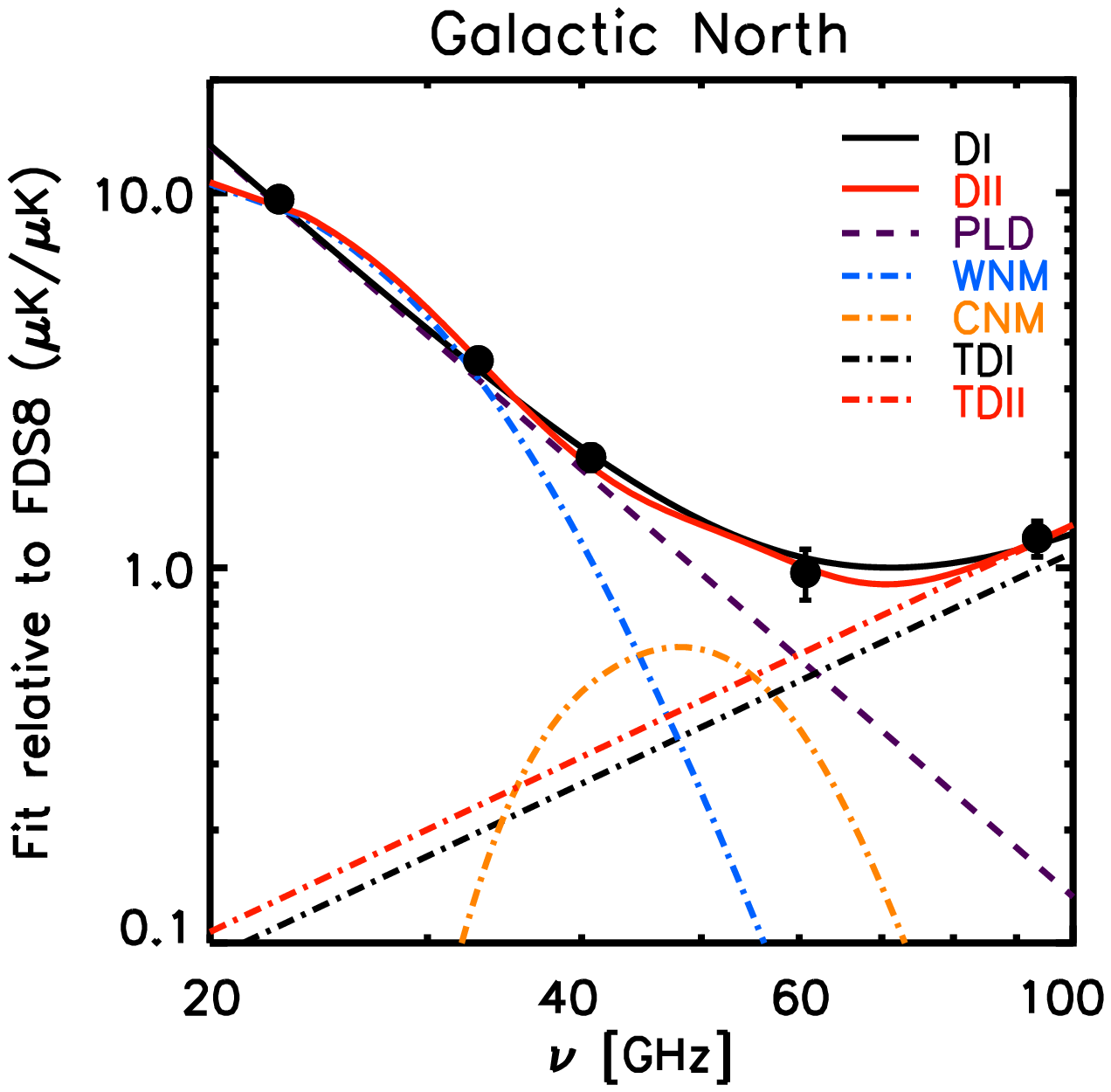,width=0.3\linewidth,angle=0,clip=} &
\epsfig{file=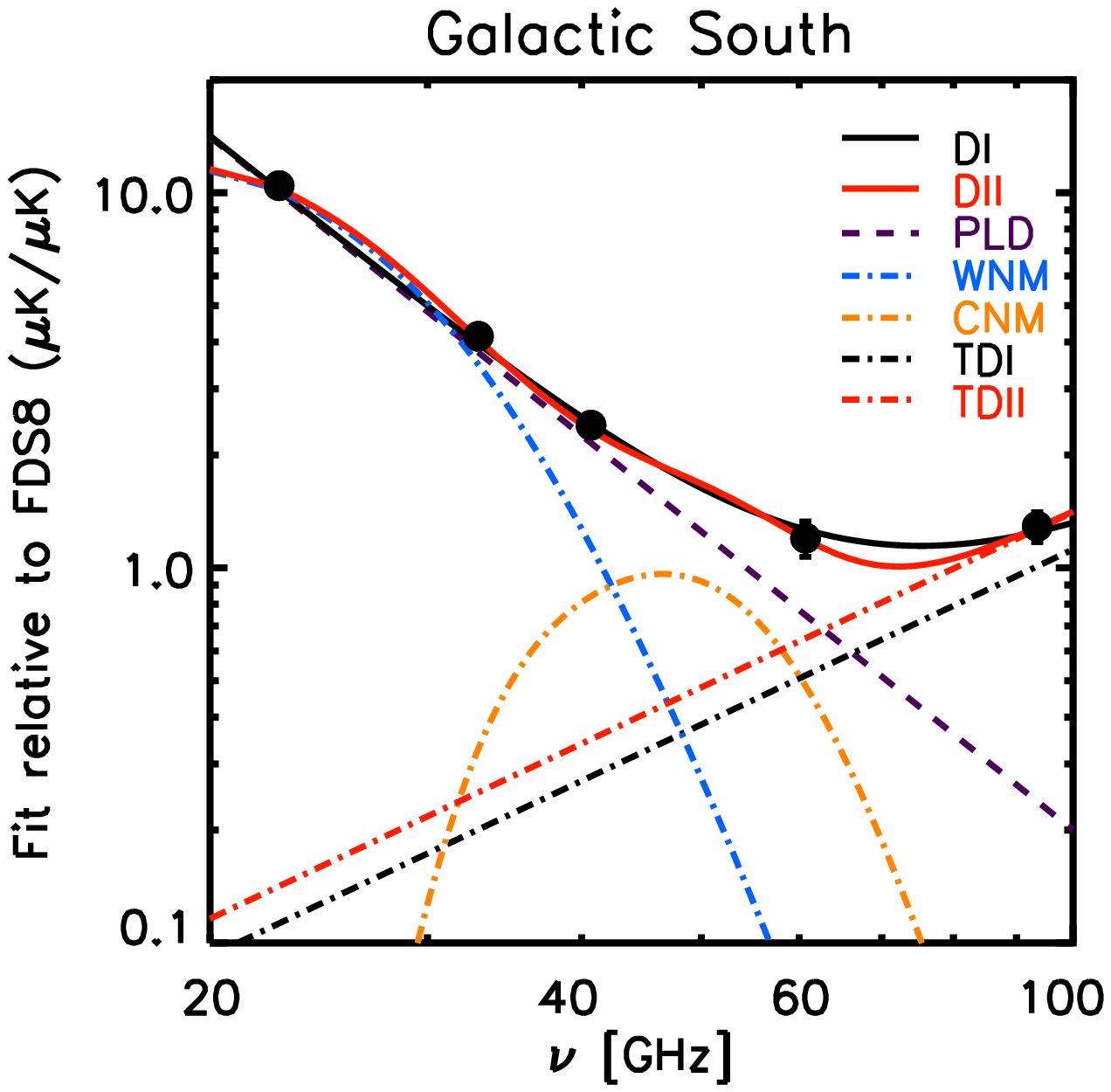,width=0.3\linewidth,angle=0,clip=} \\
\epsfig{file=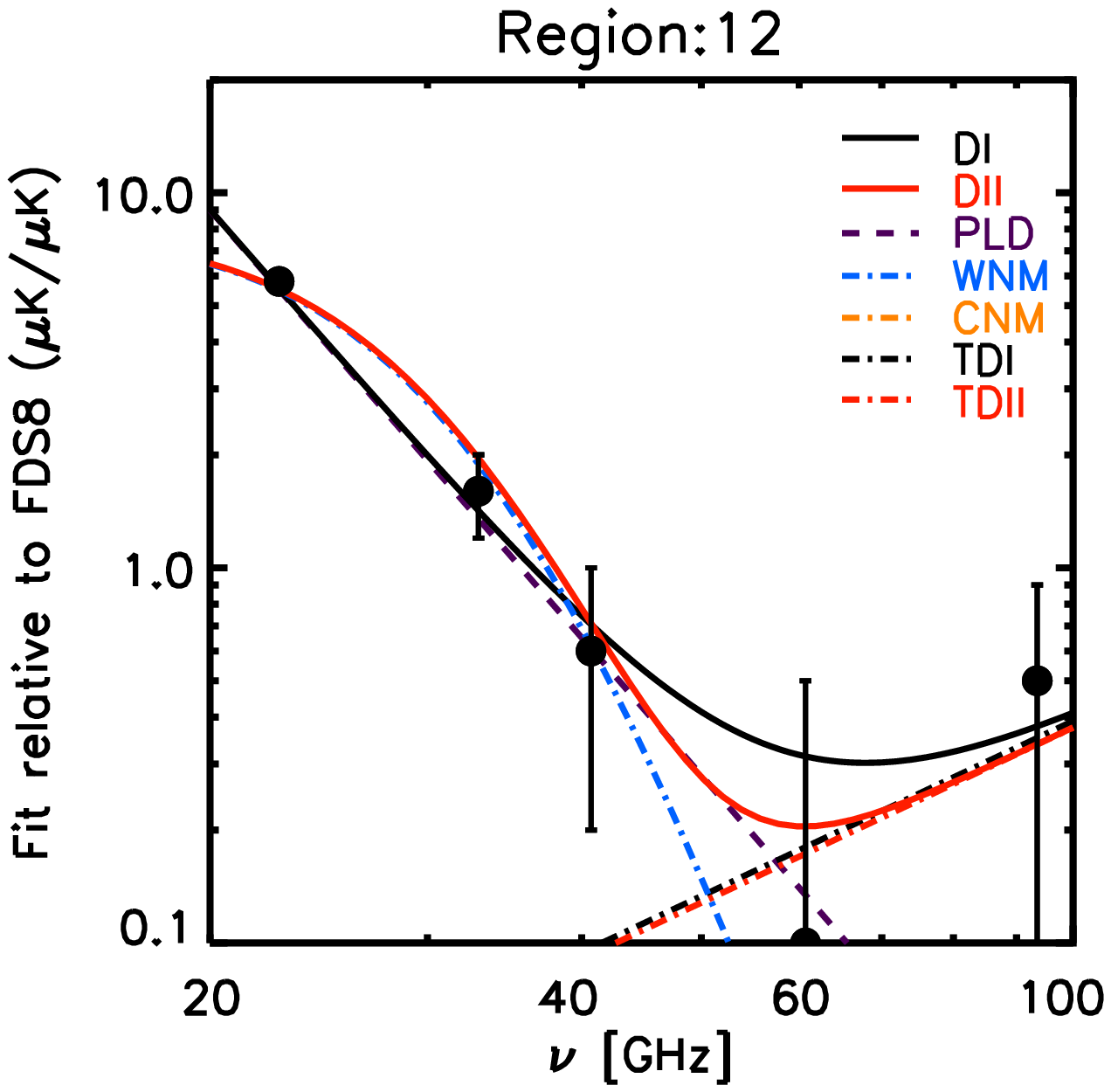,width=0.3\linewidth,angle=0,clip=} &
\epsfig{file=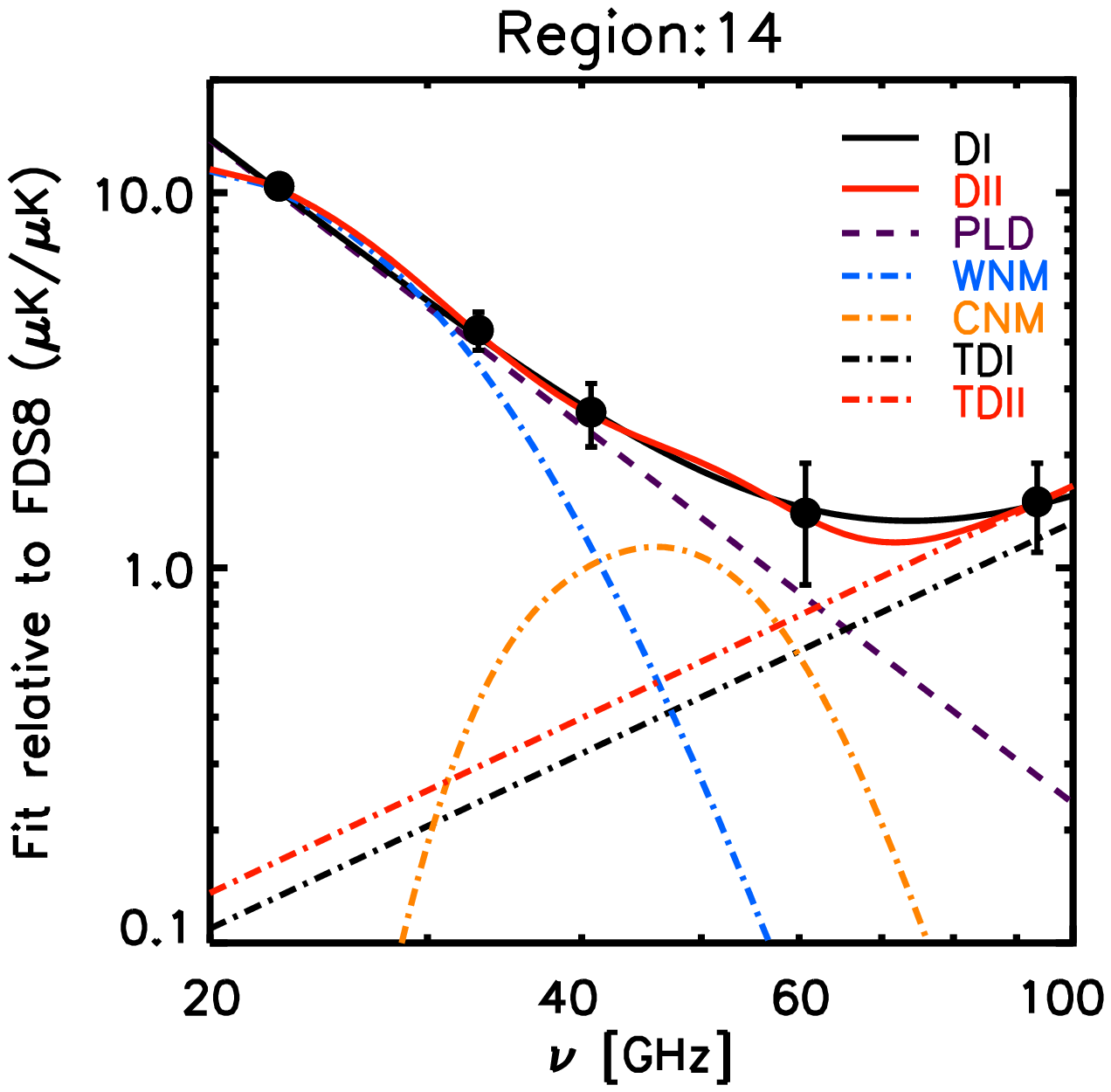,width=0.3\linewidth,angle=0,clip=} &
\epsfig{file=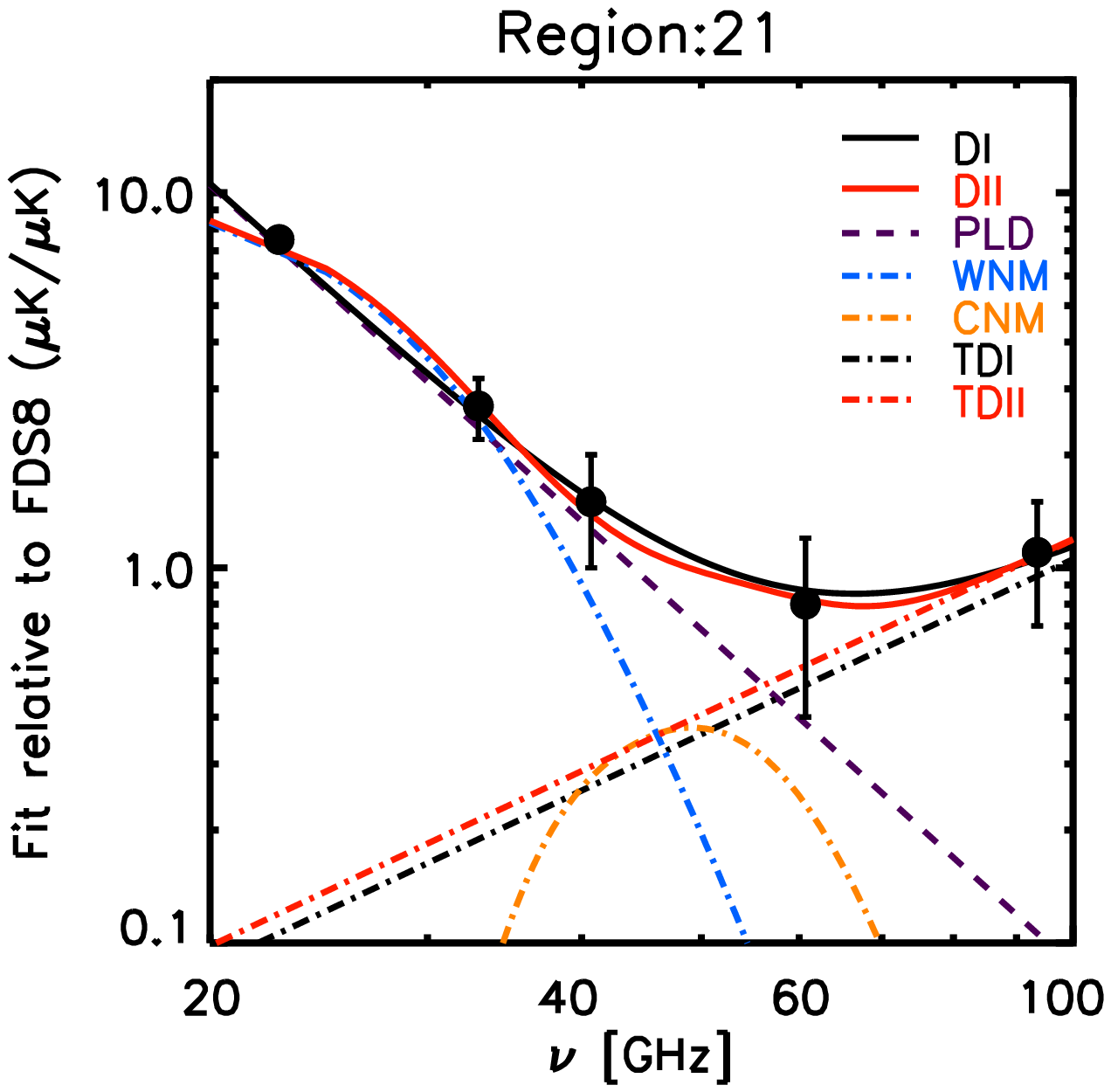,width=0.3\linewidth,angle=0,clip=} \\
\end{tabular}
\caption{Dust spectral fits for various regions of the sky. A comparison is made of the observed template fit amplitudes derived from the
5 \emph{WMAP} frequency bands and the FDS8 template with the 2 models DI and DII as defined in Section~\ref{sec:discussion_dust}. Also 
shown in the plot are the separate power-law anomalous dust (PLD) and vibrational (thermal) dust (VD) that constitute model DI, and the 
cold neutral medium (CNM) and warm neutral medium (WNM) components that, together with the vibrational dust, constitute model DII. The 
observed spectral dependencies of regions 12, 14 and 21 are representative of those seen for all regions with a significant detection at 
K-band.\label{fig:dust_model_hemispheres}}
\end{figure}
\end{center}

The results are summarised in Table~\ref{tab:dust_model_hemispheres}. Fig.~\ref{fig:dust_model_hemispheres} presents a comparison of the
model fits for three of the global masks, plus three of the regions that can be considered representative of the general results seen.

It should be apparent from the table that both models DI and DII provide adequate fits to the global masks. However, in order to
explain the emission at V-band in terms of spinning dust models, two contributions are required in combination with an enhanced thermal
dust normalisation some 20\% larger than that predicted at 94\,GHz by FDS8. Such models also require a shift in the peak CNM emission of
$\sim$22\,GHz and a ratio of WNM to CNM spinning dust emission of order 4:1.  In fact, if the CNM emission were to be omitted, or the
thermal dust amplitude kept at the canonical FDS8 value, then the V- and/or W-band amplitudes would be substantially underpredicted. In
practise, given the low number of data points sensitive to the thermal dust emission, there is an effective degeneracy between the CNM and
vibrational dust components, that may be alleviated somewhat if the spinning dust models can be enhanced in amplitude at frequencies
higher than their peak. Indeed, spinning dust can contribute significantly to emission in the $\sim$60 to 90 GHz range depending on
the local physical conditions. However, one might then expect to see a much flatter spectrum around the Q-, V-, and W-bands than is 
actually observed.  It may also be that some modification of the assumed thermal dust spectral index is required.

The fits to individual regions are less informative, as expected given the lack of clear detections for the template fit coefficients at 
V- and W-band. Given this, the analysis might be criticised for overfitting the data points. Nevertheless, both a power-law and a WNM
spinning dust component are good candidates to explain the lower frequency points. In almost all cases, a significant contribution from
CNM or thermal dust emission is not required.  This discrepancy between the regional fits and global analysis may simply reflect a
signal-to-noise issue at higher frequencies than the former, or the fact that the latter are essentially averages over many regions which
exhibit some variation in dust properties. Regions 12 and 16 are interesting in that the power-law model is clearly preferred over
model DII, although the analysis is really only constrained by the data points at K- and Ka-band.  For those regions in which a
significant vibrational dust contribution is required, the amplitude is systematically higher than the canonical FDS8 value, although 
still consistent with it. However, the required DI enhancement is lower, thus the DII results more likely again point to a deficiency in 
the specific spinning dust models used in this analysis.

The FDS8 model value for the dust spectral index used here is quite flat, whereas early \emph{Planck} papers
\citep{Planck_ERXXIV_ISM:2011,Planck_ERXXV_Molecular:2011} studying emission at frequencies higher than $\sim$100~GHz favour values 
closer to 1.8. However, \citet{Planck_ERXIX_DarkGas:2011} suggests that the dust SED flattens in the millimetre wavelength range. At 
lower frequencies, there is considerable uncertainty, and detailed modelling for candidate AME regions (including Perseus and 
$\rho$-Ophiucus) of the foreground spectra for all components and across the \emph{Planck} frequencies suggests values in the range 
1.5--2 \citep{Planck_ERXX_AME:2011}.  We have repeated the calculations above but imposing a thermal dust spectral index of 1.7 on the 
fits.  For model DI, there is a general increase in amplitude of the power-law dust component, with an associated flattening of the 
spectral index at levels below the 1-$\sigma$ error bar, and no impact on the thermal dust amplitude. Model DII shows similarly modest 
increases in the amplitudes of the two spinning dust components, with little change in the CNM peak frequency shift, and again no change 
in the thermal dust amplitude.  However, adopting the FDS8 thermal dust model at W-band and then imposing a steeper index to lower 
frequencies is not self-consistent.  Therefore, a more detailed treatment is required including specific modelling of all dust components
combined with higher frequency measurements. An obvious potential error in our analysis is the selective use of the CNM and WNM spectra 
for typical conditions in those phases of the ISM. What is clear is that the frequency range of 60--100\,GHz is a remarkably interesting 
regime for dust astrophysics.

\section{Discussion}\label{sec:conclusions}

In our study of the free-free, dust and synchrotron foreground components in the \emph{WMAP} data, we have selected fields based on the
morphological properties of three templates that trace the emission at wavelengths where the emission mechanism is (largely) dominant, 
and we assume that such partitions correlate well with spectral behaviour. Each of these components has then been quantified in terms of 
a mean value of the emissivity in each of the 5 \emph{WMAP} bands.

\begin{center}
\begin{figure}
\begin{tabular}{ccc}
\epsfig{file=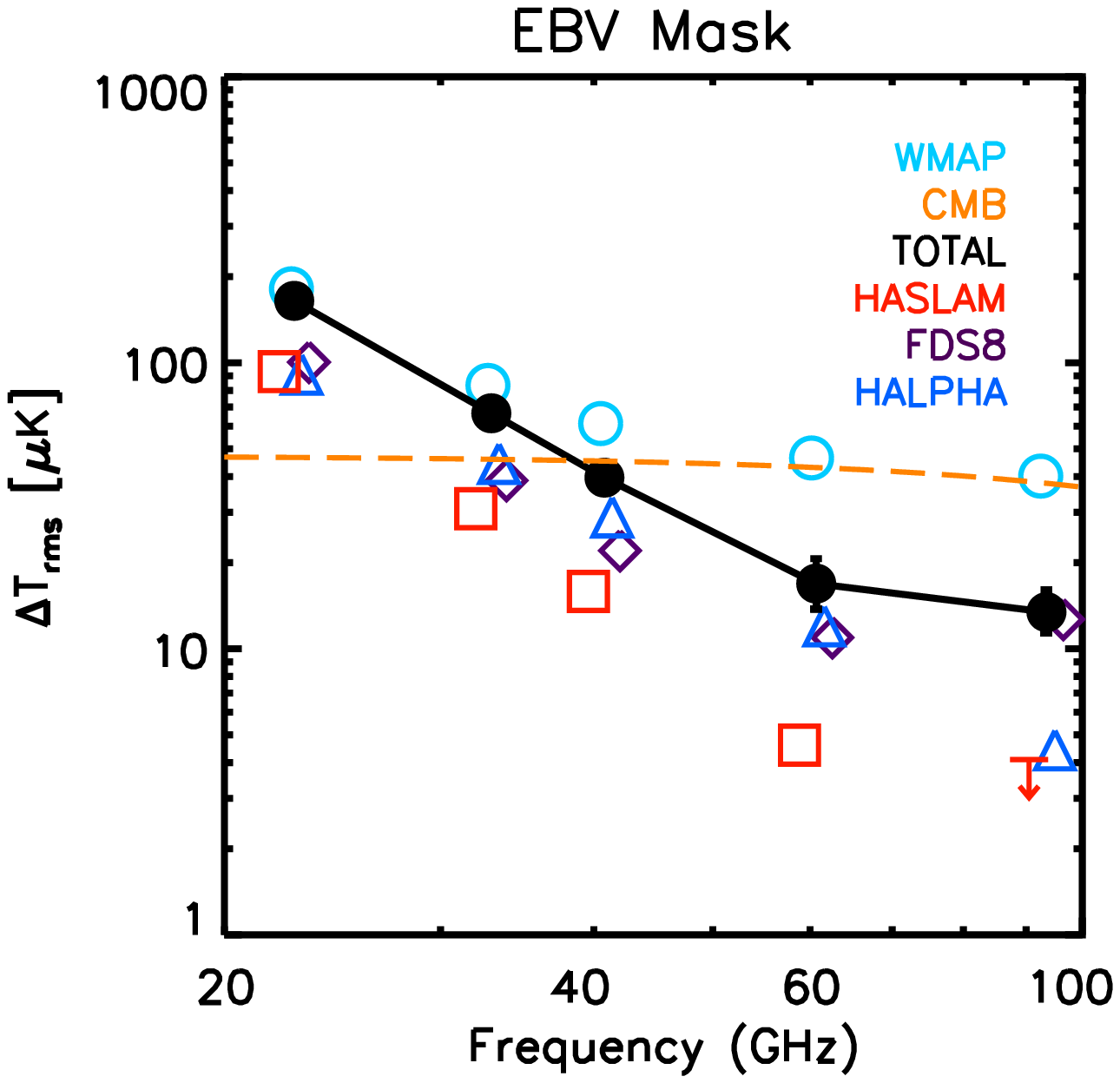,width=0.3\linewidth,angle=0,clip=} &
\epsfig{file=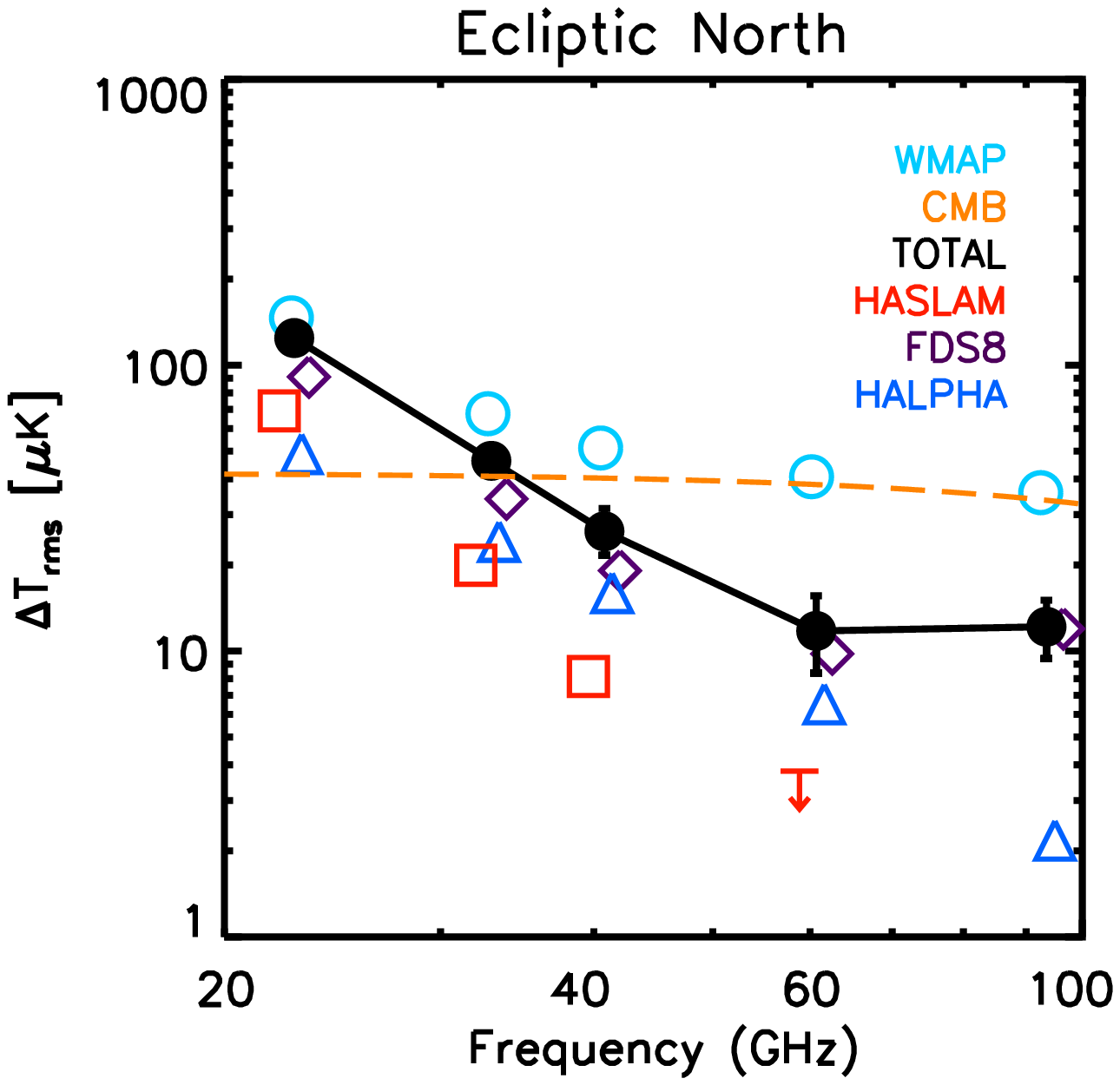,width=0.3\linewidth,angle=0,clip=} &
\epsfig{file=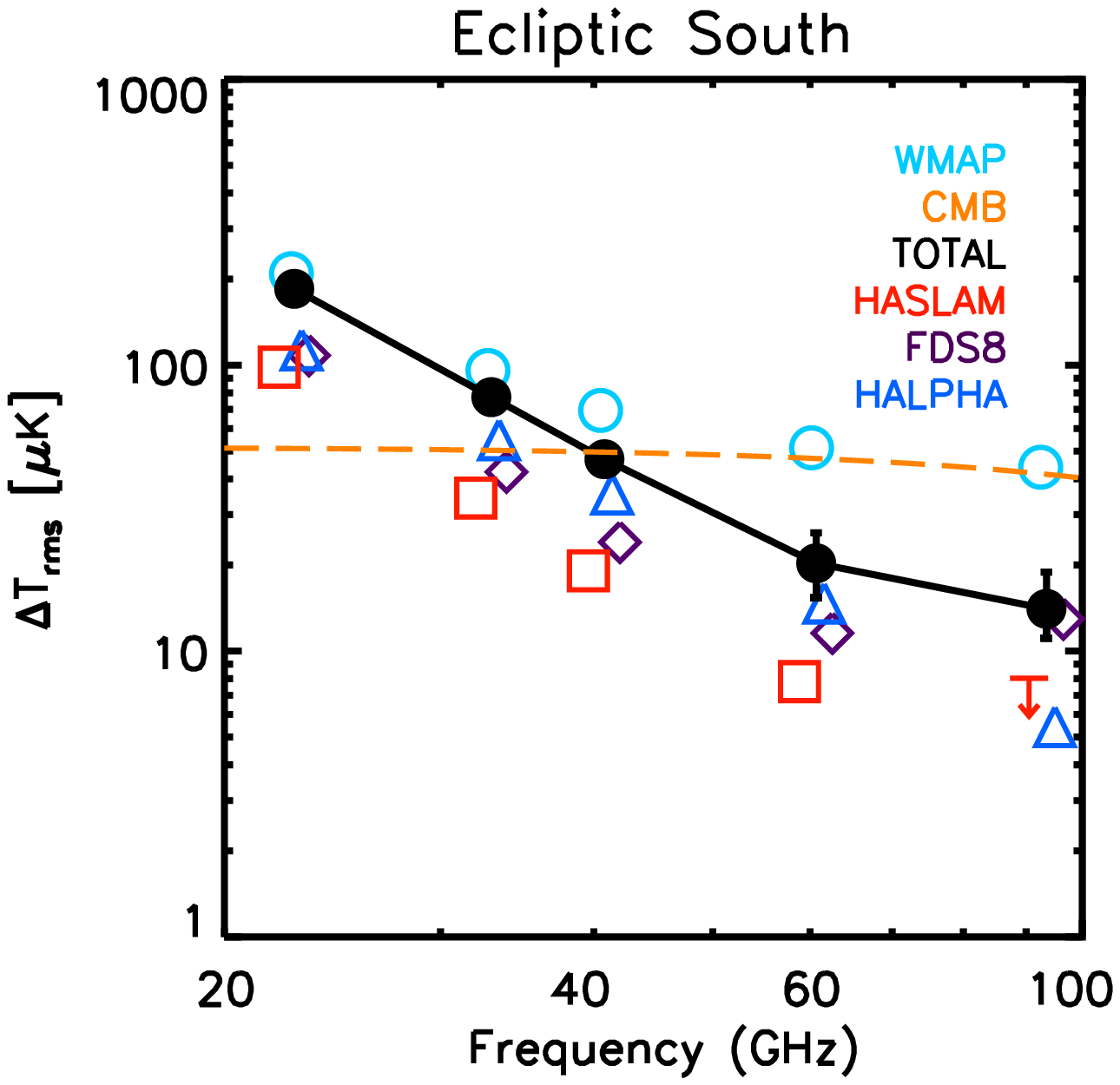,width=0.3\linewidth,angle=0,clip=} \\
\epsfig{file=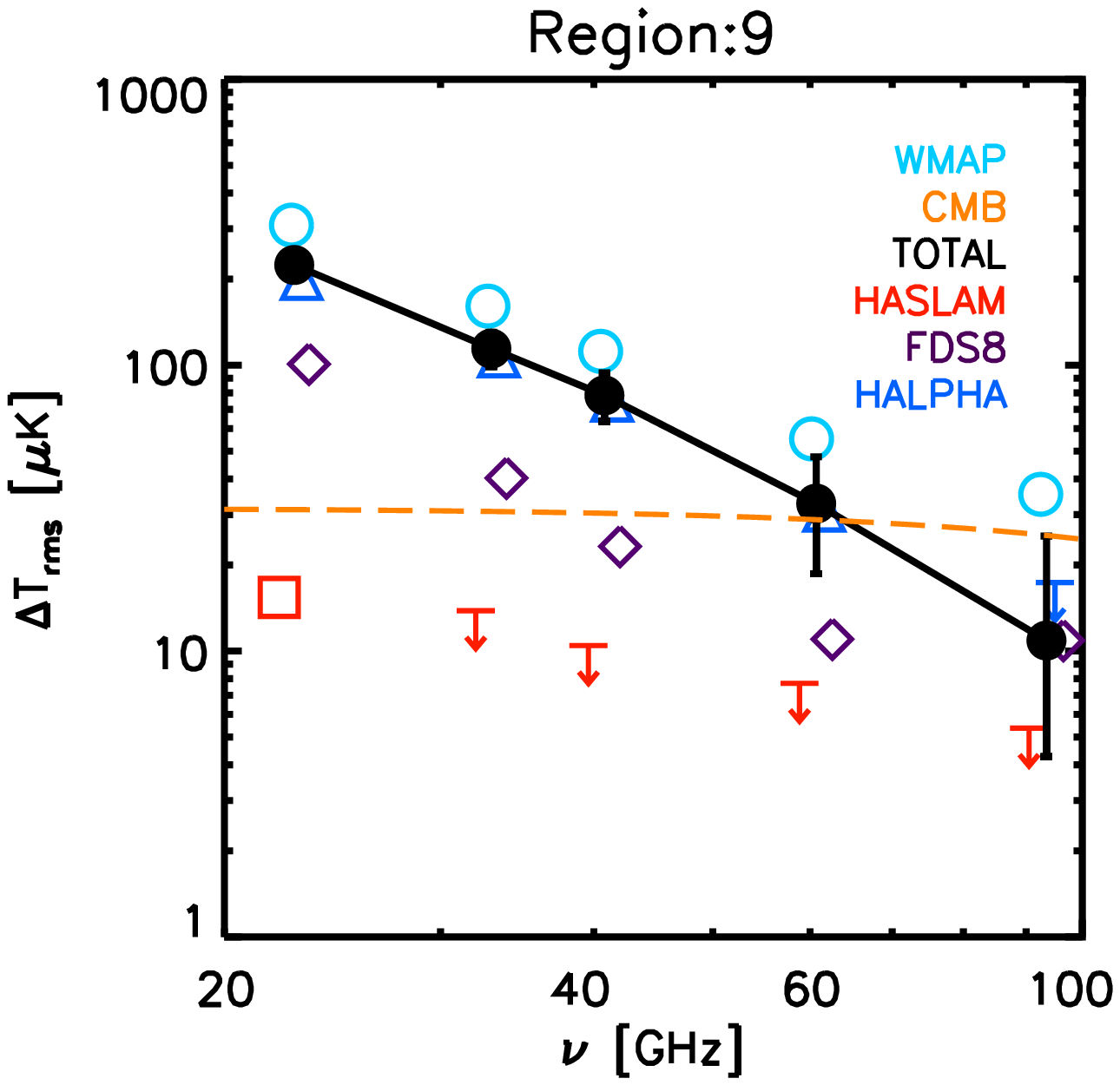,width=0.3\linewidth,angle=0,clip=} &
\epsfig{file=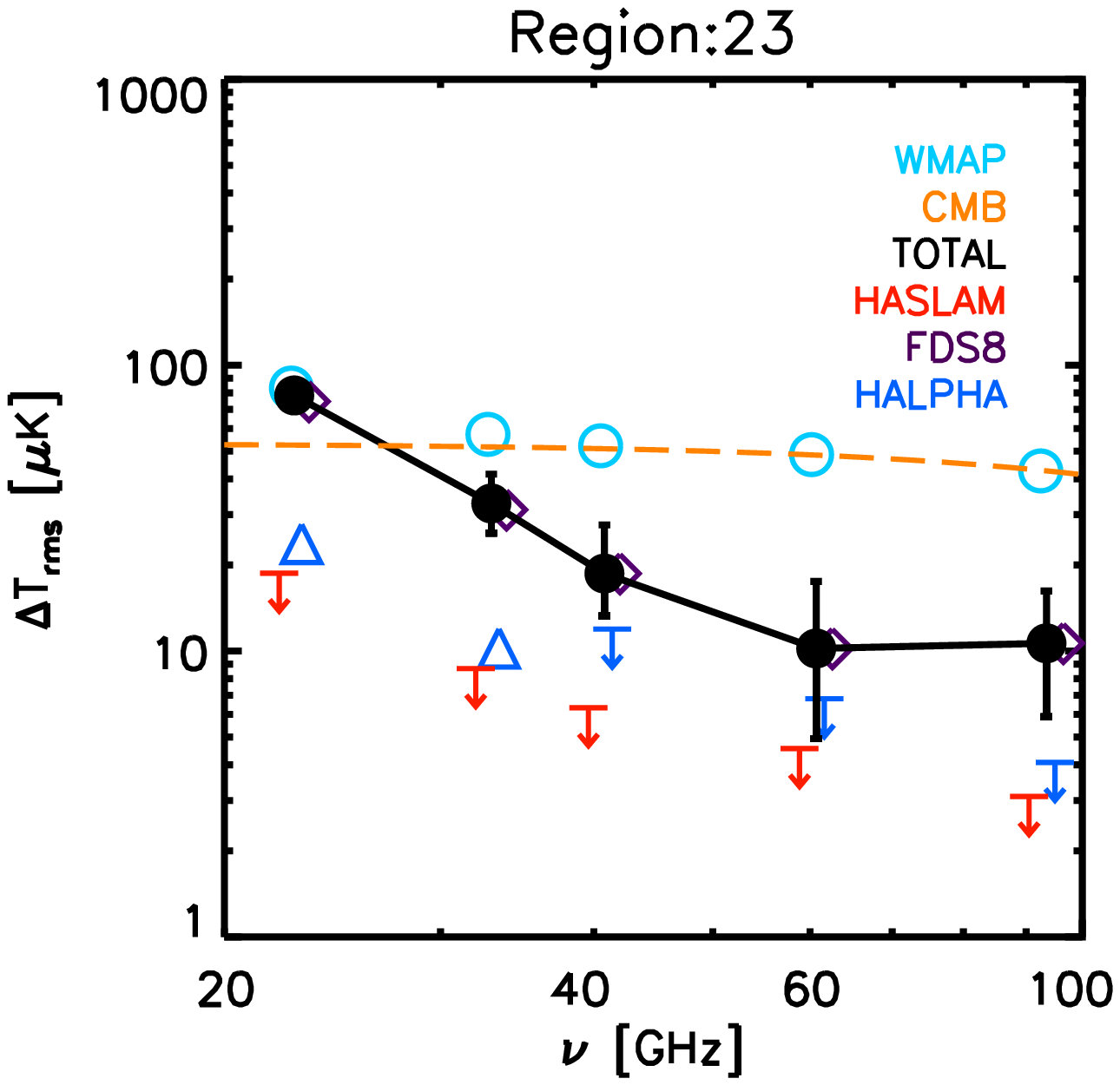,width=0.3\linewidth,angle=0,clip=} &
\epsfig{file=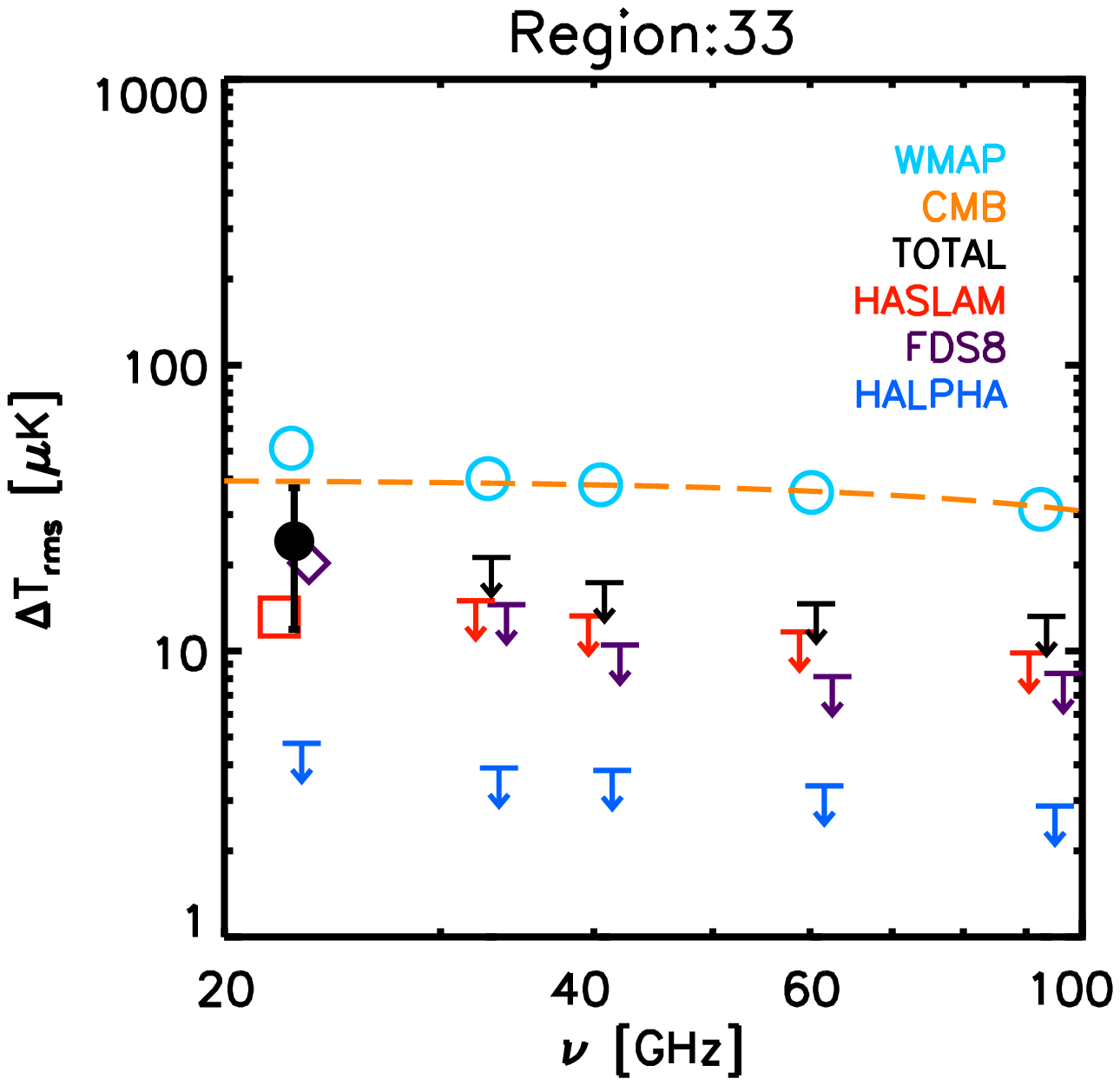,width=0.3\linewidth,angle=0,clip=} \\
\end{tabular}
\caption{R.M.S amplitude in antenna temperature observed for various regions of the sky. The signal associated with the total foreground is shown as black filled circles. The synchrotron contribution traced by the Haslam template is represented by red squares, the free-free traced by the \halpha template as dark blue triangles, and the dust contribution traced by the FDS8 template as purple diamonds. One-sided 95\% c.l. upper limits are shown for non-detections, with the appropriate colour coding for the component represented. The r.m.s. amplitude of the uncorrected WMAP frequency maps is indicated by open light blue circles. The  orange line corresponds to the CMB level estimated from the 7-year ILC map. The observed spectral shapes of regions 9, 23 and 33 are representative of those seen for all regions. \label{fig:total_fit_hemispheres}}
\end{figure}
\end{center}

In Fig.~\ref{fig:total_fit_hemispheres}, we show the r.m.s contributions of the foregrounds in antenna temperature as traced by the 
templates for three of the global fits and a representative set of the regional results. These values were derived from the template 
r.m.s values as recorded in Table~\ref{tab:template_rms_regions} and scaled appropriately by the fit coefficients from our analysis.
For comparison, we show the uncorrected r.m.s. amplitude at each \emph{WMAP} frequency, together with an estimate of the CMB
fluctuations from the ILC sky map, all determined for the same sky coverage as the foregrounds. We assume that this template based
technique traces essentially all of the foreground contribution, and this may not be the case. In particular, we recognise the 
\emph{WMAP}-haze \citep{Finkbeiner_WMAP1:2004, Dobler_WMAP3:2008a} as an important exception. Nevertheless, our interpretations should 
remain robust.

\begin{table}
\scriptsize 
\begin{center}
\begin{tabular}{lccc}
\hline
            & \multicolumn{3}{c}{r.m.s amplitude} \\ \hline
Region & 408 MHz (K) & \halpha (R) & FDS8 ($\mu$K)  \\ \hline
     EBV  &    10.91 &    9.30   &   10.21   \\
     KQ85 &     9.44 &    3.75   &   11.44   \\
     GN   &    10.74 &    6.18   &    9.79   \\
     EN   &     9.97 &    3.71   &   10.14   \\
     GS   &    11.08 &   11.56   &   10.52   \\
     ES   &    11.60 &   12.27   &   10.27   \\
\hline
      1   &     3.17 &    0.16   &    0.93   \\
      2   &     6.27 &    0.67   &    3.57   \\
      3   &    10.96 &    0.49   &    4.27   \\
      4   &     1.59 &    0.14   &    0.95   \\
      5   &     1.27 &    0.09   &    1.07   \\
      6   &     5.86 &    0.35   &    2.95   \\
      7   &     5.49 &    1.75   &    5.61   \\
      8   &    15.77 &    2.81   &    6.92   \\
      9   &     4.96 &   20.19   &    7.19   \\
     10   &     3.32 &    0.86   &    8.28   \\
     11   &     6.51 &    5.60   &    8.99   \\
     12   &     4.98 &    1.19   &    6.97   \\
     13   &     5.08 &   11.97   &    8.32   \\
     14   &     3.98 &   37.92   &   12.22   \\
     15   &     4.22 &    5.34   &    9.01   \\
     16   &    13.27 &    1.12   &    8.42   \\
     17   &     5.39 &    1.15   &    3.06   \\
     18   &     3.68 &    2.90   &    7.09   \\
     19   &     2.23 &    1.95   &    3.79   \\
     20   &     2.81 &    5.68   &    5.61   \\
     21   &     4.89 &    1.61   &    7.08   \\
     22   &     4.00 &    1.21   &    6.55   \\
     23   &     2.58 &    2.19   &    7.75   \\
     24   &     3.14 &    8.29   &    5.83   \\
     25   &     1.89 &    0.28   &    1.86   \\
     26   &     0.98 &    0.13   &    0.55   \\
     27   &     0.77 &    1.06   &    1.24   \\
     28   &     0.88 &    0.15   &    2.02   \\
     29   &     0.97 &    0.15   &    0.51   \\
     30   &     1.70 &    0.08   &    1.81   \\
     31   &     1.59 &    0.19   &    1.30   \\
     32   &     5.03 &    0.31   &    3.02   \\
     33   &     1.47 &    0.17   &    2.20   \\
\hline
\end{tabular}
\end{center}
\caption{R.m.s amplitudes of the foreground templates. These values are scaled to the \emph{WMAP} frequencies using the fit coefficients
in Tables~\ref{tab:results_synch_33regions}, \ref{tab:results_freefree_33regions} and \ref{tab:results_dust_33regions}.}
\label{tab:template_rms_regions}
\end{table}

Note that the spectral shape of the integrated foregrounds, where determined, is relatively simple when sparsely sampled by the
\emph{WMAP} frequencies, but nevertheless it is interesting that a simple power-law frequency dependence, or only modest deviations
therefrom, is needed to describe the emission from K- to V-band. This observation was central to the work of \citet{Park_SILC:2007} in
partitioning the sky to perform local CMB reconstruction.

For the EBV mask, it is clear that the foregrounds as traced by three templates are of similar amplitude at K-band, and the free-free and
dust contributions remain roughly equal to V-band, whereas the synchrotron falls off more steeply. The foregrounds remain comparable
to, or larger than, the CMB fluctuations to Q-band, and significant at V- and W-bands.  

Comparing the total foreground contributions between the North and South Ecliptic hemispheres, it can be observed that the amplitude is
higher in the South at least from K- to V-band. It also appears that the free-free component shows a particular enhancement in the South.
Whether this may be connected to problems in the \halpha template (see Appendix~\ref{app:halpha} for details) might be discussed. However,
the foreground differences are interesting in the context of the hemispherical asymmetry seen in the distribution of power in the CMB in 
Ecliptic coordinates, as noted originally by \citet{Park:2004, Eriksen_hemispheres:2004} and revisited many times in the literature 
\citep[see][for a review]{Copi_Anomalies_Review:2010}. Indeed, the r.m.s. CMB signal plotted here clearly shows this asymmetry. 
Specifically, if the foregrounds were underestimated in the Northern hemisphere (as might be the case for the free-free component), then 
correcting for this effect would only exacerbate the problem. A similar consequence results from an overestimate of the foregrounds in 
the South. Thus, it seems unlikely that the CMB asymmetry is connected with problems of Galactic foreground estimation, as noted before 
in \citet{Hansen_WIFIT:2006}.

As expected, there are also clear variations between the regions used in our study. Region 9 represents a region with strong foreground
contributions at all frequencies, the dominant contribution being from the free-free emission. Region 23 is a dust dominated region, but 
the significance of the contamination relative to the CMB is lower, although the dust is detected clearly at all frequencies. Finally,
region 33 typifies many high- and mid-latitude regions where foregrounds are only detected, if at all, at K- and possibly
Ka-band. The amplitude of the 95\% upper limits combined with the agreement between the uncorrected frequency maps and the CMB estimate
from the ILC implies that foregrounds are essentially unimportant in these regions. 

There are several interesting aspects of the study with regards to the properties of the foregrounds that should be highlighted. 
The synchrotron emission shows clear evidence of steepening relative to measures at GHz frequencies. There is a hint of this steepening 
continuing beyond K-band. The temperature of the electrons responsible for the free-free emission is now much more consistent with other
indicators such as RRLs, ie. $\sim$7000~K at high latitudes. This result is in part due to an improved understanding of the H$_{\alpha}$
template. This clearly needs to be improved and will benefit greatly from the \emph{WHAM-S} survey when available. However, uncertainties
related to the dust absorption remain. The AME is essentially ubiquitous, comparable in amplitude to the free-free emission, and
with a spectrum broader than a single component. There does appear to be an additional anomalous component correlated with the H$_{\alpha}$
emission, but there are variations in relative strength on the sky. An improved model of the AME requires an improvement 
in the understanding of the corresponding thermal dust contribution. Specifically, the emissivity of the thermal dust tail to low 
frequencies must be determined, and a detailed model of the emissivity as a function of wavelength and dust temperature needs to be 
derived from the data.

Finally, we note that studies of foregrounds at microwave wavelengths using the techniques in this paper are essentially limited by the
quality and resolution of the available templates. A comparison of synchrotron and free-free properties over the full-sky are, at best,
possible on angular scales of 1$^{\circ}$ and above, given the resolutions of the 408~MHz and \halpha data. Moreover, we have noted
some problems with the \halpha template on scales less than 3$^{\circ}$. However, this should not limit the analysis of \emph{Planck} data,
and especially that from the high-frequency instrument (HFI) operating at frequencies above 100~GHz, which will determine the R-J tail 
of thermal dust emission. Similarly, data at frequencies $\sim 5-20$\,GHz will help determine the complicated mix of low frequency 
foregrounds. For example, there could be a possible contribution from synchrotron radiation with a flatter spectral index compared to 
the values determined from correlations with 408\,MHz data. We note that a recent study by \cite{Peel:2011} who used a 2.3\,GHz survey 
of the southern sky found that the AME was relatively unaffected limiting the contribution of such a component to $\sim 7\,\%$ of the 
AME at 23\,GHz.

\section{Acknowledgements}

Some of the results in this paper have been derived using the HEALPix package \citep{Gorski_HEALPix:2005}. We acknowledge the use of the
Legacy Archive for Microwave Background Data Analysis (LAMBDA). Support for LAMBDA is provided by the NASA Office of Space Science. CD 
acknowledges an STFC Advanced Fellowship and ERC IRG grant under the FP7.

\bibliographystyle{mn2e}
\bibliography{references}

\clearpage
\newpage

\appendix

\section{\halpha Templates}\label{app:halpha}
\defcitealias{Finkbeiner:2003}{F03}

\begin{center}
\begin{figure}
\begin{tabular}{ccc}
\includegraphics[height=0.22\linewidth,angle=0]{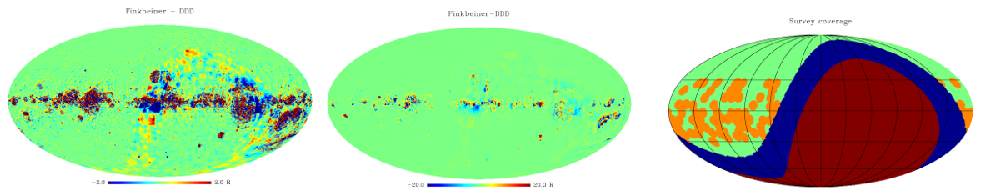} 
\end{tabular}
\caption{ {\it Left and middle:} difference map between the \citetalias{DDD:2003} and \citetalias{Finkbeiner:2003} \halpha templates. The \citetalias{Finkbeiner:2003} map has been additionally smoothed by a Gaussian beam of $(60^2-6^2)^{1/2}$ degrees so that the two sky maps
have a matched 1$^{\circ}$ resolution. The two plots are identical except for the color scale. {\it Right:} the coverage of the
different datasets used in the two templates.  In light green is where both templates use \emph{WHAM} data, and in red where both use \emph{SHASSA} data.  Orange shows where the \citetalias{Finkbeiner:2003} template uses \emph{VTSS} data. In dark blue is the region where \citetalias{DDD:2003}  uses \emph{WHAM} data while \citetalias{Finkbeiner:2003} uses \emph{SHASSA}. \label{fig:diffmaps}}
\end{figure}
\end{center}

\begin{center}
\begin{figure}
\includegraphics[width=1.0\linewidth]{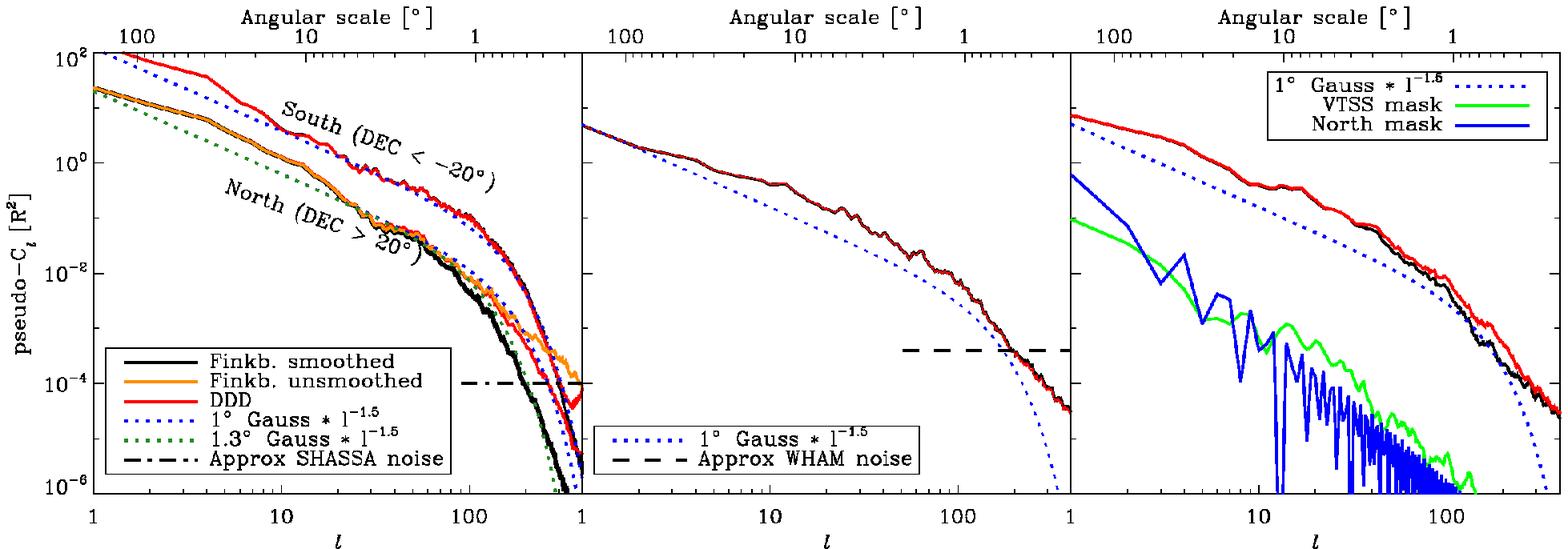} 
\caption{ Pseudo-$C_\ell$ spectra computed from the \citetalias{DDD:2003} and \citetalias{Finkbeiner:2003} \halpha templates.
{\it Left:} pseudo-$C_\ell$ spectra of discs centered on either Celestial North or South ($|\delta|>20\degr$). Dashed lines show the 
Southern spectra while
solid the Northern.  Results from the \citetalias{DDD:2003} template are shown in red, those from the
\citetalias{Finkbeiner:2003} map smoothed to 1$^{\circ}$ resolution are in black, and in orange we show results from the
\citetalias{Finkbeiner:2003} template without additional smoothing. Dotted lines give simple power law spectra smoothed with Gaussian
beams of 1$\degr$ (1.3$\degr$) in blue (green) with arbitrary power law indices and amplitudes fit by eye solely for visual comparison.
{\it Middle:} pseudo-$C_\ell$ spectra of sky regions where both \citetalias{Finkbeiner:2003} and \citetalias{DDD:2003} templates use only
\emph{WHAM} data.  The \citetalias{Finkbeiner:2003} template has been smoothed as described in the text. {\it Right:} pseudo-$C_\ell$ 
spectra of sky regions where \citetalias{Finkbeiner:2003} uses \emph{VTSS} data but \citetalias{DDD:2003} uses 
\emph{WHAM}. For comparison, the pseudo-$C_\ell$s for the masks themselves are shown, the \emph{VTSS} used for the red and black curves, 
and the North mask used in the top figure.  The effects of the highly structured \emph{VTSS} mask dominates toward high $\ell$, which 
makes the lower two plots difficult to compare with the top plot, where an azimuthally symmetric mask does not impact the power spectrum.
But it is apparent that the differences in the north are due to the use of the \emph{VTSS} data. \label{fig:pseudocls}}
\end{figure}
\end{center}

We have compared the \citetalias{DDD:2003} and \citet[hereafter F03]{Finkbeiner:2003} \halpha templates and performed template
fits to the \emph{WMAP} data using both for comparison. As in \citetalias{Davies_WMAP1:2006}, there are sometimes significant differences
in the resulting cross-correlation coefficients, that we previously ascribed to an inaccuracy in our knowledge of the {\emph{WHAM} 
effective beam. Here, we investigate the detailed differences between the templates themselves.

The \citetalias{DDD:2003} template used a combination of \emph{WHAM} data in the north and \emph{SHASSA} data in the south, with
preference for the \emph{WHAM} data down to a declination of $-10\degr$. \citetalias{Finkbeiner:2003} chose to use \emph{SHASSA} data 
wherever it was available, augmented by data from the Virginia Tech Spectral-line survey \citep[hereafter \emph{VTSS}, ][]{Dennison_VTSS:1998}, a complementary narrow-band imaging survey which covers the northern sky for $\delta \geq -15^{\circ}$, and has an
angular resolution of 1.6 arc minute. The \emph{SHASSA} data were processed differently, and different methods used to adjust the 
resolutions of these datasets for consistency and merging. The \citetalias{DDD:2003}  template claims a nominal resolution of 1$\degr$ 
FWHM, while the \citetalias{Finkbeiner:2003} template has a stated resolution of 6$\arcmin$.  

Fig.~\ref{fig:diffmaps} shows the differences between the two maps on two different color scales to bring out the small and large 
amplitude differences.  Clearly, the processing differences result in maps that differ at the roughly 1~Rayleigh (R) level over large 
regions of the sky, and of more than 20~R level near very bright regions, although the latter are generally excluded by our EBV and KQ85 
masks.

To investigate the properties of these templates further, we have computed pseudo-C$_\ell$ power spectra in the Celestial North and South
where the \citetalias{DDD:2003} template uses exclusively \emph{WHAM} or \emph{SHASSA} data, respectively. The results are shown in
Fig.~\ref{fig:pseudocls}. The noise properties of the maps are difficult to quantify because many components can contribute, including 
stellar residuals that will dominate. However, the instrumental thermal noise is approximately known. For the \emph{WHAM} survey, the 
noise level varies over the sky but is typically 0.02 R, whereas for \emph{SHASSA} the corresponding amplitude is $\sim$ 0.01 R. These 
values are overplotted in the figure. In the case of the \citetalias{DDD:2003} template, no additional smoothing has been applied, but
for the \citetalias{Finkbeiner:2003} map we have smoothed the data from the stated resolution of $6\arcmin$ to a $1^{\circ}$ effective 
FWHM. It is interesting to note the asymmetry in amplitudes between the two hemispheres. For comparison, we overplot power laws 
(with index and amplitudes chosen by eye to roughly match) as smoothed with a Gaussian beam of FWHM $1^{\circ}$. Both templates appear to 
have a roughly $1^{\circ}$ resolution in the South, where both use \emph{SHASSA} data. The \citetalias{DDD:2003} template in the North, 
where only \emph{WHAM} data are used, is also roughly consistent with $1^{\circ}$ resolution, although deviations are seen, as perhaps 
might be expected due to \emph{WHAM}'s tophat rather than Gaussian beam profile. However, the \citetalias{Finkbeiner:2003} template 
(black) does {\it not} appear to match the intended $1^{\circ}$ resolution in this region, where it uses a mix of both \emph{WHAM} and 
\emph{VTSS} data. Rather, the resolution is apparently lower, roughly following a beam of $1.3^{\circ}$ instead. This implies that the 
\citetalias{Finkbeiner:2003} template is inconsistent with the claimed $6\arcmin$ resolution and instead exhibits more complex properties
over a range of angular scales. This would appear to demonstrate the difficulties in combining and using inhomogeneous data sets.

\begin{figure}
\begin{center}
\begin{tabular}{ccc}
\epsfig{file=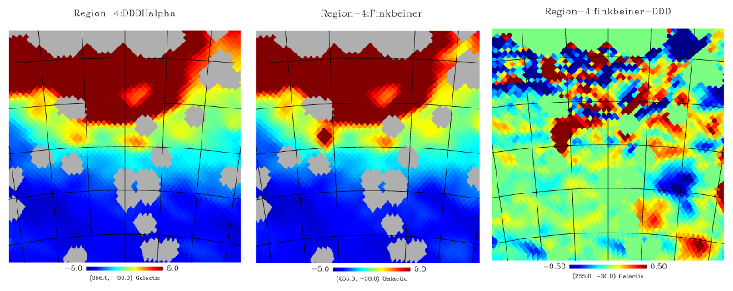,width=0.83\linewidth,angle=0,clip=} 
\end{tabular}
\caption{Comparison between the ({\it left}) \citetalias{DDD:2003} and ({\it centre}) \citetalias{Finkbeiner:2003} \halpha templates at 1$^{\circ}$ resolution, for  Region 4 of \citetalias{Davies_WMAP1:2006}. {\it Right:} the difference in structure between the two templates.
\label{fig:region4_halpha_diff}}
\end{center}
\end{figure}

Of central importance to this paper, however, is the impact of the observed differences in the templates on our template fits.
We have, therefore, performed analyses of a set of 1000 simulations, where we use one template to generate a sky with
CMB plus all foregrounds, and then we fit both the correct \halpha template and the alternate template to the data.
In many regions the cross--correlation coefficients are the same, but in equally many regions systematic differences arise due to even 
small differences in the templates. In some cases, it is clear where the difference in results likely originates; in region 4 taken from 
\citetalias{Davies_WMAP1:2006} for example, there is an obvious bright spot in the \citetalias{Finkbeiner:2003} template that is much 
fainter in the \citetalias{DDD:2003}  relative to the surrounding structure (see Fig.~\ref{fig:region4_halpha_diff}). In other cases, the
cause of the difference in results is not apparent among low-level template differences. Moreover, we have found such behaviour not only 
in regions where the data used in the two \halpha templates are different (e.g. \emph{VTSS} versus \emph{SHASSA}) but also in regions 
where the data should be consistent with each other (e.g. the common sky area covered by the \emph{SHASSA} survey), implying that even 
relatively small differences in the sky map processing can affect the results in such small regions.

With regards to the data, we find that the mean template fit coefficient to the \emph{WMAP} K-band data for the 33 regions
is $1.01\pm 0.02$ for the \citetalias{DDD:2003} template and $8.06\pm 0.05$ for the \citetalias{Finkbeiner:2003} template  (in
appropriate units). Indeed, the former results by region are systematically lower than the latter. However, assuming that the
\citetalias{Finkbeiner:2003}  results are more correct simply because the results are closer to what we expect 
(based on idealised properties of the WIM such as an electron temperature of $\sim 8000$\,K) is unwise. 
In the \citealt{Davies_WMAP1:2006} paper, partly because of the unexpectedly low cross--correlation
coefficients, we applied a small additional smoothing to the \emph{WHAM} regions.  This smoothing was an attempt to make up the difference
between a $1^{\circ}$ top-hat and a $1^{\circ}$  Gaussian, yet it appears from Fig.~\ref{fig:pseudocls} that this additional smoothing was
not appropriate.

\begin{figure}
\begin{center}
\begin{tabular}{cc}
\epsfig{file=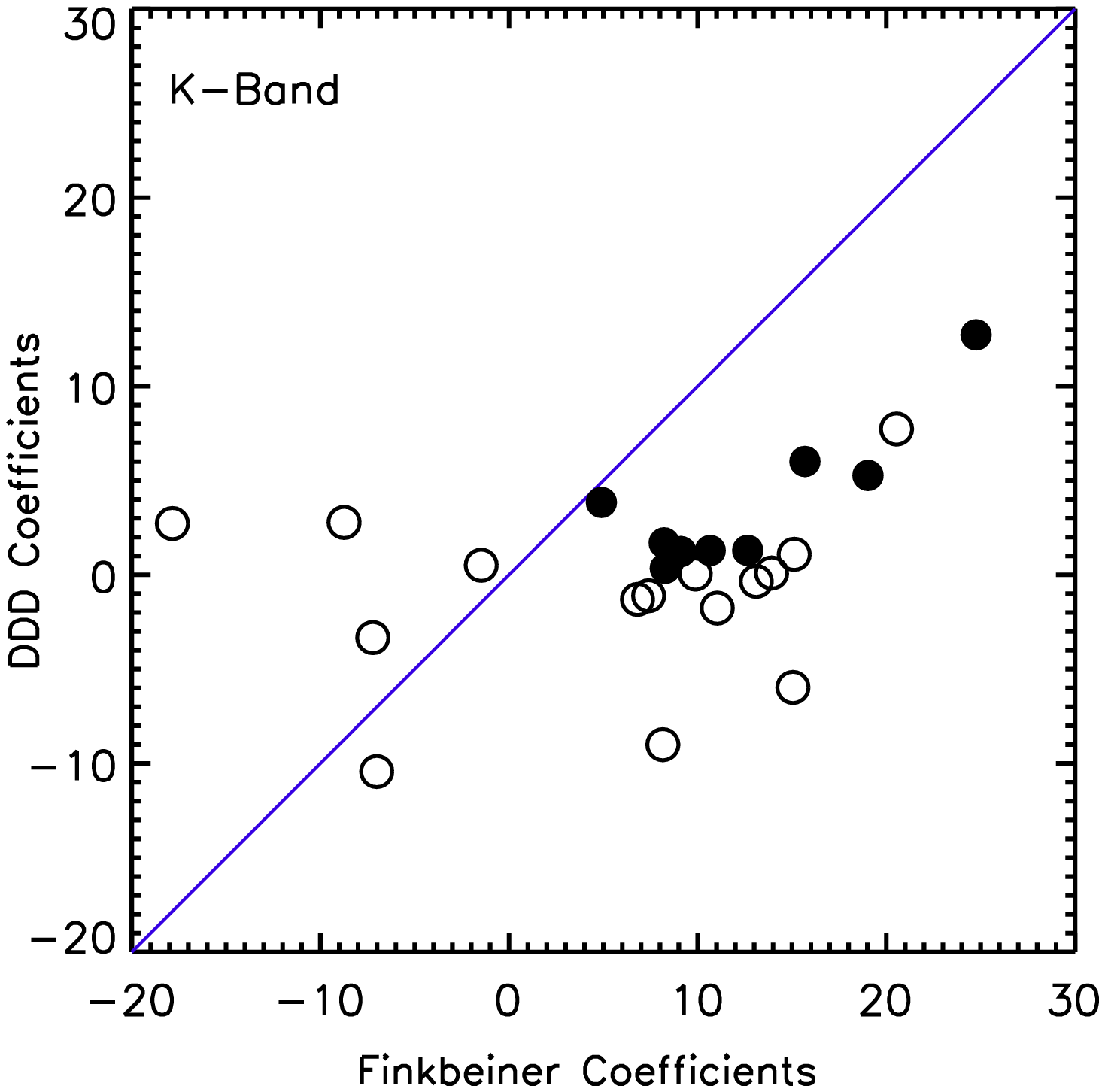,width=0.3\linewidth,angle=0,clip=} &
\epsfig{file=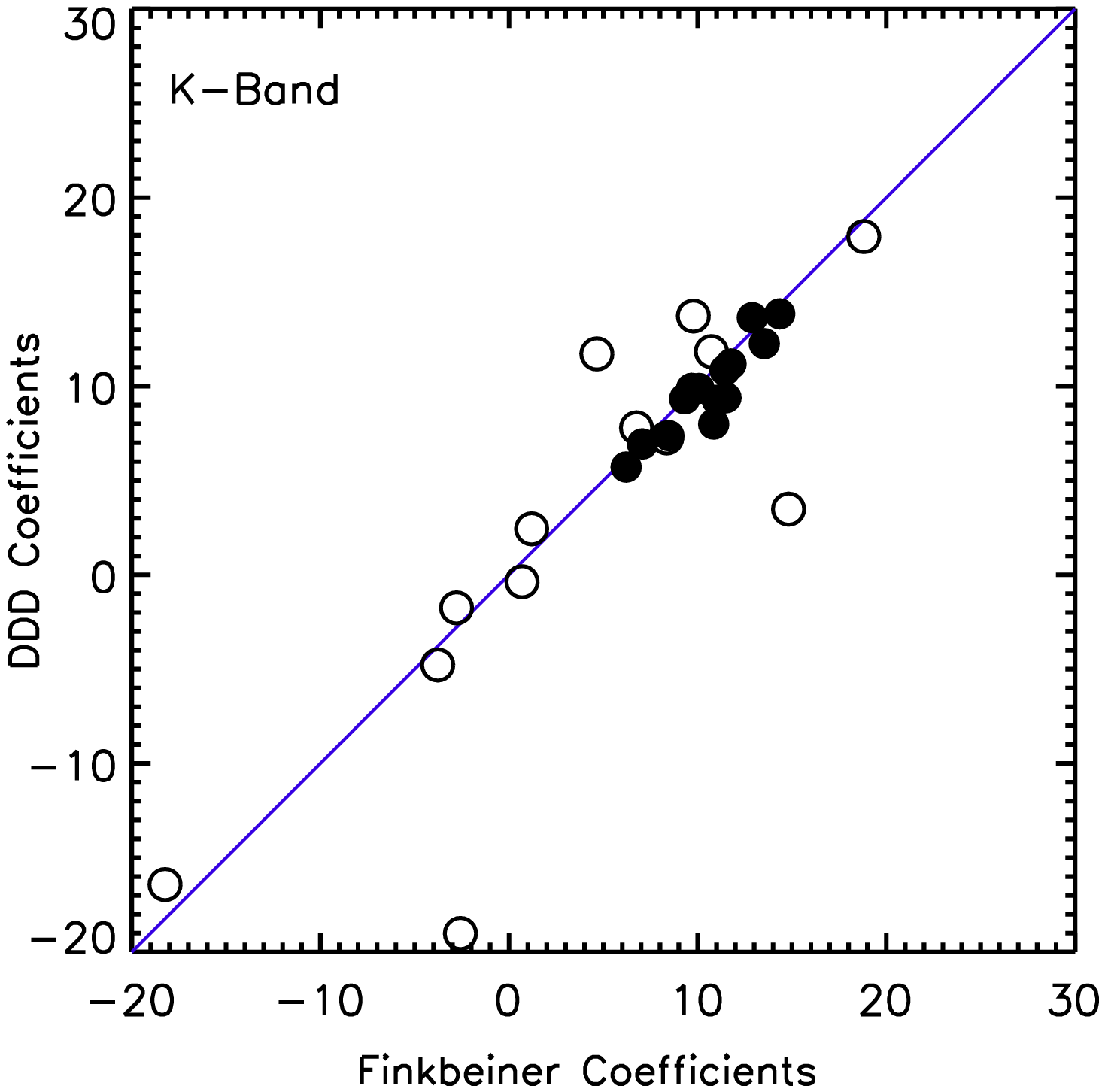,width=0.3\linewidth,angle=0,clip=} \\
\end{tabular}
\caption{Comparison of the free-free fit coefficients determined between the \emph{WMAP} K-band data and the  \citetalias{DDD:2003}
  and \citetalias{Finkbeiner:2003} templates at 1$^\circ$ (left panel) and 3$^\circ$ (right panel) resolution. \label{fig:ddd_fink_coefficients_regions}}
\end{center}
\end{figure}

Nevertheless, it is in smoothing that we find a solution to the issues observed here. After convolving the templates and \emph{WMAP} data
to an effective $3^{\circ}$ resolution, then we find statistically consistent results between the two templates -- for example,
coefficients of 9.92$\pm$  0.13 and 10.28$\pm$ 0.13 for the \citetalias{DDD:2003} and \citetalias{Finkbeiner:2003} templates,
respectively.  

The difference in results between the two templates at  $1^{\circ}$ resolution seems to be particularly difficult to reconcile with the
power spectrum plots. One would at least expect good agreement in the regions where both templates use only {\em WHAM} or {\em SHASSA} 
data. Figs~\ref{fig:ddd_fink_coefficients_regions} \& \ref{fig:ddd_fink_corr_regions} are therefore quite informative in this 
respect. It should be clear that the two templates are generally very consistent, although there are more outliers at $1^{\circ}$ 
resolution. We therefore postulate that it is the nature of the estimator $\alpha=t^TM^{-1}d/t^TM^{-1}t$ (see Section~\ref{sec:methods}) 
that amplifies the differences in the templates, as the denominator is sensitive to the square of the template. Though the numerator is 
roughly the same using both templates, the denominator differs significantly and drives the difference in $\alpha$. At $3^{\circ}$ 
resolution, the differences are suppressed sufficiently to allow excellent consistency between the results derived with either template. 

\begin{figure}
\begin{center}
\begin{tabular}{cccc}
\epsfig{file=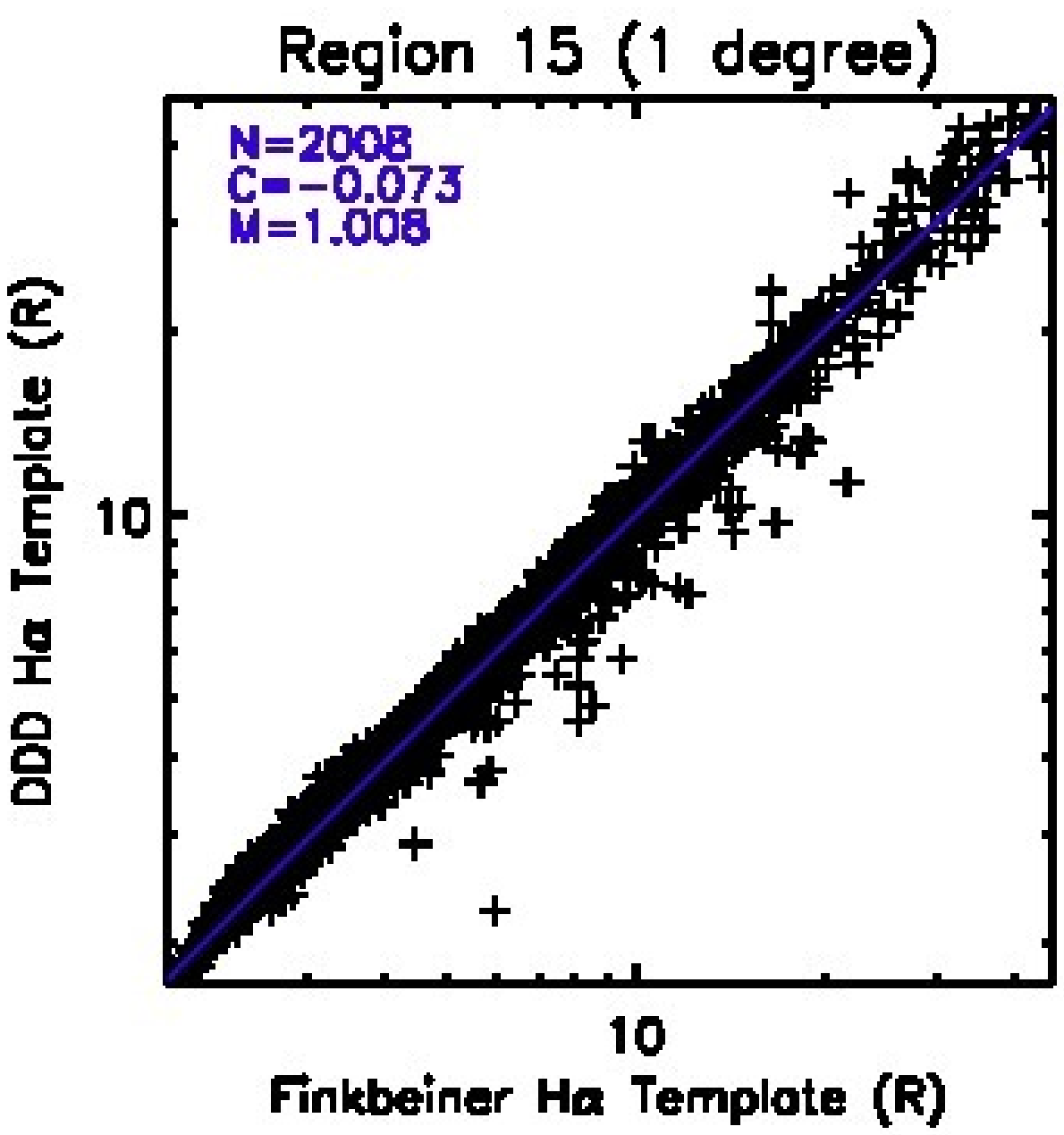,width=0.22\linewidth,angle=0,clip=} &
\epsfig{file=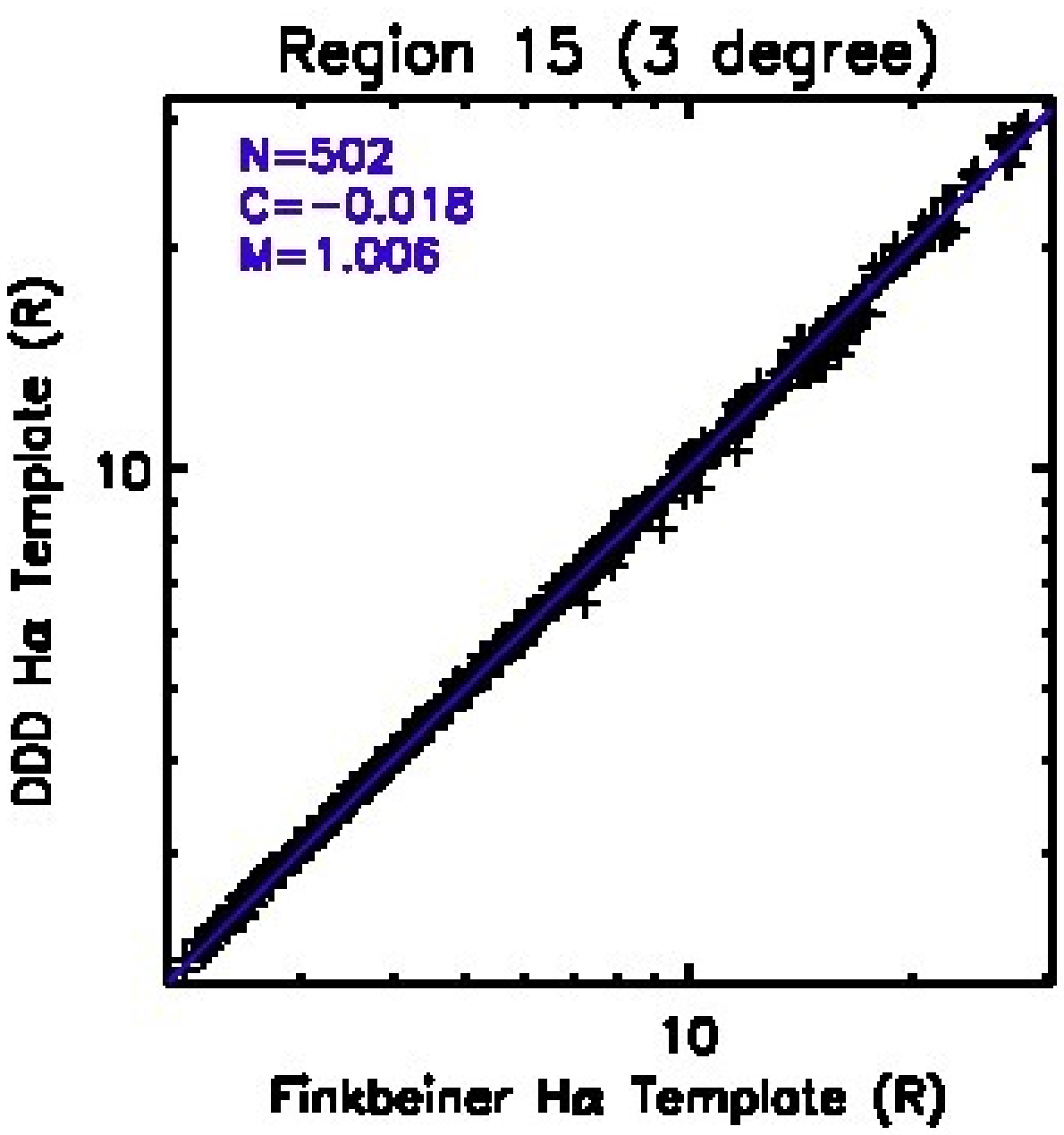,width=0.22\linewidth,angle=0,clip=} &
\epsfig{file=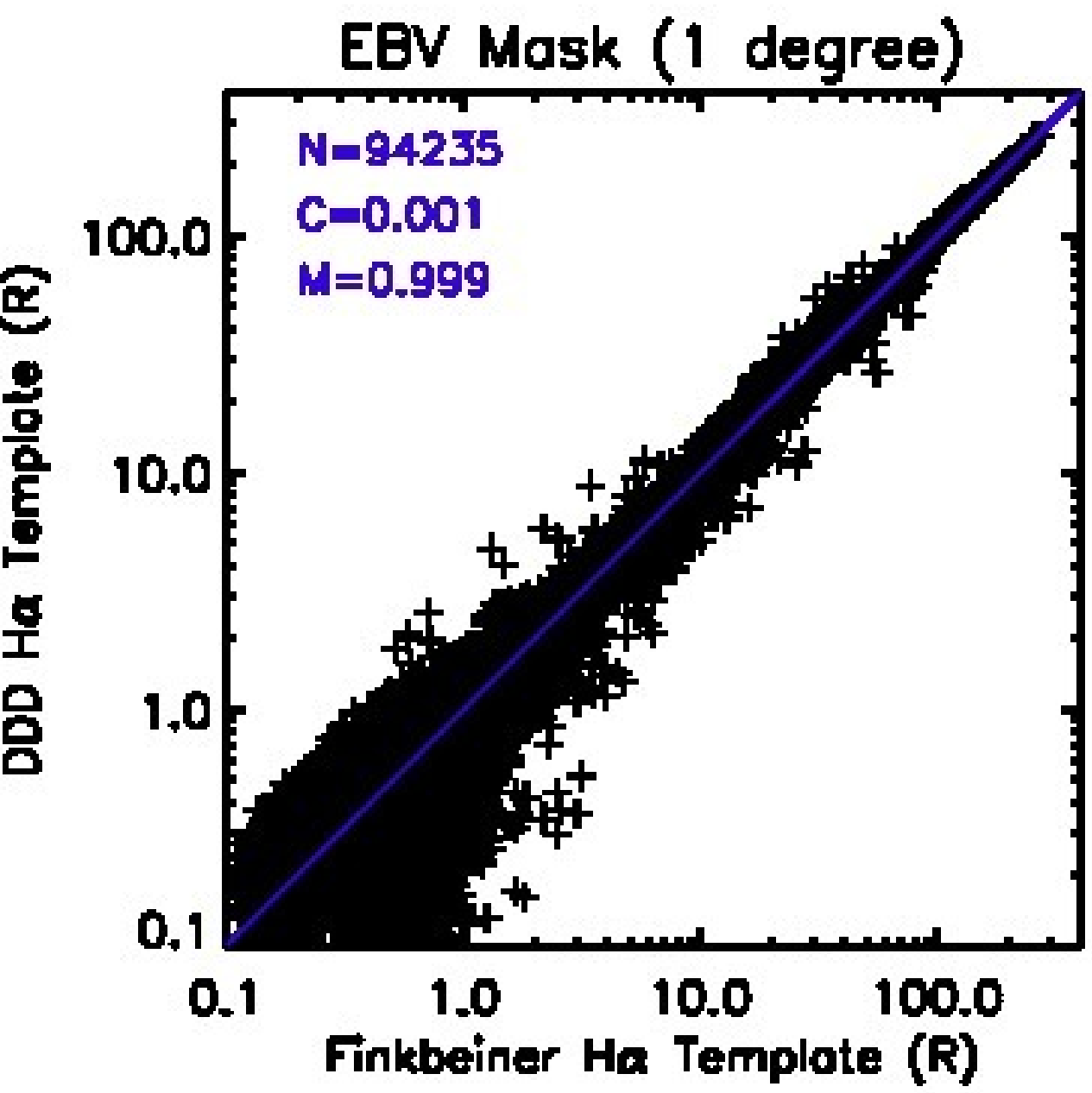,width=0.22\linewidth,angle=0,clip=} &
\epsfig{file=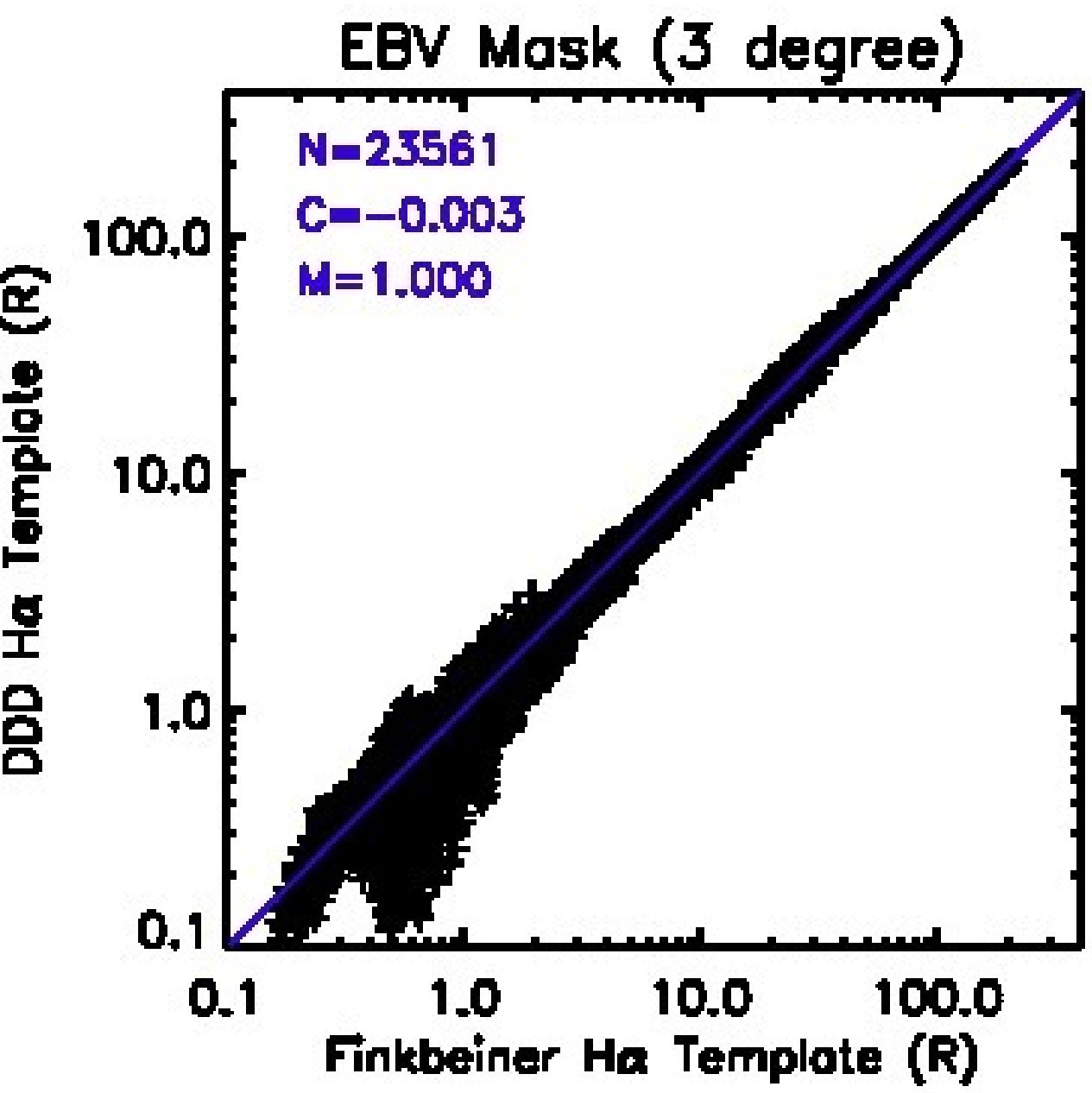,width=0.22\linewidth,angle=0,clip=} \\
\end{tabular}
\caption{Correlation between the \citetalias{DDD:2003} and \citetalias{Finkbeiner:2003} template maps at 1$^\circ$ and 3$^\circ$ resolution
for region 15 (left panels) or the EBV mask (right panels). The thick blue line represents the best fit regression with slope M and 
intercept C for the number of pixels N. Clearly the scatter of the correlation plot decreases with 3$^\circ$ smoothing. \label{fig:ddd_fink_corr_regions}}
\end{center}
\end{figure}

Though we cannot make a case for the superiority of one template over the other, we choose to utilise the \citetalias{DDD:2003} template 
since it was explicitly constructed to have a uniform resolution over the sky, consistent with the analysis above. Moreover, we wish to
emphasise that, in neither case do we consider that the templates are invalid as proxies for the Galactic free-free emission in their 
higher resolution forms, but that care must be taken as to the properties of the data in local regions. This is particularly important 
given the subtle interaction between the structure in the data and $\chi^2$ type estimators as employed in our analysis.

\clearpage
\newpage
\section{Constructing a regional definition mask}
\label{app:regions}

The \cite{Davies_WMAP1:2006} regions were selected on the basis that one of the three principal and well-known Galactic foregrounds
(free-free, dust or synchrotron emission) was dominant in each region. The regions were selected well away from the Galactic plane
where the foregrounds are inevitably confused. In this paper, we extend the analysis by defining regions over a larger fraction of the 
high latitude sky, and making more extensive use of the template morphology to achieve this. Our approach is also intended to allow a 
focus on localised areas of the sky where foreground spectral variations may be traced. To achieve this, we recognise that the overall 
Galactic emission even to high latitudes can generally be well described by a cosecant law of emission \citep{Bennett_WMAP1:2003b}, with 
the possible exception of the 408 MHz emission. Nevertheless, we wish to adopt a uniform processing of the templates. Therefore, by 
normalising the template emission by a cosecant term, we essentially flatten the structure and enhance local features. We impose a 
5$^{\circ}$ Galactic cut on the templates given that this region will be excluded by our EBV and KQ85 masks in any case, and to avoid the 
cosecant term causing problems. We then subdivide each template on the basis of intensity, a criterion which, at least in the case of 
thermal dust emission \citep{MAMD_spectra:2007}, associates variations in the physical properties of the foregrounds with mean brightness.
This is then the basis for further processing to select the individual regions. The exact algorithm to define our regions of interest is 
as follows.

\begin{description}

\item [(I)] Consider the three external foreground templates 
for synchrotron S (408 MHz), free-free F (\halpha ) 
and dust D (FDS8) as smoothed to 
 $3^{\circ}$ beam resolution. Denote the emission at a given pixel of
 component $k$ (S,F,or D) by  $T^k(\hat n)$.\\

\item [(II)] To remove the effect of the line-of-sight depth,
  we apply a sine modulation to each external template, which is defined
  as,
\begin{equation}
T^k_\rmn{sin}(\hat n) =  
\begin{cases}
T^k (\hat n) \times \text{sin}\ |b| & \text{if} \quad |b| > 5\\
0 &\text{if} \quad  |b| < 5
\end{cases}
\end{equation}

where `$b$' represents the Galactic latitude of the given position of the sky
defined by $\hat n$ , $ T^k$ and $ T^k_\rmn{sin}$ represents the
external template and the modulated template, respectively.
We do not include the $\pm
5^{\circ}$ latitude range above and below the Galactic plane.
Each modulated template has a characteristic behaviour which
helps to define the morphology of the regions of interest. \\

\item[(III)]  Sort the pixels in each template by temperature such that 
\begin{equation}
T_\rmn{sin}^k(\hat{n}_i) < T_\rmn{sin}^k(\hat{n}_{i+1}) 
\end{equation}
and then define ten intervals, each with a tenth of the ordered pixels.  
I.e. for each template, we define the set of pixels 
\begin{eqnarray}
\mathbb{S}_j^k = \{ \hat{n}_i \} \quad \text{s.t.}\quad 0.1\,N\,(j-1) < i <= 0.1\,N\,j \notag \\ 
\qquad\text{where} \quad 1<=j<=10 
\end{eqnarray}
This effectively divides the sky into ten roughly isotemperature
regions of equal total area.  We can create a map, $I^k(\hat{n})$, of
these sets by assigning the pixels the dummy temperature value of $j$
that labels the set, $\mathbb{S}_j^k$, to which the pixel belongs.
These maps are shown in the first row of Fig.~\ref{fig:defining_regions} for the three
templates.
Where the map has
a value of 10, for example, the pixel in the sine-modulated template
map has a temperature in the top 10\% of the sky emission. 

\item [(IV)] This isotemperature contour map for each template has
  distinct features and can be used to isolate morphologically
  interesting regions. We visit each pixel and determine in which
  template $I^k(\hat{n}_i)$ is maximum and then define a new region
  map, $R(\hat{n})$ as follows:
\begin{eqnarray}
 R(\hat{n}_i) = a \ \textbf{max}\{  I^k (\hat{n}_i)\} \notag \\
\qquad \text{where}\quad a = \begin{cases}
  10 & \text{if} \quad \textbf{max}\{  I^k \}=  I^{S} \\
  100 & \text{if} \quad\textbf{max}\{  I^k \}=  I^{D} \\
  1000 & \text{if} \quad\textbf{max}\{  I^k \}=  I^{F} \end{cases}
\end{eqnarray}
This map is shown with various color scalings in the 2nd row of
Fig.~\ref{fig:defining_regions}.  Levels 10-100 correspond to synchrotron regions, 100-1000 are
dust, and 1000-10000 are \halpha.

\item [(V)] The new combined map $R(\hat n)$ has up to ten regions for
  each template, and these are reduced by a factor of two into 15
  (non-contiguous) regions simply by merging regions of similar
  temperature.  This is shown in the left-hand plot of the last row of
  Fig.~\ref{fig:defining_regions}.  Lastly, the discontinuous regions with the same value label
  are then split, while very small regions are either clumped together
  or omitted, leaving a total of 35 new regions.  The resultant
  map is shown in lower-right of the figure.
  Table~\ref{tab:summary_regions} summarises some useful quantities that
  describe the regions on the sky.

\end{description}

\begin{figure*}
\begin{center}
\begin{tabular}{ccc}
\epsfig{file=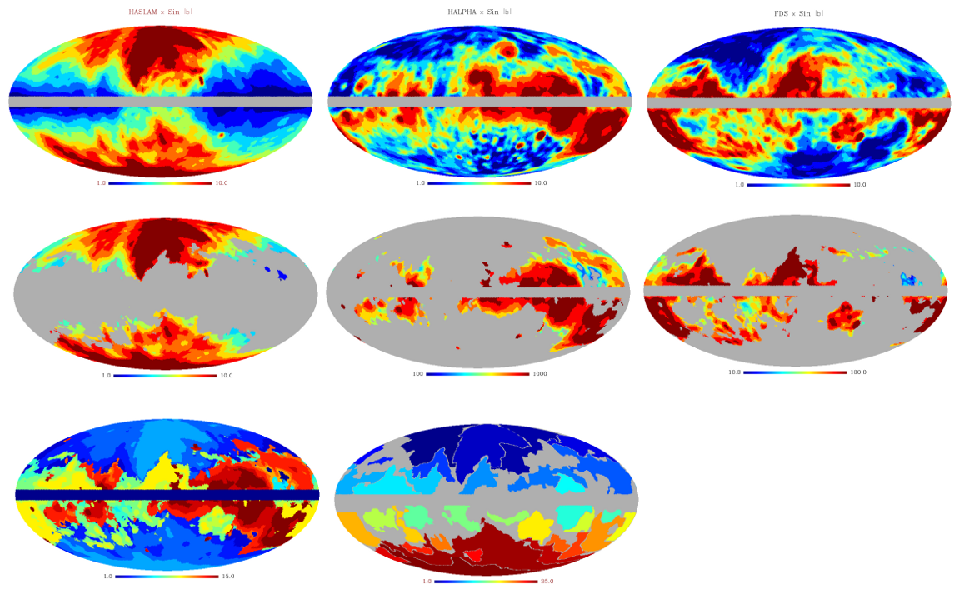,width=0.94\linewidth,angle=0,clip=} 
\end{tabular}
\caption{Definition of the new regions as
  described in \S~\ref{sec:region_def}. \textit{Top row}: the
  sine-modulated external synchrotron, free-free and dust templates at $3^{\circ}$ resolution,
  each rescaled into ten roughly isotemperature
  regions of equal area. \textit{Second row}: regions of the sky where
  where each modulated template is considered morphologically dominant.
 \textit{Last row}: the left-hand plot shows the merged map from combining and maps from
  the middle row.  The central plot defines 35 regions constructed
  from the merged maps.  These cover almost the whole sky
  excluding the $\pm 5^{\circ}$ close to the Galactic plane and some areas where
  the method resulted in many very small regions that are difficult to
  merge meaningfully. Note that two of the regions are eventually
  excluded from analysis when the KQ85 mask is also imposed on the sky.
\label{fig:defining_regions}
}
\end{center}
\end{figure*}

\begin{table*}
\scriptsize
\begin{center}
\begin{tabular}{cccrrr}
\hline
Region & \multicolumn{2}{c}{No. of Pixels (in $N_{side}=64$ map)} &  \multicolumn{2}{c}{Angular Coordinates}  & Centroid of Region \\
& Default & Surviving EBV mask & $l$ (in degrees) &  $b$ (in degrees) & $l,b$ (in degrees)\\
\hline  
  1     &  2047     &  1877     &    41$^{\circ}$.5    to   178$^{\circ}$.5     &    32$^{\circ}$.8    to    88$^{\circ}$.5    &(95$^{\circ}$.0, 62$^{\circ}$.2)\\
  2     &  1611     &  1433     &   190$^{\circ}$.7    to   353$^{\circ}$.7     &    25$^{\circ}$.3    to    87$^{\circ}$.8    &(302$^{\circ}$.2, 59$^{\circ}$.2)\\
  3     &  2825     &  2508     &   286$^{\circ}$.9    to    60$^{\circ}$.9     &    15$^{\circ}$.7    to    88$^{\circ}$.5    &(8$^{\circ}$.1, 58$^{\circ}$.3)\\
  4     &   673     &   590     &   180$^{\circ}$.8    to   250$^{\circ}$.1     &    37$^{\circ}$.9    to    67$^{\circ}$.9    &(202$^{\circ}$.1, 55$^{\circ}$.5)\\
  5     &   168     &   145     &   166$^{\circ}$.8    to   179$^{\circ}$.3     &    35$^{\circ}$.0    to    55$^{\circ}$.9    &(174$^{\circ}$.3, 45$^{\circ}$.2)\\
  6     &   670     &   585     &    39$^{\circ}$.4    to    95$^{\circ}$.6     &    17$^{\circ}$.6    to    52$^{\circ}$.0    &(67$^{\circ}$.6, 33$^{\circ}$.2)\\
  7     &  1624     &  1507     &   181$^{\circ}$.4    to   251$^{\circ}$.0     &     6$^{\circ}$.0    to    56$^{\circ}$.6    &(208$^{\circ}$.5, 30$^{\circ}$.4)\\
  8     &  1026     &   920     &   239$^{\circ}$.1    to   333$^{\circ}$.3     &     8$^{\circ}$.4    to    34$^{\circ}$.2    &(287$^{\circ}$.6, 22$^{\circ}$.3)\\
  9     &   775     &   260     &   345$^{\circ}$.9    to    31$^{\circ}$.6     &     6$^{\circ}$.0    to    41$^{\circ}$.8    &(7$^{\circ}$.7, 20$^{\circ}$.9)\\
 10     &   702     &   575     &   128$^{\circ}$.7    to   179$^{\circ}$.3     &     6$^{\circ}$.0    to    34$^{\circ}$.2    &(160$^{\circ}$.4, 18$^{\circ}$.9)\\
 11     &   696     &   594     &    55$^{\circ}$.5    to   111$^{\circ}$.1     &     6$^{\circ}$.0    to    35$^{\circ}$.0    &(75$^{\circ}$.7, 18$^{\circ}$.2)\\
 12     &  1093     &   279     &    79$^{\circ}$.5    to   158$^{\circ}$.9     &     6$^{\circ}$.0    to    33$^{\circ}$.5    &(121$^{\circ}$.0, 14$^{\circ}$.2)\\
 13     &   421     &   373     &   243$^{\circ}$.3    to   277$^{\circ}$.0     &     6$^{\circ}$.0    to    25$^{\circ}$.3    &(260$^{\circ}$.8, 13$^{\circ}$.9)\\
 14     &   704     &   551     &   234$^{\circ}$.1    to   277$^{\circ}$.0     &   -28$^{\circ}$.6    to    -6$^{\circ}$.0    &(257$^{\circ}$.5,-14$^{\circ}$.8)\\
 15     &   598     &   502     &    66$^{\circ}$.1    to    99$^{\circ}$.8     &   -28$^{\circ}$.0    to    -6$^{\circ}$.0    &(84$^{\circ}$.6,-16$^{\circ}$.6)\\
 16     &   597     &   479     &     9$^{\circ}$.1    to    53$^{\circ}$.4     &   -30$^{\circ}$.7    to    -6$^{\circ}$.0    &(31$^{\circ}$.3,-17$^{\circ}$.2)\\
 17     &   290     &   269     &   308$^{\circ}$.7    to   358$^{\circ}$.6     &   -24$^{\circ}$.0    to   -10$^{\circ}$.8    &(330$^{\circ}$.2,-18$^{\circ}$.0)\\
 18     &   627     &   573     &   109$^{\circ}$.7    to   149$^{\circ}$.1     &   -30$^{\circ}$.7    to    -9$^{\circ}$.0    &(130$^{\circ}$.0,-18$^{\circ}$.9)\\
 19     &   330     &   275     &   229$^{\circ}$.9    to   258$^{\circ}$.7     &   -30$^{\circ}$.7    to   -10$^{\circ}$.2    &(238$^{\circ}$.8,-22$^{\circ}$.5)\\
 20     &   286     &   135     &   181$^{\circ}$.4    to   192$^{\circ}$.7     &   -39$^{\circ}$.5    to   -12$^{\circ}$.6    &(186$^{\circ}$.2,-26$^{\circ}$.1)\\
 21     &   899     &   703     &   276$^{\circ}$.3    to   319$^{\circ}$.2     &   -44$^{\circ}$.2    to   -10$^{\circ}$.8    &(297$^{\circ}$.7,-27$^{\circ}$.5)\\
 22     &   392     &   354     &    87$^{\circ}$.2    to   111$^{\circ}$.1     &   -41$^{\circ}$.8    to    -6$^{\circ}$.0    &(101$^{\circ}$.2,-29$^{\circ}$.2)\\
 23     &   862     &   496     &   148$^{\circ}$.4    to   179$^{\circ}$.3     &   -45$^{\circ}$.8    to   -12$^{\circ}$.6    &(166$^{\circ}$.8,-28$^{\circ}$.6)\\
 24     &   918     &   812     &   180$^{\circ}$.8    to   228$^{\circ}$.5     &   -51$^{\circ}$.3    to   -12$^{\circ}$.6    &(208$^{\circ}$.0,-34$^{\circ}$.4)\\
 25     &   242     &   213     &    54$^{\circ}$.5    to    77$^{\circ}$.9     &   -49$^{\circ}$.7    to   -34$^{\circ}$.2    &(66$^{\circ}$.0,-43$^{\circ}$.5)\\
 26     &   258     &   231     &   235$^{\circ}$.2    to   272$^{\circ}$.2     &   -52$^{\circ}$.0    to   -34$^{\circ}$.2    &(255$^{\circ}$.2,-44$^{\circ}$.4)\\
 27     &   395     &   348     &   197$^{\circ}$.4    to   239$^{\circ}$.1     &   -61$^{\circ}$.2    to   -30$^{\circ}$.0    &(225$^{\circ}$.1,-46$^{\circ}$.7)\\
 28     &   180     &   160     &    87$^{\circ}$.3    to   123$^{\circ}$.9     &   -55$^{\circ}$.1    to   -42$^{\circ}$.6    &(106$^{\circ}$.5,-50$^{\circ}$.2)\\
 29     &   159     &   144     &     7$^{\circ}$.9    to    35$^{\circ}$.8     &   -63$^{\circ}$.4    to   -50$^{\circ}$.5    &(19$^{\circ}$.8,-56$^{\circ}$.8)\\
 30     &   195     &   182     &   158$^{\circ}$.1    to   179$^{\circ}$.1     &   -64$^{\circ}$.9    to   -48$^{\circ}$.1    &(168$^{\circ}$.7,-56$^{\circ}$.9)\\
 31     &   579     &   523     &   181$^{\circ}$.0    to   305$^{\circ}$.2     &   -67$^{\circ}$.2    to   -38$^{\circ}$.7    &(250$^{\circ}$.2,-59$^{\circ}$.9)\\
 32     &  3988     &  3607     &   181$^{\circ}$.2    to   148$^{\circ}$.6     &   -76$^{\circ}$.8    to   -18$^{\circ}$.2    &(357$^{\circ}$.3,-67$^{\circ}$.5)\\
 33     &  1507     &  1363     &   182$^{\circ}$.6    to   178$^{\circ}$.5     &   -89$^{\circ}$.3    to   -4$^{\circ}$5.0    &(109$^{\circ}$.1,-82$^{\circ}$.4)\\
\hline
\end{tabular}
\end{center}
\caption{Definition of regions for the template fit analysis. The
  number of pixels contained in the region by default, and after
  application of the EBV mask are shown, together with the range
  of latitude and longitude values spanning a given region, and the
  angular coordinates of the region centroid. \label{tab:summary_regions}}
\end{table*}

\clearpage
\newpage
\section{Full Fit Results}
\label{app:complete_results}

Tables~\ref{tab:results_synch_33regions}, \ref{tab:results_freefree_33regions} \& \ref{tab:results_dust_33regions}
list the solutions for all 33 regions, for the template fit coefficients of the three
Galactic foreground components of synchrotron, free-free and dust
emission respectively. Discussions of these basic results can be found in
Section~\ref{sec:results}, and model fits to the coefficients are detailed and
discussed in Section~\ref{sec:modelfits}.

\begin{table*}
\scriptsize 
\begin{tabular}{|lccccc|}
\hline
 & \multicolumn{5}{|c|}{\bf Synchrotron} \\
\hline
Region & K &  Ka & Q & V & W \\
\hline
EBV        &    8.53      $^{\pm    0.20         }$&    2.83      $^{\pm    0.20         }$&    1.45      $^{\pm    0.19         }$&    0.42      $^{\pm    0.19         }$&    0.10      $^{\pm    0.17       }$\\
KQ85     &    8.20      $^{\pm    0.23         }$&    2.63      $^{\pm    0.23         }$&    1.32      $^{\pm    0.23         }$&    0.28      $^{\pm    0.22         }$&   -0.02      $^{\pm    0.19       }$\\
NPS          &    5.06      $^{\pm    0.62         }$&    1.48$^{\pm   0.61         }$&    0.63      $^{\pm    0.60         }$& 0.08      $^{\pm    0.58         }$&   -0.02      $^{\pm    0.52       }$\\
GN          &    8.73      $^{\pm    0.28         }$&    2.92      $^{\pm    0.28         }$&    1.53      $^{\pm    0.27         }$&    0.49      $^{\pm    0.26         }$&    0.13      $^{\pm    0.24       }$\\
GN$_{\rm reduced}$ &    9.46      $^{\pm    0.31         }$&    3.26      $^{\pm    0.30         }$&    1.77      $^{\pm    0.30         }$&    0.64      $^{\pm    0.29         }$&    0.22      $^{\pm    0.26       }$\\
EN   &    6.94      $^{\pm    0.29         }$&    2.00      $^{\pm    0.28         }$&    0.82      $^{\pm    0.28         }$&   -0.06      $^{\pm    0.27         }$&   -0.28      $^{\pm    0.24       }$\\
EN$_{\rm reduced}$&    7.38      $^{\pm    0.32         }$&    2.19      $^{\pm    0.31         }$&    0.93      $^{\pm    0.31         }$&   -0.01      $^{\pm    0.30         }$&   -0.27      $^{\pm    0.26       }$\\
GS  &    8.39      $^{\pm    0.29         }$&    2.81      $^{\pm    0.28         }$&    1.43      $^{\pm    0.28         }$&    0.36      $^{\pm    0.27         }$&    0.06      $^{\pm    0.24       }$\\
ES  &    8.49      $^{\pm    0.25         }$&    2.96      $^{\pm    0.25         }$&    1.65      $^{\pm    0.24         }$&    0.67      $^{\pm    0.23         }$&    0.35      $^{\pm    0.21       }$\\
\hline 
 1           &     3.7      $^{\pm     1.9         }$&     0.1      $^{\pm     1.9         }$&    -0.5      $^{\pm     1.8         }$&    -0.8      $^{\pm     1.8         }$&    -0.9      $^{\pm     1.6       }$\\
 2           &     7.3      $^{\pm     1.4         }$&     2.5      $^{\pm     1.4         }$&     1.1      $^{\pm     1.4         }$&     0.3      $^{\pm     1.3         }$&    -0.1      $^{\pm     1.2       }$\\
 3           &     5.0      $^{\pm     0.6         }$&     1.6      $^{\pm     0.6         }$&     0.8      $^{\pm     0.6         }$&     0.3      $^{\pm     0.5         }$&     0.1      $^{\pm     0.5       }$\\
 4           &     7.7      $^{\pm     4.2         }$&     4.2      $^{\pm     4.1         }$&     3.4      $^{\pm     4.0         }$&     2.2      $^{\pm     3.9         }$&     2.0      $^{\pm     3.5       }$\\
 5           &     5.3      $^{\pm     6.3         }$&     0.1      $^{\pm     6.2         }$&    -0.0      $^{\pm     6.0         }$&    -0.9      $^{\pm     5.9         }$&    -1.8      $^{\pm     5.3       }$\\
 6           &     3.7      $^{\pm     1.7         }$&     0.6      $^{\pm     1.7         }$&    -0.1      $^{\pm     1.7         }$&    -0.5      $^{\pm     1.6         }$&    -0.7      $^{\pm     1.4       }$\\
 7           &     4.5      $^{\pm     2.0         }$&     0.6      $^{\pm     2.0         }$&    -0.4      $^{\pm     1.9         }$&    -1.0      $^{\pm     1.9         }$&    -1.2      $^{\pm     1.7       }$\\
 8           &     9.9      $^{\pm     0.6         }$&     3.2      $^{\pm     0.6         }$&     1.7      $^{\pm     0.6         }$&     0.5      $^{\pm     0.5         }$&     0.1      $^{\pm     0.5       }$\\
 9           &     3.1      $^{\pm     1.3         }$&     0.7      $^{\pm     1.3         }$&     0.1      $^{\pm     1.2         }$&    -0.4      $^{\pm     1.2         }$&    -0.6      $^{\pm     1.1       }$\\
10           &     8.4      $^{\pm     2.5         }$&     5.5      $^{\pm     2.5         }$&     5.2      $^{\pm     2.4         }$&     4.6      $^{\pm     2.3         }$&     3.9      $^{\pm     2.1       }$\\
11           &    13.9      $^{\pm     1.4         }$&     3.2      $^{\pm     1.4         }$&     0.8      $^{\pm     1.4         }$&    -0.7      $^{\pm     1.3         }$&    -1.2      $^{\pm     1.2       }$\\
12           &     7.8      $^{\pm     2.1         }$&     3.2      $^{\pm     2.1         }$&     2.2      $^{\pm     2.0         }$&     1.1      $^{\pm     1.9         }$&     0.8      $^{\pm     1.7       }$\\
13           &     2.6      $^{\pm     1.9         }$&    -0.5      $^{\pm     1.9         }$&    -1.1      $^{\pm     1.9         }$&    -1.2      $^{\pm     1.8         }$&    -1.3      $^{\pm     1.6       }$\\
14           &    -2.3      $^{\pm     2.1         }$&    -3.2      $^{\pm     2.1         }$&    -3.0      $^{\pm     2.1         }$&    -2.2      $^{\pm     2.0         }$&    -1.6      $^{\pm     1.8       }$\\
15           &    11.0      $^{\pm     1.8         }$&     2.1      $^{\pm     1.7         }$&    -0.3      $^{\pm     1.7         }$&    -1.8      $^{\pm     1.6         }$&    -2.2      $^{\pm     1.5       }$\\
16           &     6.3      $^{\pm     0.7         }$&     2.1      $^{\pm     0.7         }$&     1.1      $^{\pm     0.7         }$&     0.3      $^{\pm     0.7         }$&     0.1      $^{\pm     0.6       }$\\
17           &     7.2      $^{\pm     1.4         }$&     1.4      $^{\pm     1.4         }$&     0.1      $^{\pm     1.4         }$&    -1.1      $^{\pm     1.3         }$&    -1.1      $^{\pm     1.2       }$\\
18           &     8.5      $^{\pm     2.2         }$&     4.0      $^{\pm     2.2         }$&     3.0      $^{\pm     2.2         }$&     1.7      $^{\pm     2.1         }$&     1.5      $^{\pm     1.8       }$\\
19           &     2.3      $^{\pm     5.0         }$&    -0.2      $^{\pm     4.9         }$&    -0.9      $^{\pm     4.8         }$&    -2.7      $^{\pm     4.7         }$&    -2.3      $^{\pm     4.2       }$\\
20           &     4.9      $^{\pm     4.2         }$&     2.1      $^{\pm     4.1         }$&     2.0      $^{\pm     4.1         }$&     1.1      $^{\pm     3.9         }$&     0.7      $^{\pm     3.5       }$\\
21           &     6.2      $^{\pm     1.9         }$&     1.9      $^{\pm     1.8         }$&     1.0      $^{\pm     1.8         }$&     0.2      $^{\pm     1.7         }$&    -0.0      $^{\pm     1.5       }$\\
22           &     6.1      $^{\pm     2.7         }$&     2.9      $^{\pm     2.7         }$&     2.3      $^{\pm     2.6         }$&     1.6      $^{\pm     2.5         }$&     1.2      $^{\pm     2.3       }$\\
23           &     2.6      $^{\pm     2.9         }$&    -1.2      $^{\pm     2.8         }$&    -2.1      $^{\pm     2.8         }$&    -2.6      $^{\pm     2.7         }$&    -2.7      $^{\pm     2.4       }$\\
24           &    12.8      $^{\pm     2.5         }$&     6.8      $^{\pm     2.4         }$&     5.0      $^{\pm     2.4         }$&     3.5      $^{\pm     2.3         }$&     2.6      $^{\pm     2.1       }$\\
25           &     2.0      $^{\pm     3.6         }$&    -0.3      $^{\pm     3.5         }$&    -0.5      $^{\pm     3.5         }$&    -1.0      $^{\pm     3.4         }$&    -0.9      $^{\pm     3.0       }$\\
26           &    12.0      $^{\pm     5.7         }$&     5.4      $^{\pm     5.7         }$&     3.9      $^{\pm     5.4         }$&     2.9      $^{\pm     5.3         }$&     2.6      $^{\pm     4.8       }$\\
27           &    14.5      $^{\pm     5.3         }$&     7.8      $^{\pm     5.2         }$&     6.3      $^{\pm     5.1         }$&     4.3      $^{\pm     4.9         }$&     3.8      $^{\pm     4.4       }$\\
28           &     2.8      $^{\pm     6.3         }$&    -1.6      $^{\pm     6.2         }$&    -2.6      $^{\pm     6.1         }$&    -4.0      $^{\pm     5.9         }$&    -3.5      $^{\pm     5.3       }$\\
29           &    13.0      $^{\pm     6.3         }$&     7.0      $^{\pm     6.2         }$&     5.7      $^{\pm     6.1         }$&     4.2      $^{\pm     5.9         }$&     3.1      $^{\pm     5.3       }$\\
30           &     2.3      $^{\pm     5.9         }$&    -1.3      $^{\pm     5.8         }$&    -2.1      $^{\pm     5.7         }$&    -2.4      $^{\pm     5.4         }$&    -2.4      $^{\pm     4.8       }$\\
31           &     9.7      $^{\pm     3.8         }$&     0.8      $^{\pm     3.8         }$&    -1.5      $^{\pm     3.7         }$&    -2.8      $^{\pm     3.6         }$&    -2.8      $^{\pm     3.2}$\\
32           &     7.1      $^{\pm     1.1         }$&     2.4      $^{\pm     1.1         }$&     1.3      $^{\pm     1.1         }$&     0.3      $^{\pm     1.0         }$&     0.2      $^{\pm     0.9       }$\\
33           &     8.9      $^{\pm     3.1         }$&     5.1      $^{\pm     3.1         }$&     4.0      $^{\pm     3.0         }$&     3.1      $^{\pm     2.9         }$&     2.4      $^{\pm     2.6       }$\\
\hline
 Average &    6.92      $^{\pm    0.26         }$&    2.11      $^{\pm    0.26         }$&    1.00      $^{\pm    0.25         }$&    0.15      $^{\pm    0.24         }$&   -0.08      $^{\pm    0.22       }$\\
\hline
\end{tabular}
\caption{Template fit coefficients between the \emph{WMAP} data and the \citet{Haslam:1982} 408~MHz data, used as a proxy for Galactic 
emission due to synchrotron radiation, in units of $\mathrm {\mu K\ K^{-1}_{408~MHz}}$. Results are provided for the global regions defined
in Section~\ref{sec:masks} and for the 33 sky regions defined in Section~\ref{sec:region_def}. Monopole and dipole terms are also fitted 
simultaneously. For the global fits, NPS -- North Polar Spur, GN -- Galactic North, GN$_{\rm reduced}$ -- Galactic North with the NPS removed, EN -- Ecliptic North, EN$_{\rm reduced}$ -- Ecliptic North with the NPS removed, GS -- Galactic South, ES -- Ecliptic South. The average
of the 33 regions is  also tabulated for convenience. 
\label{tab:results_synch_33regions}
}
\end{table*}

\begin{table*}
\scriptsize 
\begin{tabular}{|lccccc|}
\hline
 & \multicolumn{5}{|c|}{\bf Free-Free} \\
\hline
Region & K &  Ka & Q & V & W \\
\hline
EBV        &    9.80      $^{\pm    0.10         }$&    4.78      $^{\pm    0.09         }$&    3.11      $^{\pm    0.09         }$&    1.29      $^{\pm    0.09         }$&    0.48      $^{\pm    0.08       }$\\
KQ85            &   10.43      $^{\pm    0.20         }$&    5.14      $^{\pm    0.20         }$&    3.40      $^{\pm    0.19         }$&    1.45      $^{\pm    0.19         }$&    0.62      $^{\pm    0.17       }$\\
NPS         &   20.99      $^{\pm   12.11         }$&   11.61$^{\pm   11.94         }$&   10.35      $^{\pm   11.74         }$& 7.14      $^{\pm   11.33         }$&    5.29      $^{\pm   10.14       }$\\
GN  &   10.31      $^{\pm    0.18         }$&    4.99      $^{\pm    0.18         }$&    3.20      $^{\pm    0.17         }$&    1.16      $^{\pm    0.17         }$&    0.27      $^{\pm    0.15       }$\\
GN$_{\rm reduced}$ &   10.27      $^{\pm    0.18         }$&    4.97      $^{\pm    0.18         }$&    3.18      $^{\pm    0.17         }$&    1.15      $^{\pm    0.17         }$&    0.26      $^{\pm    0.15       }$\\
EN  &   13.22      $^{\pm    0.26         }$&    6.52      $^{\pm    0.26         }$&    4.32      $^{\pm    0.25         }$&    1.76      $^{\pm    0.24         }$&    0.59      $^{\pm    0.22       }$\\
EN$_{\rm reduced}$ &   13.12      $^{\pm    0.26         }$&    6.48      $^{\pm    0.26         }$&    4.29      $^{\pm    0.25         }$&    1.75      $^{\pm    0.24         }$&    0.58      $^{\pm    0.22       }$\\
GS  &    9.81      $^{\pm    0.12         }$&    4.79      $^{\pm    0.12         }$&    3.13      $^{\pm    0.12         }$&    1.37      $^{\pm    0.11         }$&    0.57      $^{\pm    0.10       }$\\
ES  &    9.22      $^{\pm    0.11         }$&    4.47      $^{\pm    0.11         }$&    2.89      $^{\pm    0.10         }$&    1.20      $^{\pm    0.10         }$&    0.44      $^{\pm    0.09       }$\\
\hline
 1           &   -16.4      $^{\pm    15.6         }$&   -15.4      $^{\pm    15.4         }$&   -15.1      $^{\pm    15.0         }$&   -14.9      $^{\pm    14.7         }$&   -13.9      $^{\pm    13.3     } $\\
 2           &    11.2      $^{\pm     5.4         }$&     7.2      $^{\pm     5.3         }$&     6.7      $^{\pm     5.2         }$&     4.3      $^{\pm     5.0         }$&     3.3      $^{\pm     4.4       }$\\
 3           &    11.8      $^{\pm     7.7         }$&     4.1      $^{\pm     7.6         }$&     2.8      $^{\pm     7.5         }$&    -0.4      $^{\pm     7.2         }$&    -1.4      $^{\pm     6.5       }$\\
 4           &    17.9      $^{\pm    23.8         }$&     8.2      $^{\pm    23.5         }$&     3.7      $^{\pm    23.0         }$&    -1.2      $^{\pm    22.2         }$&    -4.3      $^{\pm    19.9       }$\\
 5           &    62.3      $^{\pm    45.0         }$&    63.8      $^{\pm    44.4         }$&    65.4      $^{\pm    43.2         }$&    65.0      $^{\pm    42.3         }$&    60.1      $^{\pm    38.2       }$\\
 6           &    13.7      $^{\pm     9.4         }$&     9.0      $^{\pm     9.3         }$&     6.6      $^{\pm     9.0         }$&     5.0      $^{\pm     8.7         }$&     4.2      $^{\pm     7.9       }$\\
 7           &     9.4      $^{\pm     2.1         }$&     2.7      $^{\pm     2.1         }$&     0.7      $^{\pm     2.1         }$&    -1.4      $^{\pm     2.1         }$&    -2.0      $^{\pm     1.9       }$\\
 8           &     6.9      $^{\pm     1.6         }$&     3.8      $^{\pm     1.5         }$&     2.2      $^{\pm     1.5         }$&     0.6      $^{\pm     1.5         }$&    -0.1      $^{\pm     1.3       }$\\
 9           &     9.9      $^{\pm     0.3         }$&     5.3      $^{\pm     0.3         }$&     3.7      $^{\pm     0.3         }$&     1.5      $^{\pm     0.3         }$&     0.4      $^{\pm     0.3       }$\\
10           &    -1.8      $^{\pm     5.5         }$&    -4.5      $^{\pm     5.4         }$&    -6.2      $^{\pm     5.3         }$&    -6.5      $^{\pm     5.2         }$&    -5.8      $^{\pm     4.6       }$\\
11           &    25.0      $^{\pm     0.8         }$&    11.1      $^{\pm     0.8         }$&     7.0      $^{\pm     0.8         }$&     3.3      $^{\pm     0.8         }$&     1.9      $^{\pm     0.7       }$\\
11 (KQ85)&    15.8      $^{\pm    1.3         }$&    7.1      $^{\pm    1.3         }$&    4.4      $^{\pm    1.3         }$&    1.8      $^{\pm    1.3         }$&    1.0      $^{\pm    1.1       }$\\
12           &     9.9      $^{\pm     2.5         }$&     3.3      $^{\pm     2.4         }$&     1.0      $^{\pm     2.4         }$&    -1.3      $^{\pm     2.3         }$&    -1.9      $^{\pm     2.1       }$\\
13           &     9.3      $^{\pm     0.4         }$&     4.3      $^{\pm     0.4         }$&     2.7      $^{\pm     0.4         }$&     1.1      $^{\pm     0.4         }$&     0.4      $^{\pm     0.3       }$\\
14           &     9.8      $^{\pm     0.2         }$&     4.8      $^{\pm     0.2         }$&     3.2      $^{\pm     0.2         }$&     1.4      $^{\pm     0.2         }$&     0.6      $^{\pm     0.1       }$\\
15           &    12.3      $^{\pm     0.8         }$&     6.2      $^{\pm     0.7         }$&     4.3      $^{\pm     0.7         }$&     2.1      $^{\pm     0.7         }$&     1.2      $^{\pm     0.6       }$\\
16           &     7.3      $^{\pm     5.2         }$&     3.8      $^{\pm     5.1         }$&     2.3      $^{\pm     5.0         }$&     0.3      $^{\pm     4.8         }$&     0.2      $^{\pm     4.3       }$\\
17           &    -0.4      $^{\pm     6.0         }$&     0.8      $^{\pm     5.9         }$&     0.7      $^{\pm     5.8         }$&     0.4      $^{\pm     5.6         }$&    -0.3      $^{\pm     5.0       }$\\
18           &    13.8      $^{\pm     2.0         }$&     8.1      $^{\pm     2.0         }$&     5.8      $^{\pm     2.0         }$&     3.6      $^{\pm     1.9         }$&     2.2      $^{\pm     1.7       }$\\
19           &     5.7      $^{\pm     2.6         }$&     2.9      $^{\pm     2.6         }$&     1.4      $^{\pm     2.5         }$&     0.0      $^{\pm     2.5         }$&    -0.4      $^{\pm     2.2       }$\\
20           &     8.0      $^{\pm     0.9         }$&     4.4      $^{\pm     0.9         }$&     3.6      $^{\pm     0.9         }$&     2.2      $^{\pm     0.9         }$&     1.5      $^{\pm     0.8       }$\\
21           &    13.6      $^{\pm     3.2         }$&     8.9      $^{\pm     3.1         }$&     7.6      $^{\pm     3.0         }$&     5.3      $^{\pm     2.9         }$&     3.8      $^{\pm     2.6       }$\\
22           &     7.8      $^{\pm     5.8         }$&     1.6      $^{\pm     5.7         }$&     0.4      $^{\pm     5.6         }$&    -1.5      $^{\pm     5.4         }$&    -2.7      $^{\pm     4.9       }$\\
23           &    10.8      $^{\pm     1.7         }$&     4.7      $^{\pm     1.6         }$&     2.8      $^{\pm     1.6         }$&     0.6      $^{\pm     1.6         }$&    -0.4      $^{\pm     1.4       }$\\
24           &     7.4      $^{\pm     0.4         }$&     3.7      $^{\pm     0.4         }$&     2.4      $^{\pm     0.4         }$&     1.0      $^{\pm     0.4         }$&     0.5      $^{\pm     0.4       }$\\
25           &     2.4      $^{\pm    13.2         }$&     1.2      $^{\pm    13.0         }$&    -1.1      $^{\pm    12.8         }$&    -1.6      $^{\pm    12.3         }$&    -3.1      $^{\pm    11.0       }$\\
26           &    11.7      $^{\pm    22.5         }$&     0.1      $^{\pm    22.1         }$&    -3.4      $^{\pm    21.3         }$&    -8.2      $^{\pm    20.8         }$&    -5.4      $^{\pm    18.7       }$\\
27           &    -4.8      $^{\pm     6.2         }$&    -7.8      $^{\pm     6.1         }$&    -8.5      $^{\pm     6.0         }$&    -9.2      $^{\pm     5.8         }$&    -9.1      $^{\pm     5.2       }$\\
28           &   -25.7      $^{\pm    44.5         }$&   -30.0      $^{\pm    43.9         }$&   -30.8      $^{\pm    43.0         }$&   -30.3      $^{\pm    41.8         }$&   -33.9      $^{\pm    37.7       }$\\
29           &   -42.4      $^{\pm    23.4         }$&   -40.6      $^{\pm    23.0         }$&   -35.0      $^{\pm    22.6         }$&   -35.1      $^{\pm    22.2         }$&   -34.3      $^{\pm    20.2       }$\\
30           &    36.6      $^{\pm    48.8         }$&    27.6      $^{\pm    48.2         }$&    23.3      $^{\pm    47.2         }$&    18.0      $^{\pm    46.5         }$&    10.3      $^{\pm    42.4       }$\\
31           &   -19.0      $^{\pm    17.4         }$&   -27.9      $^{\pm    17.1         }$&   -36.1      $^{\pm    16.7         }$
&   -33.0      $^{\pm    16.1         }$&   -27.3      $^{\pm    14.4       }$\\
32           &     9.3      $^{\pm     3.0         }$&     5.9      $^{\pm     3.0         }$&     4.4      $^{\pm     2.9         }$&     3.1      $^{\pm     2.9         }$&     2.4      $^{\pm     2.6       }$\\
33           &     3.5      $^{\pm    14.9         }$&    -1.2      $^{\pm    14.7         }$&    -1.3      $^{\pm    14.5         }$&    -3.0      $^{\pm    13.9         }$&    -3.5      $^{\pm    12.4       }$\\
\hline
 Average &    9.92      $^{\pm    0.13         }$&    4.88      $^{\pm    0.13         }$&    3.23      $^{\pm    0.13         }$&    1.42      $^{\pm    0.12         }$&    0.58      $^{\pm    0.11       }$\\
(a) &    9.54      $^{\pm    0.13         }$&    4.72      $^{\pm    0.13         }$&    3.14      $^{\pm    0.13         }$&    1.37      $^{\pm    0.12         }$&    0.55      $^{\pm    0.11       }$\\
(b) &    9.60      $^{\pm    0.13         }$&    4.74      $^{\pm    0.13         }$&    3.15      $^{\pm    0.13         }$&    1.38      $^{\pm    0.12         }$&    0.56      $^{\pm    0.11       }$\\
\hline
\end{tabular}
\caption{Template fit coefficients between the \emph{WMAP} data and \citetalias{DDD:2003}  \halpha data, used as a proxy for Galactic 
emission due to free-free radiation, in units of $\mathrm {\mu K\ R^{-1}}$.  Details as per Table \ref{tab:results_synch_33regions}. Note
that, as described in Section~\ref{sec:dust_extinct},  region 11 shows evidence of anomalous behaviour in terms of its correlation with 
the \halpha template. We therefore additionally show the corresponding result when the KQ85 mask is imposed instead of the EBV mask. The 
average of the 33 regions is  also tabulated for convenience. (a) is a variant on this in which we  completely exclude the region 11 
contribution; (b) includes the region 11 (KQ85) values instead of the standard ones. \label{tab:results_freefree_33regions}}
\end{table*}

\begin{table*}
\scriptsize 
\begin{tabular}{|lccccc|}
\hline
 & \multicolumn{5}{|c|}{\bf Dust} \\
\hline
Region & K &  Ka & Q & V & W \\
\hline
EBV        &    9.84      $^{\pm    0.10         }$&    3.80      $^{\pm    0.10         }$&    2.16      $^{\pm    0.10         }$&    1.07      $^{\pm    0.09         }$&    1.24      $^{\pm    0.08       }$\\
KQ85            &    8.23      $^{\pm    0.09         }$&    3.15      $^{\pm    0.09         }$&    1.85      $^{\pm    0.09         }$&    1.06      $^{\pm    0.08         }$&    1.31      $^{\pm    0.08       }$\\
NPS           &    8.58      $^{\pm    1.57         }$&    3.16$^{\pm    1.54         }$&    1.75      $^{\pm    1.52         }$& 1.02      $^{\pm    1.46         }$&    1.61      $^{\pm    1.30       }$\\
GN   &    9.61      $^{\pm    0.16         }$&    3.57      $^{\pm    0.16         }$&    1.97      $^{\pm    0.15         }$&    0.97      $^{\pm    0.15         }$&    1.20      $^{\pm    0.13       }$\\
GN$_{\rm reduced}$ &    9.55      $^{\pm    0.16         }$&    3.55      $^{\pm    0.16         }$&    1.96      $^{\pm    0.15         }$&    0.96      $^{\pm    0.15         }$&    1.20      $^{\pm    0.13       }$\\
EN  &    8.97      $^{\pm    0.15         }$&    3.36      $^{\pm    0.15         }$&    1.88      $^{\pm    0.15         }$&    0.96      $^{\pm    0.14         }$&    1.18      $^{\pm    0.13       }$\\
EN$_{\rm reduced}$ &    8.85      $^{\pm    0.15         }$&    3.30      $^{\pm    0.15         }$&    1.86      $^{\pm    0.15         }$&    0.95      $^{\pm    0.14         }$&    1.17      $^{\pm    0.13       }$\\
GS  &   10.44      $^{\pm    0.14         }$&    4.14      $^{\pm    0.14         }$&    2.40      $^{\pm    0.14         }$&    1.20      $^{\pm    0.13         }$&    1.29      $^{\pm    0.12       }$\\
ES &   10.53      $^{\pm    0.14         }$&    4.12      $^{\pm    0.13         }$&    2.34      $^{\pm    0.13         }$&    1.13      $^{\pm    0.13         }$&    1.26      $^{\pm    0.11       }$\\
\hline
 1           &    10.1      $^{\pm     3.4         }$&     4.2      $^{\pm     3.4         }$&     2.7      $^{\pm     3.3         }$&     1.2      $^{\pm     3.2         }$&     1.6      $^{\pm     2.9       }$\\
 2           &     8.8      $^{\pm     1.8         }$&     2.2      $^{\pm     1.8         }$&     0.5      $^{\pm     1.8         }$&    -0.4      $^{\pm     1.7         }$&    -0.2      $^{\pm     1.5       }$\\
 3           &     9.5      $^{\pm     1.1         }$&     3.9      $^{\pm     1.0         }$&     2.5      $^{\pm     1.0         }$&     1.9      $^{\pm     1.0         }$&     2.2      $^{\pm     0.9       }$\\
 4           &     6.6      $^{\pm     5.2         }$&     3.6      $^{\pm     5.2         }$&     3.2      $^{\pm     5.1         }$&     2.9      $^{\pm     4.9         }$&     3.2      $^{\pm     4.3       }$\\
 5           &     5.1      $^{\pm     9.3         }$&     1.9      $^{\pm     9.1         }$&     0.5      $^{\pm     8.9         }$&     0.7      $^{\pm     8.7         }$&     1.4      $^{\pm     7.8       }$\\
 6           &     8.1      $^{\pm     2.6         }$&     2.9      $^{\pm     2.6         }$&     1.4      $^{\pm     2.5         }$&     0.7      $^{\pm     2.4         }$&     1.0      $^{\pm     2.2       }$\\
 7           &    10.5      $^{\pm     0.7         }$&     4.3      $^{\pm     0.7         }$&     2.6      $^{\pm     0.7         }$&     1.3      $^{\pm     0.7         }$&     1.5      $^{\pm     0.6       }$\\
 8           &     7.3      $^{\pm     0.7         }$&     2.8      $^{\pm     0.7         }$&     1.8      $^{\pm     0.7         }$&     1.3      $^{\pm     0.6         }$&     1.5      $^{\pm     0.6       }$\\
 9           &    14.0      $^{\pm     0.5         }$&     5.6      $^{\pm     0.5         }$&     3.2      $^{\pm     0.5         }$&     1.5      $^{\pm     0.5         }$&     1.5      $^{\pm     0.5       }$\\
10           &     7.1      $^{\pm     0.8         }$&     2.4      $^{\pm     0.7         }$&     1.2      $^{\pm     0.7         }$&     0.5      $^{\pm     0.7         }$&     0.8      $^{\pm     0.6       }$\\
11           &     4.5      $^{\pm     0.7         }$&     2.1      $^{\pm     0.7         }$&     1.2      $^{\pm     0.6         }$&     0.7      $^{\pm     0.6         }$&     1.0      $^{\pm     0.6       }$\\
12           &     5.8      $^{\pm     0.4         }$&     1.6      $^{\pm     0.4         }$&     0.6      $^{\pm     0.4         }$&     0.1      $^{\pm     0.4         }$&     0.5      $^{\pm     0.4       }$\\
13           &     6.1      $^{\pm     0.8         }$&     2.1      $^{\pm     0.7         }$&     1.1      $^{\pm     0.7         }$&     0.5      $^{\pm     0.7         }$&     0.9      $^{\pm     0.6       }$\\
14           &    10.4      $^{\pm     0.5         }$&     4.3      $^{\pm     0.5         }$&     2.6      $^{\pm     0.5         }$&     1.4      $^{\pm     0.5         }$&     1.5      $^{\pm     0.4       }$\\
15           &     8.9      $^{\pm     0.6         }$&     3.9      $^{\pm     0.6         }$&     2.5      $^{\pm     0.6         }$&     1.7      $^{\pm     0.6         }$&     1.9      $^{\pm     0.5       }$\\
16           &     7.0      $^{\pm     0.7         }$&     1.4      $^{\pm     0.7         }$&     0.2      $^{\pm     0.7         }$&    -0.4      $^{\pm     0.7         }$&    -0.0      $^{\pm     0.6       }$\\
17           &     7.2      $^{\pm     2.0         }$&     2.0      $^{\pm     2.0         }$&     0.5      $^{\pm     2.0         }$&    -0.2      $^{\pm     1.9         }$&     0.4      $^{\pm     1.7       }$\\
18           &     7.6      $^{\pm     1.0         }$&     2.4      $^{\pm     1.0         }$&     1.0      $^{\pm     1.0         }$&     0.4      $^{\pm     0.9         }$&     0.7      $^{\pm     0.8       }$\\
19           &    10.1      $^{\pm     2.1         }$&     5.1      $^{\pm     2.1         }$&     3.9      $^{\pm     2.1         }$&     3.5      $^{\pm     2.0         }$&     3.3      $^{\pm     1.8       }$\\
20           &    10.6      $^{\pm     0.8         }$&     3.9      $^{\pm     0.8         }$&     2.0      $^{\pm     0.8         }$&     0.9      $^{\pm     0.8         }$&     1.1      $^{\pm     0.7       }$\\
21           &     7.5      $^{\pm     0.5         }$&     2.7      $^{\pm     0.5         }$&     1.5      $^{\pm     0.5         }$&     0.8      $^{\pm     0.4         }$&     1.1      $^{\pm     0.4       }$\\
22           &     9.4      $^{\pm     0.9         }$&     4.2      $^{\pm     0.9         }$&     2.6      $^{\pm     0.9         }$&     1.7      $^{\pm     0.9         }$&     2.0      $^{\pm     0.8       }$\\
23           &     9.6      $^{\pm     0.4         }$&     4.0      $^{\pm     0.4         }$&     2.4      $^{\pm     0.3         }$&     1.3      $^{\pm     0.3         }$&     1.4      $^{\pm     0.3       }$\\
24           &     8.4      $^{\pm     0.6         }$&     2.4      $^{\pm     0.6         }$&     0.9      $^{\pm     0.6         }$&    -0.0      $^{\pm     0.6         }$&     0.2      $^{\pm     0.5       }$\\
25           &     4.9      $^{\pm     2.2         }$&     0.4      $^{\pm     2.1         }$&    -0.5      $^{\pm     2.1         }$&    -1.2      $^{\pm     2.0         }$&    -0.9      $^{\pm     1.8       }$\\
26           &     4.4      $^{\pm     7.1         }$&     3.8      $^{\pm     7.0         }$&     3.4      $^{\pm     6.7         }$&     3.6      $^{\pm     6.7         }$&     2.9      $^{\pm     6.0       }$\\
27           &    12.5      $^{\pm     4.4         }$&     6.3      $^{\pm     4.3         }$&     4.8      $^{\pm     4.2         }$&     3.9      $^{\pm     4.1         }$&     3.8      $^{\pm     3.7       }$\\
28           &     9.2      $^{\pm     2.0         }$&     4.4      $^{\pm     2.0         }$&     3.3      $^{\pm     2.0         }$&     2.9      $^{\pm     1.9         }$&     2.8      $^{\pm     1.7       }$\\
29           &     6.3      $^{\pm    13.4         }$&    -1.1      $^{\pm    13.2         }$&    -5.3      $^{\pm    13.0         }$&    -5.3      $^{\pm    12.6         }$&    -5.3      $^{\pm    11.3       }$\\
30           &    10.5      $^{\pm     3.5         }$&     5.0      $^{\pm     3.5         }$&     3.9      $^{\pm     3.4         }$&     2.8      $^{\pm     3.3         }$&     2.8      $^{\pm     2.9       }$\\
31           &    12.9      $^{\pm     3.6         }$&     6.4      $^{\pm     3.6         }$&     4.8      $^{\pm     3.5         }$
&     3.5      $^{\pm     3.4         }$&     3.4      $^{\pm     3.0       }$\\
32           &     8.8      $^{\pm     0.9         }$&     3.1      $^{\pm     0.9         }$&     1.9      $^{\pm     0.9         }$&     0.9      $^{\pm     0.9         }$&     1.0      $^{\pm     0.8       }$\\
33           &     9.2      $^{\pm     2.1         }$&     3.3      $^{\pm     2.0         }$&     1.5      $^{\pm     2.0         }$&     0.6      $^{\pm     1.9         }$&     1.0      $^{\pm     1.7       }$\\
\hline
Average &    8.62      $^{\pm    0.14         }$&    3.24      $^{\pm    0.14         }$&    1.81      $^{\pm    0.14         }$&    0.93      $^{\pm    0.13         }$&    1.13      $^{\pm    0.12       }$\\
\hline
\end{tabular}
\caption{Template fit coefficients between the \emph{WMAP} data and the \citetalias{FDS:1999} model 8 data, used as a proxy for Galactic 
emission due to dust radiative processes, in units of $\mathrm {\mu K\ \mu K^{-1}_{FDS8}}$. Details as per Table 
\ref{tab:results_synch_33regions}. \label{tab:results_dust_33regions}}
\end{table*}

\end{document}